\documentclass[aps,prb,twocolumn,superscriptaddress,nofootinbib]{revtex4-2}

\usepackage{amssymb}
\usepackage{amsmath}
\usepackage{mathrsfs}
\usepackage{mathtools}
\mathtoolsset{showonlyrefs=true}
\usepackage{bm}
\usepackage{braket}
\usepackage[T1]{fontenc}
\usepackage{newtxtext,newtxmath}
\usepackage{multirow}
\usepackage{booktabs}
\usepackage{diagbox}
\usepackage{enumitem}
\usepackage[svgnames]{xcolor}
\usepackage{float}
\usepackage{tikz}
\usepackage{graphicx}
\usepackage[
    setpagesize=false,
    bookmarks=false,
    colorlinks=true,
    allcolors=magenta
]{hyperref} 
\usepackage[titletoc]{appendix}

\newcommand{\eq}{\mathrm{eq}}
\newcommand{\ketbra}[2]{\ket{#1}\!\!\bra{#2}}
\newcommand{\sqrtiswap}{$\sqrt{\mbox{iSWAP}^\dagger}$}

\begin{document}
\title{Entanglement dynamics and performance of two-qubit gates \\ for superconducting qubits under non-Markovian effects}
\author{Kiyoto Nakamura}
\email{kiyoto.nakamura@uni-ulm.de}
\author{Joachim Ankerhold}
\affiliation{Institute for Complex Quantum Systems and IQST, Ulm University, D-89069 Ulm, Germany}

\date{\today}

\begin{abstract}
Within a numerically exact simulation technique, the dissipative dynamics of a two-qubit architecture is considered in which each qubit couples to its individual noise source (reservoir). The goal is to reveal the role of subtle qubit--reservoir correlations including non-Markovian processes as a prerequisite to guide further improvements of quantum computing devices.
This paper addresses the following three topics.
First, we examine the validity of the rotating wave approximation imposed previously on the qubit--reservoir coupling with respect to the disentanglement dynamics.
Second, generation of the entanglement as well as destruction are analyzed by monitoring the reduced dynamics during and after application of a \sqrtiswap{} gate, also focusing on memory effects caused by reservoirs.
Finally, the performance of a Hadamard + CNOT sequence is analyzed for different gate decomposition schemes.
In all  three cases, various types of noise sources and qubit parameters are considered.
\end{abstract}
\maketitle

\section{Introduction} \label{sec:introduction}
The performance of quantum computing devices has improved remarkably in the last decade.
Especially for superconducting qubits, much longer coherence times~\cite{PlaceNATCOMMUN2021,WangNPJQI2022,WangPhysRevAppl2025,TuokkolaNATCOMMUN2025}, substantially enhanced fidelities of gate operations, and efficient readout~\cite{NegirneacPRL2021,SungPRX2021,KandalaPRL2021,LiPRX2024,WangPRL2025} have been achieved.
From the perspective of information science, quantum supremacy has been demonstrated under certain conditions~\cite{GoogleNATURE2019, GoogleNATURE2023, IBMNATURE2023}.
Stimulated by those advances, quantum algorithms for noisy intermediate-scale quantum devices have been implemented~\cite{HavlicekNATURE2019,KandalaNATURE2017,RossmannekJPCL2023,WillschQIP2020,PerezNPJQI2020,DanJCTC2025}.

These achievements result to a large extent from the considerable reduction of relevant noise sources, including, for example, thermal fluctuations, $1/f$ type of noise originating from two-level fluctuators (impurities)~\cite{IthierPRB2005}, and quasiparticle noise~\cite{RisteNATCOMMUN2013,CardaniNATCOMMUN2021,PanNATCOMMUN2022}.
However, with growing fidelities the need to further understand and mitigate or control even minute effects of residual degrees of freedom (reservoirs) has turned into the focus of current activities. 
For this purpose, high-precision numerical simulations for devices of a modest number of qubits and for sequences of single- and two-qubit operations are one prerequisite to guide further progress.
It has become clear by now that for subtle aspects of qubit--environment interaction including time-retarded effects (memory effects), conventionally adopted methods to describe the time evolution of dissipative quantum systems, such as, for example, the Bloch--Redfield and the Lindblad equation, are insufficient.
Hence, studies with more advanced methodologies have been conducted in recent years.
For example, gate performance~\cite{TuorilaPRR2019,BabuNPJQI2021,PapicARXIV2023,NakamuraPRR2024,GulacsiPRR2025}, non-Markovianity~\cite{GulasciPRB2022,NakamuraPRB2024}, and dynamical decoupling schemes~\cite{NakamuraPRB2025} have been studied for single-qubit systems. 
For multiple-qubit systems, error correction codes in the presence of noise~\cite{BabuPRR2023}, spatially correlated noise effects~\cite{GulacsiPRR2025-2}, and performance of elemental two-qubit gates~\cite{PapicARXIV2023} and realistic quantum algorithms~\cite{ZhangJCP2024} have been investigated.

\textcolor{black}{For dissipative two-qubit gate operations, a general and accurate dynamical description at low temperatures with the focus on entanglement properties in the presence of retarded reservoir feedback and effects beyond the rotating wave approximation (RWA) is still lacking.
Since the discovery of the ``entanglement sudden death'' (ESD)~\cite{YuPRL2004,YuOPTCOMMUN2006,YuPRL2006,LopezPRL2008} twenty years ago, a phenomenon where the entanglement between two qubits abruptly decays to zero, theoretical research on the disentanglement dynamics of two-qubit systems has been conducted actively.
However, in most of these studies, the RWA was imposed on the system--reservoir coupling~\cite{YuPRL2004,YuPRL2006,BellomoPRL2007,BellomoPRA2008,BellomoPHYSSCR2010,LopezPRL2008}.
Notably, the general importance of counter-rotating terms has been revealed within a perturbative treatment~\cite{CaoPRA2008} and based on more accurate methods including quasiadiabatic path-integral and hierarchical equations of motion (HEOM) approaches~\cite{MaPRA2012,WuNJP2013,DuanJCP2013,WangNJP2013,WuPRA2017,SunPRA2018}.
Now, these studies need to be extended to actual gate operations.}

\textcolor{black}{Noise effects on the entanglement of two-qubit systems beyond the Born--Markov approximation have been addressed~\cite{DijkstraPRL2010,WangPHYSICAA2011}, but the main focus was on overall trends in the long-time domain~\cite{PaladinoPHYSICAE2010,PaladinoNJP2011,D'ArrigoNJP2012} instead of microscopic characteristics during gate operations.
Performance of sequences consisting of elemental single- and two-qubit gates has also been studied~\cite{LossPRA1998,ThorwartPRA2001,ChengPRA2004,StorczPRA2005}, but most of those rely on some assumptions, for example, the Markov approximation and the weak-coupling approximation between the system and reservoir implying sufficiently elevated temperatures.
More rigorous simulations for realistic models, for example, two-levels systems with a fixed direction of the Zeeman splitting and switchable external fields for single- and two-qubit gates, are needed.
}

\textcolor{black}{Here, to tackle those problems, we utilize a numerically rigorous method, the free-pole hierarchical equations of motion (FP-HEOM)~\cite{XuPRL2022,Xu2026}, and study entanglement/disentanglement dynamics of two-qubit systems and the performance of two-qubit gates in the presence of a broad class of reservoirs.
The treatment of dissipation during the quantum dynamics method is system agnostic and does not rely on any assumption about the nature of the noise, except that it is predominantly Gaussian.
The strength of the method was reported in Ref.~\cite{NakamuraPRR2024}, in which the performance of sequences consisting of single-qubit gates and idlings has been analyzed thoroughly.
The respective two-qubit gates constitute along with the general single-qubit gates considered in Ref.~\cite{NakamuraPRR2024} a universal set of gate operation for general quantum information processing. While our findings are of general nature, we specifically have superconducting platforms in mind.}

\begin{figure}
    \centering
    \includegraphics[scale=0.7]{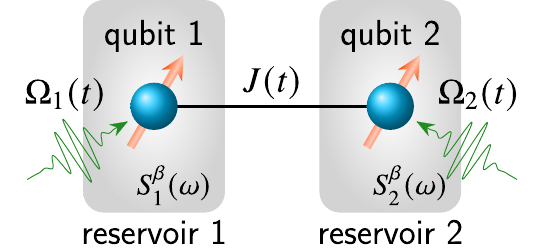}
    \caption{Schematic of the model we consider in this study.
    The qubits are directly coupled with each other with the strength $J (t)$, but each qubit couples with a different reservoir (gray rectangle). 
    Each reservoir is characterized by the spectral noise power $S_j^{\beta} (\omega)$.
    For single-qubit gates, external rotating fields are applied to each qubit [$\Omega_j (t)$].
    The external fields and the qubit--qubit coupling can be switched on and off according to gate operation.
    \label{fig:schematicModel}}
\end{figure}

Figure~\ref{fig:schematicModel} shows a schematic of the setting:
Each qubit couples independently to its private reservoir, characterized by a spectral noise power $S_j^\beta(\omega)$ ($j \in \{1, 2\}$) defined below.
While it is known that two qubits embedded in a common reservoir may become entangled~\cite{ManiscalcoPRL2008,D'ArrigoNJP2008,MazzolaPRA2009,KastPRB2014,HartmannQUANTUM2020}, we here consider the usual case where local noise sources have detrimental effect on global quantum correlations. Single-qubit gates are realized by imposing external rotating fields [$\Omega_j(t)$], while two-qubit gates are realized by a direct time-dependent coupling between the qubits [$J(t)$]
(For more details, see Sec.~\ref{sec:model}).

In more detail, we address the following three topics.
(I)~First, we revisit the \emph{disentanglement dynamics} by utilizing the FP-HEOM method.
Following previous studies, we prepare maximally entangled initial states and monitor the correlated dynamics with both the external field and the qubit--qubit coupling being off [$\Omega_1(t) = \Omega_2(t) = J(t) = 0$].
Compared to previous studies, we consider a broad class of noise sources.

(II)~Second, we focus on processes for the \emph{generation of entanglement} prior to disentanglement processes.
In those simulations, we first switch on the coupling $J$ and wait until the operation of a \sqrtiswap{} gate is completed.
Then, the direct coupling is switched off, and the decay of entanglement during the idle phase is monitored.
The dynamics during and after the application of the gate turn out to be correlated owing to the retarded reservoir feedback.


(III) Finally, a realistic sequence, namely, a \emph{Hadamard + CNOT sequence}, is considered.
In this case, both the single- [$\Omega_j (t)$] and two-qubit [$J (t)$] gates are switched on and off to realize the sequence.
Two different schemes to decompose this sequence into  elemental gates are compared by analyzing the fidelity and the entanglement.

This paper is organized as follows.
In Sec.~\ref{sec:model}, the Hamiltonian corresponding to the model in Fig.~\ref{fig:schematicModel} is introduced.
The method we use in this study, FP-HEOM, is also briefly explained.
Sections~\ref{sec:resultRWA}--\ref{sec:resultHCNOT} are devoted to numerical results and corresponding discussions.
The simulation settings in those sections correspond to the case (I)--(III) discussed above, respectively.
Finally, we draw conclusions in Sec.~\ref{sec:conclusion}.

\section{Model and method} \label{sec:model}
The two-qubit systems are described by the following Hamiltonian (Fig.~\ref{fig:schematicModel}):
\begin{align}
    \hat{H}_\mathrm{tot}(t) = & \hat{H}_1(t) + \hat{H}_2(t) + \hbar J(t) (\hat{\sigma}_1^{+}\hat{\sigma}_2^{-} + \hat{\sigma}_1^{-}\hat{\sigma}_2^{+})\, ,
    \label{eq:totH}
\end{align}
where $\hat{H}_j(t)$ ($j \in \{1, 2\}$) includes both system and reservoir Hamiltonian and reads
\begin{align}
    \hat{H}_j(t) = 
    & \frac{\hbar \Omega_j(t)}{2} [\hat{\sigma}_j^{x}\cos(\omega_j^{\mathrm{ex}}t + \phi_j) + \hat{\sigma}_j^{y} \sin(\omega_j^{\mathrm{ex}}t + \phi_j)] \\
    & + \frac{\hbar \omega_j^{q}}{2} \hat{\sigma}_j^z
    + \hat{H}_{R, j} - \hat{H}_{I, j} \\
    = & \hat{H}_{S, j}(t) + \hat{H}_{R, j} - \hat{H}_{I, j}  \ , \ \ \ j = 1, 2\, .
    \label{eq:eachH}
\end{align}
All the noise effects are described through the dynamics of independent reservoirs $\hat{H}_{R, j},$ and the interaction between qubit $j$ and its reservoir is given by  $\hat{H}_{I, j}$ [for details of $\hat{H}_{R, j}$ and $\hat{H}_{I, j}$, see Eqs.~\eqref{eq:HI_RWA} and \eqref{eq:HI_noRWA} below and Appendix~\ref{sec:appHEOMandRWA}]. 
The operators $\hat{\sigma}_j^\alpha$ ($\alpha \in \{x, y, z\}$) indicate the Pauli matrices for the $j$th qubit, and the ladder operators are defined accordingly as $\hat{\sigma}_j^{\pm} = (\hat{\sigma}_j^x \pm i \hat{\sigma}_j^y)/2$.
The frequency $\omega_j^q$ in Eq.~\eqref{eq:eachH} is the transition frequency of the $j$th qubit, and the first term in Eq.~\eqref{eq:eachH} and the third term in Eq.~\eqref{eq:totH} are introduced for single- and two-qubit gate operations, respectively.
For the single-qubit gate, a rotating external field is introduced with amplitude, frequency, and initial phase given by $\Omega_j(t)$, $\omega_j^{\mathrm{ex}}$, and $\phi_{j}$, respectively, while the two-qubit gate is implemented via a direct coupling between the two qubits.
Both qubits are considered to be in resonance ($\omega_1^q = \omega_2^q = \omega_q$) so that counter-rotating terms in the weak and time-dependent qubit--qubit coupling with amplitude $J(t)$, i.e., $\hat{\sigma}_1^{+} \hat{\sigma}_2^{+}$ and $\hat{\sigma}_1^{-} \hat{\sigma}_2^{-}$, are negligible.
For the simulations of the qubit dynamics, we switch on and off the rotating external field $\Omega_j(t)$ as well as the qubit--qubit coupling $J(t)$ to realize specific single- and two-qubit gates.


In previous studies~\cite{YuPRL2004,BellomoPRL2007,BellomoPRA2008,BellomoPHYSSCR2010}, for the system--reservoir interaction $\hat{H}_{I, j}$, the RWA was imposed in order to obtain analytical expressions for the disentanglement dynamics at zero temperature. Under this assumption, the coupling Hamiltonian is expressed as
\begin{subequations}
\begin{align}
    \hat{H}_{I, j} = \hbar(\hat{\sigma}_j^{+} \hat{B}_j + \hat{\sigma}_j^{-} \hat{B}_j^\dagger)\, , 
    \label{eq:HI_RWA}
\end{align}
where $\hat{B}_j$ ($\hat{B}_j^\dagger$) is the annihilation (creation) operator of the $j$th reservoir. By contrast, within a full description (Caldeira--Leggett model~\cite{CLModel}) counter-rotating terms are taken into account so that one has
\begin{align}
    \hat{H}_{I, j} = \hbar(\hat{\sigma}_j^{+} \hat{B}_j + \hat{\sigma}_j^{-} \hat{B}_j^\dagger + \hat{\sigma}_j^{+} \hat{B}_j^\dagger + \hat{\sigma}_j^{-}\hat{B}_j)
    = \hbar \hat{\sigma}_{j}^x \hat{X}_j\, ,
    \label{eq:HI_noRWA}
\end{align}
\end{subequations}
where $\hat{X}_j = \hat{B}_j + \hat{B}_j^\dagger$ is the collective position of the reservoir.
It is an interesting topic on its own to analyze the impact of the counter-rotating terms in Eq.~\eqref{eq:HI_noRWA} in the dynamics of two-qubit systems~\cite{WangNJP2013,DuanJCP2013}.
We stress that here the term ``RWA'' indicates solely the approximation in Eq.~\eqref{eq:HI_RWA} in the following unless otherwise stated.

For Gaussian fluctuations, all the information on the noise is encoded in the autocorrelation function
\begin{align}
    C_j (t) = \braket{\hat{X}_j(t) \hat{X}_j(0)}_{R, j}\, ,
\end{align}
where $\braket{\bullet}_{R, j} = \mathrm{tr}\{\bullet \hat{\rho}_{R, j}^\eq \}$ denotes the expectation value with respect to the equilibrium state of the $j$th reservoir $\hat{\rho}_{R, j}^\eq = e^{-\beta \hat{H}_{R, j}} / \mathrm{tr}\{e^{-\beta \hat{H}_{R, j}}\}$ at the temperature $\beta = 1 / (k_\mathrm{B} T)$; the mean value $\braket{\hat{X}_j(t)}_{R, j}$ is chosen to be always zero.
Because the reservoirs are independent, one has
\begin{align}
    \braket{\hat{X}_1(t) \hat{X}_2(0)}_R = \braket{\hat{X}_2(t) \hat{X}_1(0)}_R = 0\, ,
\end{align}
where $\braket{\bullet}_R = \mathrm{tr}\{\bullet \hat{\rho}_{R, 1}^\eq \otimes \hat{\rho}_{R, 2}^\eq\}$.
For the sake of convenience, we introduce the spectral noise power
\begin{align}
    S_j^{\beta} (\omega) = \frac{1}{2\pi}\int_{-\infty}^{\infty} dt  C_j(t) e^{i\omega t}\, ,
\end{align}
which is the Fourier transform of $C_j(t)$, and related to the temperature-independent spectral density via $S_j^{\beta}(\omega) = \hbar J_j(\omega) [1 + n^\beta(\omega)]$ with
$n^\beta(\omega) = 1 / [\exp(\beta \hbar \omega) - 1]$ being the Bose distribution function. We assume identical temperatures for both reservoirs.

In this study, we consider the two classes of spectral densities, namely, 
\begin{enumerate}
    \item Lorentzian noise spectrum
    \begin{align}
        J_j(\omega) = \frac{\kappa_j \omega_j^q \lambda_j^2}{(\omega_j^q - \Delta_j - \omega)^2 + \lambda_j^2}\, , 
        \label{eq:sdLorentz}
    \end{align}
    \item broadband noise spectrum
    \begin{align}
        J_j(\omega) = \mathrm{sgn}(\omega) \; \omega_{\mathrm{ph}, j}^{1-s_j}
        \frac{\kappa_j  |\omega|^{s_j}}
        {\bigl[1+(\omega/\omega_j^c)^2\bigr]^2}\, ,
        \label{eq:sdOhmic}
    \end{align}
\end{enumerate}
characterized by spectral exponents $s_j$ with 
 $\mathrm{sgn}(\omega)$ being the sign function.
A qubit coupled to a reservoir with a Lorentzian spectrum corresponds to a damped Rabi model (with RWA it is simplified to the Jaynes--Cummings model)~\cite{Breuer2002}, where a two-level system interacts with an electromagnetic field inside an imperfect cavity with frequency $\omega_j^q - \Delta _j$~\cite{BellomoPRL2007,BellomoPHYSSCR2010}.
The line width of the Lorentzian peak $\lambda_j$ determines the rate for leakage out of the cavity.
Instead, the broadband noise represents interaction between a qubit and a macroscopic reservoir in its proximity.
In case of transmon qubits, for example, relevant environmental degrees of freedom (noise source) include (i) fluctuations of electromagnetic fields, (ii) diffusing quasiparticles in the superconductor, and (iii) charge defects in the circuitry (two-level fluctuators).
\textcolor{black}{Each of these situations can be modeled by considering spectral exponents $s_j$ in specific ranges, for example, (i) $s_j \approx 1$~\cite{Weiss2012,Barone1982,WendinARXIV2005}, (ii) $s_j \approx 1/2$~\cite{GlazmanSPPLN2021}, and (iii) $1 \gg s_j \simeq 0.1$~\cite{MachlupJAP1954,Weiss2012,PaladinoRMP2014,MullerRPP2019,IthierPRB2005,BylanderNP2011,PlaceNATCOMMUN2021}, respectively.}
In the context of open quantum systems, a reservoir with $s_j = 1$ is referred to as an Ohmic reservoir, while one with $0 < s_j < 1$ as a sub-Ohmic reservoir.
The bandwidth of the broadband spectrum is restricted by the cutoff frequency $\omega_j^c$.


For both noise spectra, the \textcolor{black}{dimensionless} quantity $\hbar\kappa_j$ determines the coupling strength between the system and the reservoir.
The quantity $\omega_{\mathrm{ph}, j}$ in Eq.~\eqref{eq:sdOhmic} has been introduced in order to fix the dimension of $\kappa_j$ irrespective of $s_j$. Here, we set $\kappa_1 = \kappa_2 = \kappa$ for the sake of simplicity of the discussion.
Within the Born--Markov approximation (second-order perturbation theory), the relaxation time of the qubit population $T_{1}$ in the absence of any drives [$\Omega_j(t) = J(t) = 0$] is related to this coupling constant via (see Appendix~\ref{sec:appHEOMandRWA})
\begin{align}
    T_{1} \; \omega_j^q \simeq \frac{1}{2\pi \hbar \kappa} \, ,
    \label{eq:approxT1}
\end{align}
\textcolor{black}{so that equivalently $2 \pi \hbar \kappa \omega_j^q$  characterizes the relaxation rate.} This figure is often used when characterizing qubits together with the phase decoherence time $T_2$ (or $T_2^*$ when external noise is included).

In the following, we study the dynamics of the reduced density operator (RDO) of two-qubit systems $\hat{\rho}_S(t) = \mathrm{tr}_R\{\hat{\rho}_\mathrm{tot}(t)\}$, where $\mathrm{tr}_R\{\bullet\}$ is the partial trace over the reservoir degrees of freedom.
\textcolor{black}{When we impose the RWA on the system--reservoir coupling [Eq.~\eqref{eq:HI_RWA}], the dynamics of the RDOs at zero temperature in the absence of any gate operations can be expressed in analytical form.}
For example, the population $\rho_{11} (t) = \braket{11|\hat{\rho}_S(t)|11}$ is then described as~\cite{Breuer2002,BellomoPRL2007}
\begin{align}
    \rho_{11} (t) = \rho_{11}(0) |h_1(t)|^2 |h_2(t)|^2 \, ,
    \label{eq:DynamicsRWA}
\end{align}
where the functions $h_j(t)$ are specified together with the other matrix elements of $\hat{\rho}_S(t)$ in Appendix~\ref{sec:appRWA}.
As basis states for the $2\times 2$ Hilbert space, we have introduced states $\ket{ab} = \ket{a}_1 \otimes \ket{b}_2$ ($a, b \in \{0, 1\}$) with $\ket{1}_j$ ($\ket{0}_j$) being the excited (ground) state of the $j$th qubit in the $\hat{\sigma}_j^z$ basis.

The full quantum dynamics is simulated with a numerically rigorous method, the FP-HEOM method~\cite{XuPRL2022}. In this method, auxiliary density operators [ADOs, $\hat{\rho}_{\vec{m}, \vec{n}}(t)$] are introduced in addition to the RDO [$\hat{\rho}_S(t) = \hat{\rho}_{\vec{m} = \vec{0}, \vec{n} = \vec{0}}(t)$] according to a proper decomposition of the reservoir correlators in terms of auxiliary modes (see below). Multi-indices $\vec{m} = [m_{1, 1}, \ldots, m_{1, K_1}, m_{2, 1}, \ldots m_{2, K_2}]$ and $\vec{n} = [n_{1, 1}, \ldots, n_{1, K_1}, n_{2, 1}, \ldots n_{2, K_2}]$ distinguish ADOs, and the time evolution equation in extended state space reads 
\begin{align}
    \frac{\partial \hat{\rho}_{\vec{m},\vec{n}}(t)}{\partial t} =& 
    -i\mathcal{L}_S (t) \hat{\rho}_{\vec{m},\vec{n}}(t) \\
    & -\sum_{j=1}^{2} \Biggl\{\sum_{{k}=1}^{K_j} \bigl[(m_{j, k} z_{j, k}+n_{j, k} z_{j, k}^*) \hat{\rho}_{\vec{m},\vec{n}}(t) \\
    & +i \mathcal{L}_{j, k}^{+} \hat{\rho}_{\vec{m},\vec{n}}(t) +i \mathcal{L}_{j, k}^{-} \hat{\rho}_{\vec{m},\vec{n}}(t) \bigr] \Biggr\}\, ,
    \label{eq:HEOM}
\end{align}
where $\mathcal{L}_S (t) \bullet = [\hat{H}_S (t), \bullet]/ \hbar$ with $\hat{H}_S (t) \equiv \hat{H}_{S, 1} (t) + \hat{H}_{S, 2} (t) + \hbar J(t) (\hat{\sigma}_1^{+}\hat{\sigma}_2^{-} + \hat{\sigma}_1^{-}\hat{\sigma}_2^{+})$ is the commutator of the system Hamiltonian, and the raising and lowering superoperators $\{\mathcal{L}_{j, k}^{\pm}\}$ describe interaction between different ADOs (see Appendix~\ref{sec:appHEOM}).
Factorized initial states $\hat{\rho}_\mathrm{tot}(0) = \hat{\rho}_S(0) \otimes \hat{\rho}_{R, 1}^\eq\otimes \hat{\rho}_{R, 2}^\eq$ correspond to an initial RDO $\hat{\rho}_{\vec{0}, \vec{0}}(0) = \hat{\rho}_S(0)$ with the other ADOs $\hat{\rho}_{(\vec{m} , \vec{n}) \neq (\vec{0}, \vec{0})}(0) = 0$.
Complex-valued frequencies of the auxiliary modes $z_{j, k} = i \omega_{j, k} + \gamma_{j, k}$ ($\gamma_{j, k} > 0$) are obtained through a mathematically systematic procedure together with complex-valued amplitudes $d_{j, k}$ in the decomposition of the reservoir correlators
\begin{align}
    C_j(t) = \sum_{k=1}^{K_j} d_{j, k} e^{-i\omega_{j, k} t-\gamma_{j, k} t} 
    = \sum_{k=1}^{K_j} d_{j, k} e^{-z_{j, k} t}
    \\ (t > 0)\, .
    \label{eq:CF}    
\end{align}

We use dimensionless quantities through the paper with the qubit frequency $\omega_q$ as the frequency unit. All simulations use a set of fixed parameter values: $\lambda_1 = \lambda_2 = 0.01 \omega_q$ in Eq.~\eqref{eq:sdLorentz}, and $\omega_{\mathrm{ph}, 1} = \omega_{\mathrm{ph}, 2} = \omega_q$ and $\omega_1^c = \omega_2^c = 50 \omega_q$ in Eq.~\eqref{eq:sdOhmic}.
The other parameters for the reservoirs are varied, i.e., detuning $\Delta\equiv \Delta_1 = \Delta_ 2 $ in Eq.~\eqref{eq:sdLorentz}, spectral exponent $s_j$ in Eq.~\eqref{eq:sdOhmic}, temperature $\beta$, and coupling strength $\kappa$. Parameter values for the gate operations in Eqs.~\eqref{eq:totH} and \eqref{eq:eachH} are determined by the specific quantum circuits that we consider below.

To facilitate the numerical simulations, we utilize tensor-train (TT) representation (Appendix~\ref{sec:appTT}).
For the parameter values of the FP-HEOM and TT methods and related discussions, see Appendix~\ref{sec:appHEOMParams}.

\subsection*{Figure of merits}
In this paper, we characterize the performance of gate operations by means of the following quantities:

1. \emph{Concurrence}: 
\textcolor{black}{Entanglement cannot be generated through local operations and classical communication, which are more easily realized than two-qubit-gate operations.
In this sense, the degree of entanglement is one of the most important quantities to characterize the performance of two-qubit gates.}
For two two-level systems, it is consistently characterized by the concurrence $\mathcal{C} (t)$ with
\begin{align}
    \mathcal{C}(t) = \max\{0, \Lambda_1 - \Lambda_2 - \Lambda_3 - \Lambda_4\}\, ,
    \label{eq:conc}
\end{align}
where $\Lambda_j$ are the square roots of the eigenvalues of $\hat{\rho}_S(t) \tilde{\rho}_S(t)$ in descending order.
The quantity $\tilde{\rho}_S(t)$ is evaluated as $\tilde{\rho}_S(t) = \hat{\sigma}_1^{y}\hat{\sigma}_2^y \hat{\rho}_S^*(t) \hat{\sigma}_1^{y}\hat{\sigma}_2^y$ with $\hat{\rho}_S^*(t)$ being the complex conjugate of the matrix $\hat{\rho}_S(t)$ in the standard product basis $\mathcal{B} = \{\ket{1}, \ket{2}, \ket{3}, \ket{4}\} \equiv \{\ket{11}, \ket{10}, \ket{01}, \ket{00}\}$.
The concurrence takes values in the range $0 \leq \mathcal{C} \leq 1$, and the two-qubit system is maximally entangled when $\mathcal{C} = 1$, while totally disentangled when $\mathcal{C} = 0$.
Note that when the system is expressed as an ``$X$ state,'' which is given by
\begin{align}
    \hat{\rho}_S = 
    \begin{bmatrix}
        \rho_{11} & 0 & 0 & \rho_{14} \\
        0 & \rho_{22} & \rho_{23} & 0 \\
        0 & \rho_{32} & \rho_{33} & 0 \\
        \rho_{41} & 0 & 0 & \rho_{44}
    \end{bmatrix}\, ,
\end{align}
the concurrence is evaluated as
\begin{align}
    \mathcal{C}(t) = &2\max\{0, \quad |\rho_{23}(t)| - \sqrt{\rho_{11}(t) \rho_{44}(t)}, \\
    & |\rho_{14}(t)| - \sqrt{\rho_{22}(t)\rho_{33}(t)}\}
    \label{eq:concXState}
\end{align}
(here, $\hat{\rho}_S$ is a Hermitian matrix, of course).

2. \emph{Fidelity}: 
\textcolor{black}{Another important figure of merit is the fidelity, which evaluates the closeness of two quantum states.}
Dissipative $\hat{\rho}_S(t)$ are compared with ideal states $\hat{\rho}_\mathrm{iso}(t)$ obtained with the von Neumann equation for isolated systems ($\kappa = 0$).
The fidelity is then defined as
\begin{align}
    F(t) = \left(\mathrm{tr}\left\{\sqrt{\sqrt{\hat{\rho}_\mathrm{iso}(t)}
    \hat{\rho}_S(t) \sqrt{\hat{\rho}_\mathrm{iso}(t)}}
    \right\}\right)^2\, .
    \label{eq:fidelity}
\end{align}
The quantity is bounded by $0 \leq F(t) \leq 1$ and takes the value $1$ if and only if $\hat{\rho}_S(t) = \hat{\rho}_\mathrm{iso}(t)$:
Smaller fidelity implies worse gate performance.

\section{Effects of counter-rotating terms in system--reservoir coupling} \label{sec:resultRWA}
First, we investigate the effects of the counter-rotating terms in the qubit--reservoir interaction, see Eq.~\eqref{eq:HI_noRWA}, in order to explore the validity of this approximation imposed in most of the previous studies.
For this purpose, we compare the time traces of the concurrence obtained with the two coupling forms in Eqs.~\eqref{eq:HI_RWA} and \eqref{eq:HI_noRWA}.
Note that the temperature of the reservoirs is set to zero in the results illustrated in this section.

More specifically, we consider as initial states $\hat{\rho}_S(0) \equiv \hat{\rho}_0$ two Bell states according to 
\begin{align}
    \ket{\Phi} = &\frac{\ket{10} + \ket{01}}{\sqrt{2}}\, , \\
    \ket{\Psi} = &\frac{\ket{11} + \ket{00}}{\sqrt{2}}\, ,
    \label{eq:Bell}
\end{align}
and monitor free relaxation processes via $\rho_S(t)$ in the absence of external driving and inter-qubit coupling [$\Omega_1(t) = \Omega_2(t) = J(t) = 0$].
Thus, in both cases, the RDO is in $X$ states at any time, so that we can utilize Eq.~\eqref{eq:concXState}, which holds no matter whether the RWA is imposed or not under the above condition as proven in Appendix~\ref{sec:appXState}.

\begin{figure}[t]
    \centering
    \includegraphics[width=0.94\linewidth]{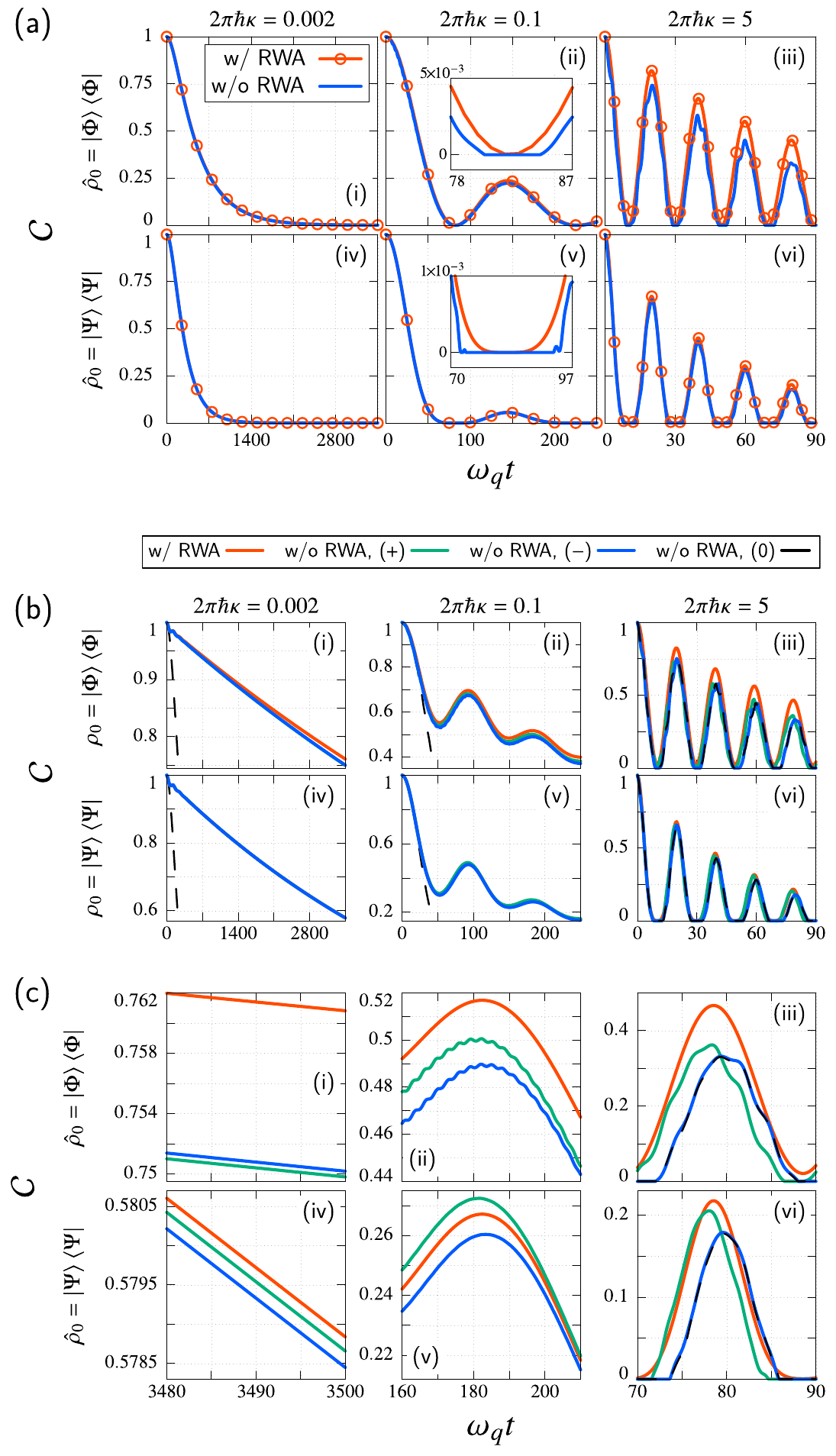}
    \caption{Time traces of the concurrence $\mathcal{C}(t)$ of two-qubit systems coupled with Lorentzian noise sources for various coupling strengths $\kappa$ and initial states $\hat{\rho}_0$.
    (a) Results without detuning, $\Delta = 0$.
    Results obtained with (w/ RWA, red curves and circles) and without (w/o RWA, blue curves) the RWA are depicted.
    Insets of (ii) and (v): Magnifications of $\mathcal{C}(t)$ around certain time periods.
    (b), (c) Results with detuning $\Delta\neq 0$.
    Dynamics in a whole time domains (b) and magnifications of those in long-time domains (c) are displayed.
    Subpanels (i)--(vi) in panels (b) and (c) display the same dynamics in different time domains.
    For the case without RWA (w/o RWA), the values $\Delta = 0.05 \omega_q$ [($+$), green curves] and $-0.05 \omega_q$ [($-$), blue curves] are taken into account.
    For the case with RWA (w/ RWA), both results with $\Delta = 0.05 \omega_q$ and $-0.05 \omega_q$ are exactly the same, and hence only one red curve is depicted in each panel.
    The dashed curves are the results without detuning (0), corresponding to the blue curves in (a).
    \label{fig:RWA_Lorentz}}
\end{figure}

\subsection{Lorentzian noise spectrum}
Time traces for the concurrence are shown in Fig.~\ref{fig:RWA_Lorentz} for the two-qubit systems with individual Lorentzian noise sources.
Various qubit detunings $\Delta$ (qubit transition frequency and noise peak) and reservoir coupling strengths $\kappa$  are considered while the line width of the Lorentzian function is fixed to $0.01 \omega_q$ for both reservoirs.

We start with the resonant case ($\Delta = 0$) in Fig.~\ref{fig:RWA_Lorentz}(a) as previously explored in Ref.~\cite{WangNJP2013}.
Note that while the ratio $\kappa_j / \lambda_j$ was kept fixed in this previous study, here, we fix $\lambda_j$ and vary $\kappa_j$, which in turn allows us to tune its ratio.

\begin{figure}
    \centering
    \includegraphics[width=0.85\linewidth]{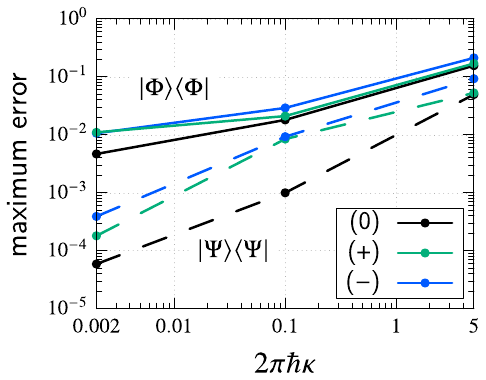}
    \caption{\textcolor{black}{Log--log plot of the maximum absolute error between the results with and without the RWA for various coupling strengths between the system and reservoir $2 \pi \hbar \kappa$.
    The two initial states, $\hat{\rho}_0 = \ketbra{\Phi}{\Phi}$ (solid curves) and $\ketbra{\Psi}{\Psi}$ (dashed curves), and three detunings, $\Delta = 0$ [(0), black curves], $0.05\omega_q$ [($+$), green curves], and $-0.05\omega_q$ [($-$×), blue curves], are considered.
    }
    \label{fig:RWA_LorentzErr}}
\end{figure}

In Fig.~\ref{fig:RWA_Lorentz}(a), one realizes that significant discrepancies between the results with (red curves and circles) and without the RWA (blue curves) appear at stronger coupling.
This is as expected since the RWA is based on a weak-coupling assumption:
The rotating terms appear as leading-order corrections within the framework of a perturbative treatment.
When different initial states are compared, one finds that the discrepancy is more pronounced for $\ketbra{\Phi}{\Phi}$ than $\ketbra{\Psi}{\Psi}$.
\textcolor{black}{This discrepancy can be easily confirmed through the solid and dashed black curves in Fig.~\ref{fig:RWA_LorentzErr}, which illustrates the maximum error between the results with and without the RWA.}
The reason is the two-photon processes (double excitation) caused by the counter-rotating term $\hat{\sigma}_j^+ \hat{B}_j^\dagger$ in Eq.~\eqref{eq:HI_noRWA}.
Since the reservoirs are initially in the ground state (i.e., zero temperature), the qubit state $\ket{11}$ is populated from the initial state $\ketbra{\Phi}{\Phi}$ only if the above term acts on the system and reservoirs.
\textcolor{black}{For example, the qubit state $\ket{11}$ can be populated through the process $\ket{01}_S\otimes \ket{0}_{R, 1}\otimes \ket{0}_{R, 2} \to \ket{11}_S\otimes \ket{1}_{R, 1}\otimes \ket{0}_{R, 2}$, but this process can be achieved only by $\hat{\sigma}_1^+ \hat{B}_1^\dagger$, which is the two-photon excitation process.} 
Accordingly, when the RWA is imposed on the interaction Hamiltonian, the population $\braket{11|\hat{\rho}_S(t)|11}$ is always zero, so that the concurrence in Eq.~\eqref{eq:concXState} is given by $|\rho_{23}(t)|$.
By contrast, the contribution $\sqrt{\rho_{11}(t) \rho_{44}(t)}$ is finite for the case without the RWA.
Note that the dynamics of the coherence $|\rho_{23}(t)|$ with and without the RWA are similar to each other so that the difference between those results with $\ketbra{\Phi}{\Phi}$ [blue and red curves in the upper panels of Fig.~\ref{fig:RWA_Lorentz}(a)] are mainly attributed to the population $\rho_{11}(t)$.
We note in passing that the importance of the population of the most excited state for the entanglement has been already discussed in previous studies~\cite{IkramPRA2007}.
\textcolor{black}{For the second initial state $\ketbra{\Psi}{\Psi}$, the decay dynamics of the matrix element $\braket{11|\hat{\rho}_S(t)|11}$ to $\braket{10|\hat{\rho}_S(t)|10}$ and $\braket{01|\hat{\rho}_S(t)|01}$ can be observed for both Hamiltonians with and without the RWA.}
Consequently, differences of the concurrence with $\ketbra{\Psi}{\Psi}$ as initial state are smaller compared to those with $\ketbra{\Phi}{\Phi}$ [note that the concurrence is given by $2 \max \{0, |\rho_{14}(t)| - \sqrt{\rho_{22}(t)\rho_{33}(t)}\}$ in the case with $\ketbra{\Psi}{\Psi}$].

Another major difference related to the RWA is the occurrence of so-called dark periods~\cite{BellomoPRA2008}
with zero concurrence.
Namely, one can easily find horizontal lines with $\mathcal{C} = 0$ between two adjacent peaks for the strong-coupling case in Figs.~\ref{fig:RWA_Lorentz}(a)-(iii) and (vi), and even in the case with the intermediate coupling strength [$2\pi \hbar \kappa = 0.1$, insets of Figs.~\ref{fig:RWA_Lorentz}(a)-(ii) and (v)], a closer look reveals a finite dark period for the non-RWA situation.
Since the concurrence with RWA is expressed as the second and fourth power of sinusoidal functions with the initial state $\ketbra{\Phi}{\Phi}$ and $\ketbra{\Psi}{\Psi}$, respectively~\cite{BellomoPRL2007}, the red curves show infinitesimal dark periods only.
This contrasts with the exact results as discussed earlier~\cite{WangNJP2013}.
In this previous study, the oscillatory behavior of the concurrence is suppressed with increasing coupling strength $\kappa$, in contrast to our results where they are present even in the strong-coupling region ($2\pi\hbar\kappa = 5$).
We discuss this difference in more detail in Sec.~\ref{sec:resultRWABroadband}.

Next, we turn to the off-resonant case ($\Delta_1 = \Delta_2 = \Delta\neq 0$)  depicted in Fig.~\ref{fig:RWA_Lorentz}(b) for $\Delta = \pm 0.05 \omega_q$.

Before we explore interplay of the detuning and the RWA, we compare the exact results with and without the detuning with respect to the qubit frequencies.
Figure~\ref{fig:RWA_Lorentz}(b) displays time traces of the concurrence for both cases.
One finds that for weak and moderate coupling ($2\pi\hbar\kappa=0.002$, $0.1$) the concurrence tends to decay faster in the resonant than in the off-resonant situation.
A qualitative understanding can be obtained within the framework of the Born--Markov approximation and the second-order perturbation theory, in which the relaxation and decoherence time are given by
\begin{align}
    T_1 \omega_q = \frac{T_2 \omega_q}{2} = \frac{\omega_q}{2\pi \hbar J(\omega_q)} = \frac{1}{2 \pi \hbar \kappa} \frac{\Delta^2 + \lambda^2}{\lambda^2}
    \label{eq:T1Detuned}
\end{align}
at zero temperature \textcolor{black}{(see also Appendix~\ref{sec:appHEOMandRWA})}. Here, we omit the subscript $j$ in Eq.~\eqref{eq:sdLorentz} for convenience.
Owing to the rightmost expression, detuning induces a larger relaxation time ($\Delta \neq 0$).
Of course, strictly speaking, the above approximations are not valid for moderate coupling, where the concurrence oscillates.
By contrast, for the strong coupling ($2\pi \hbar \kappa = 5$), the concurrences for the resonant and detuned noise sources are almost identical.
Since $2 \pi \hbar \kappa$ is much greater than $\Delta / \omega_q$ and $\lambda / \omega_q$ in this case, the system--reservoir coupling dominantly affects the dynamics.

Let us now turn to the impact of the RWA.
Similar to the resonant case, the deviation of the concurrence with RWA from the exact results is larger for the initial state $\ketbra{\Phi}{\Phi}$ compared to $\ketbra{\Psi}{\Psi}$ owing to the double excitation process.
One can confirm this with Fig.~\ref{fig:RWA_Lorentz}(b), where we can distinguish two curves in the upper panels, while the curves are almost indistinguishable from each other in the lower panels.
\textcolor{black}{Figure~\ref{fig:RWA_LorentzErr} also indicates how the initial states affect the errors caused by the RWA.}

Looking more closely at the concurrence in the long-time domain [Fig.~\ref{fig:RWA_Lorentz}(c)], one sees that its time trace depends on the sign of the detuning.
\textcolor{black}{As mentioned in Eq.~\eqref{eq:T1Detuned} and Appendix~\ref{sec:appHEOMandRWA}}, the value of the spectral noise power at the qubit frequency $S^\beta(\omega_q)$ plays a dominant role in the dynamics of reduced systems, which is the same irrespective of the sign of the detuning.
When higher-order corrections are taken into account, the peak values, $S^\beta(\omega_q - \Delta)$ and $S^\beta(\omega_q + \Delta)$, were found to contribute to the dynamics in an asymmetric manner.
This sign effect cannot be detected within the RWA, in which the concurrence is exactly the same for $\Delta = 0.05 \omega_q$ and $-0.05 \omega_q$.
One can confirm this by noting that the concurrence is determined by the absolute value of $h_j(t)$ in Eq.~\eqref{eq:DynamicsRWA} (see Appendix~\ref{sec:appRWA} for more details).
Another interesting phenomenon relating to this asymmetric contribution can be seen in Figs.~\ref{fig:RWA_Lorentz}(c)-(iii) and (vi):
The concurrence with the detuning $\Delta = -0.05\omega_q$ is almost the same as that with the resonant noise sources ($\Delta = 0$), while the concurrence with $\Delta = 0.05\omega_q$ is obviously different from those curves.

Since the autocorrelation function of Eq.~\eqref{eq:CF} is expressed as $C_j (t) \propto \exp[-\lambda_j|t|-i(\omega_j^q-\Delta_j)t]$ [see Eq.~\eqref{eq:cfLorentz} in Appendix~\ref{sec:appRWA} for details], it oscillates more slowly with positive detuning.
Taking this into account, we conclude that in the strong-coupling regime, quantum correlations in the two-qubit system are more sensitive to low-frequency modes of the noise sources compared to faster modes.
This appears to be analogous to the fact that noise modes with much higher frequencies than the cutoff frequency $\omega_c$ merely have an almost time-independent renormalization effect on the reduced-system dynamics~\cite{LeggettRMP87A,Weiss2012}.
It is worth noting that when we set $\pm 0.1 \omega_q$ to $\Delta$, both detuned results are different from the resonant results (results are not shown) so that the above discussion is restricted to sufficiently small detunings.

It is interesting that for the intermediate coupling with the initial state $\ketbra{\Psi}{\Psi}$ [Fig.~\ref{fig:RWA_Lorentz}(c)-(v)], the concurrence for the positive detuning takes the largest local maximum among the three cases.
Again, we attribute this phenomenon to the two-photon excitation process caused by the counter-rotating term:
The term $-\sqrt{\rho_{11}(t)\rho_{44}(t)}$ causes a dominant difference between the results with and without the RWA for the initial state $\ketbra{\Phi}{\Phi}$, as mentioned above, see the upper panel of Fig.~\ref{fig:RWA_Lorentz}(c).
By contrast, both $|\rho_{14}(t)|$ and $-\sqrt{\rho_{22}(t)\rho_{33}(t)}$ contribute to the different dynamics of $\mathcal{C} (t)$ in a complicated manner for the cases with the initial state$\ketbra{\Psi}{\Psi}$.

Next, we comment on the decay speed of the entanglement.
With the initial state $\ketbra{\Phi}{\Phi}$, one can see that the decay rate is smaller for the weak-coupling case [monotonic decay, Figs.~\ref{fig:RWA_Lorentz}(b)-(i) and (iv)], and the amplitude of the recovery is larger for strong-coupling [damped oscillations, Figs.~\ref{fig:RWA_Lorentz}(b)-(iii) and (vi)], compared to $\ketbra{\Psi}{\Psi}$.
We found that the coherence $|\rho_{23}(t)|$ in the case $\ketbra{\Phi}{\Phi}$ and $|\rho_{14}(t)|$ in $\ketbra{\Psi}{\Psi}$ are almost the same in the cases with and without RWA, and $\sqrt{\rho_{11}(t)\rho_{44}(t)}$ in $\ketbra{\Phi}{\Phi}$ is quite smaller compared to $\sqrt{\rho_{22}(t)\rho_{33}(t)}$ in $\ketbra{\Psi}{\Psi}$, although the double excitation processes occur in $\ketbra{\Phi}{\Phi}$.
The diagonal elements play a crucial role in the concurrence. 

\textcolor{black}{Finally, we analyze how errors caused by the RWA depend on the coupling strength between system and reservoir.
In Fig.~\ref{fig:RWA_LorentzErr}, the maximum absolute error between the results with and without RWA during the time period considered in Fig.~\ref{fig:RWA_Lorentz} is depicted.
As discussed above, the error for the initial state $\ketbra{\Phi}{\Phi}$ is substantially larger than for $\ketbra{\Psi}{\Psi}$ for all the coupling strengths.
For the smallest couplings considered here, the error is in the range of $10^{-2}$ and grows to more than $10^{-1}$ towards larger $\kappa$.
Even for the more benevolent state $\ketbra{\Psi}{\Psi}$, does the maximum error lie in the range of $10^{-3}$ for weak couplings. 
This value must be compared to gate fidelities on the order of $0.999$ required for the state-of-the-art quantum computing.
Hence, one concludes that counter-rotating terms play a crucial role in actual qubit operations and must be included when providing high-precision predictions through numerical simulations.
Note that for multilevel systems, such as transmon qubits, effects of the counter-rotating terms are generally even more pronounced.}

\begin{figure}[t]
    \centering
    \includegraphics[width=\linewidth]{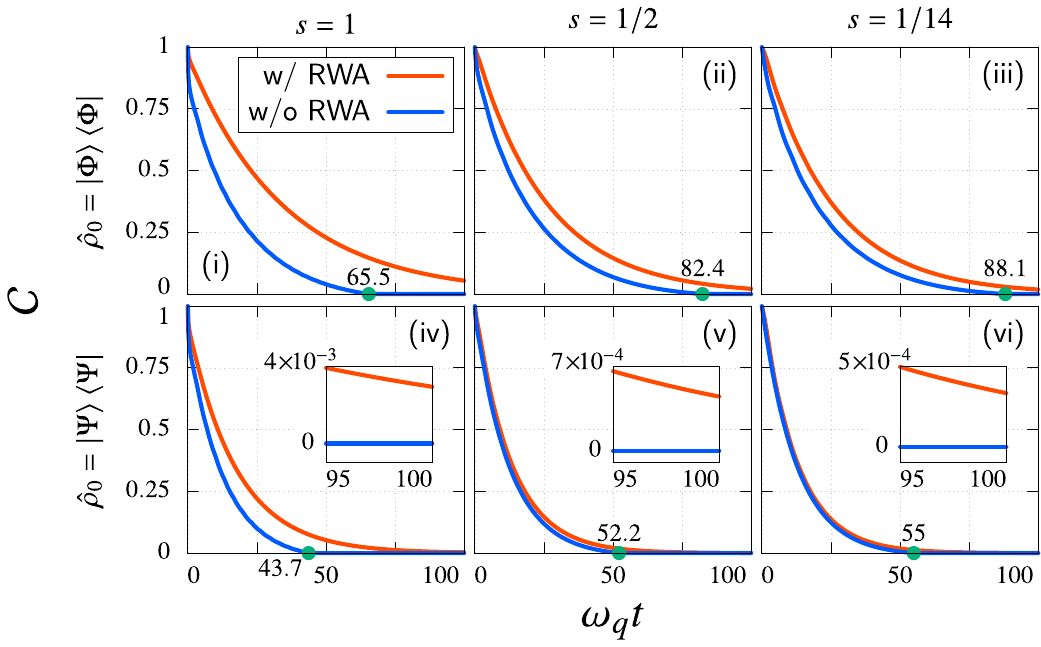}
    \caption{Time traces of the concurrence $\mathcal{C}(t)$ of two-qubit systems coupled with broadband noise sources for various spectral exponents $s$ and initial states $\hat{\rho}_0$.
    \textcolor{black}{The coupling strength between the qubits and reservoirs is fixed to $2 \pi \hbar \kappa = 0.04$.}
    Results obtained with (w/ RWA, red curve) and without (w/o RWA, blue curve) the RWA are depicted.
    The green filled circles indicate the times at which the concurrence reaches zero (ESD).
    The times of the ESD are indicated by the associated numbers.
    \textcolor{black}{Insets of (iv)--(vi): Magnifications of the main plots around the time $\omega_q t \simeq 100$.}
    \label{fig:RWA_subOhmic}}
\end{figure}

\subsection{Broadband noise spectrum} \label{sec:resultRWABroadband}
In this subsection, we investigate effects of the counter-rotating terms on the systems in the presence of broadband noise sources given by Eq.~\eqref{eq:sdOhmic}.
Here, we fix the coupling strength to $2\pi \hbar \kappa = 0.04$ and instead vary the spectral exponent $s$.
Note that the exponents of each reservoir take the same values in these simulations ($s_1 = s_2 = s$); the temperature is set to zero.

In Fig.~\ref{fig:RWA_subOhmic}, the time traces of the concurrence are depicted.
In contrast to the results for the Lorentzian noise spectrum, only monotonic decay is observed.
This is because a larger number of modes of the noise source contributes to the system dynamics in this case compared to the Lorentzian noise source, in which the mode with the frequency $\omega_q - \Delta$ dominantly affects the system.
The tendency of the slower decay of the concurrence and the more significant difference between dynamics with and without the RWA for the initial state $\ketbra{\Phi}{\Phi}$ than those for the state $\ketbra{\Psi}{\Psi}$ also holds in this case, which again reveals the crucial role of the diagonal elements.

Comparing the concurrences with different spectral exponents, one finds that the effects of the counter-rotating terms are more significant for larger $s$.
For an Ohmic reservoir ($s = 1$) without the RWA, the coherence dynamics $|\rho_{23}(t)|$ starting with $\ketbra{\Phi}{\Phi}$ and $|\rho_{14}(t)|$ with $\ketbra{\Psi}{\Psi}$ already differ from those with the RWA.
This contrasts with the Lorentzian-spectrum case, and this difference decreases as the exponent $s$ gets smaller.
The difference in the Ohmic case is mainly attributed to the steep decay in the short-time domain $\omega_q t \simeq 0$.
A large portion of the high-frequency modes of $S^\beta(\omega)$ results in fast decoherence in a short-time domain owing to the fast reconfiguration process of the environment, and in this short-time region, both rotating and counter-rotating terms contribute to the system dynamics almost equivalently.
For the smaller $s$, those fast dynamics of the reservoirs are suppressed, and the effects of the counter-rotating terms on the off-diagonal elements are also suppressed accordingly.
The diagonal elements $-\sqrt{\rho_{11}(t)\rho_{44}(t)}$ and $-\sqrt{\rho_{22}(t)\rho_{33}(t)}$ are also affected by the counter-rotating terms:
The double excitation process mainly contributes to the difference of the dynamics with and without the RWA in Figs.~\ref{fig:RWA_subOhmic}(ii) and \ref{fig:RWA_subOhmic}(iii).
The steep decay of the concurrence found in Fig.~\ref{fig:RWA_subOhmic}(i) results from the steep decay of both $|\rho_{23}(t)|$ and $-\sqrt{\rho_{11}(t)\rho_{44}(t)}$.

\begin{figure*}
    \centering
    \includegraphics[width=0.9\linewidth]{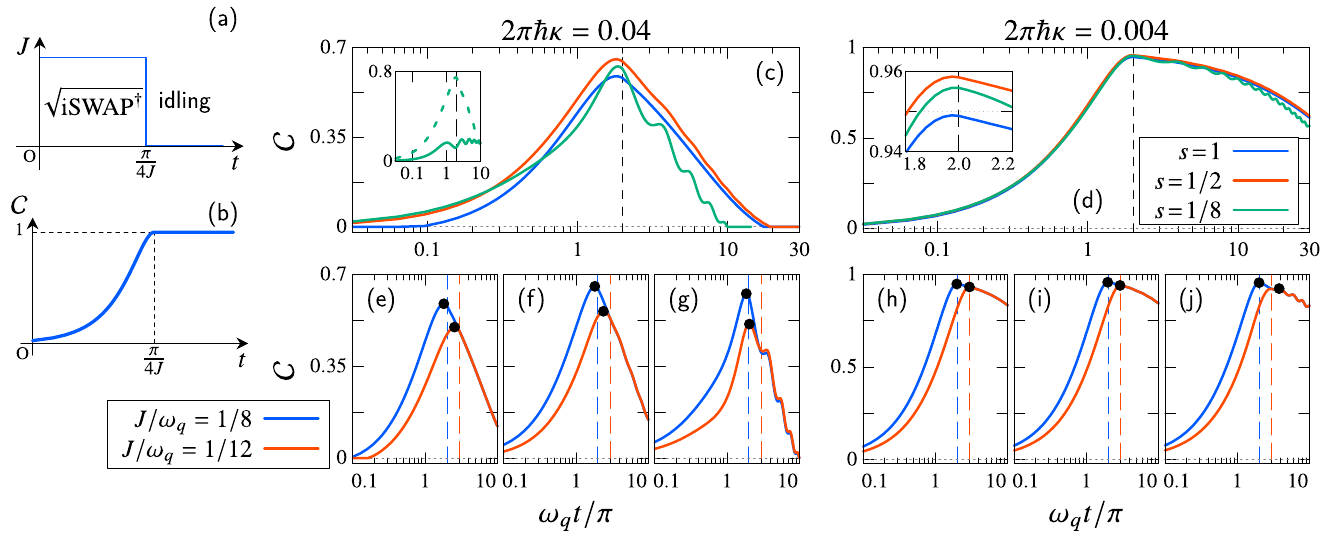}
    \caption{(a) Schematic of the sequence of the qubit--qubit coupling strength $J(t)$.
    \textcolor{black}{(b) Schematic of the time trace of the concurrence $\mathcal{C}(t)$ for an isolated system during and after the application of a \sqrtiswap{} gate (linear--log plots).}
    (c)--(j) Linear--log plots of the time traces of the concurrence $\mathcal{C}(t)$ during and after application of a \sqrtiswap{} gate in homogeneous environments, $s_1 = s_2 = s$.
    The vertical dashed lines indicate the end time of the gate operation.
    The system--reservoir coupling strength is $2\pi\hbar \kappa = 0.04$ in panels (c) and (e)--(g), and $0.004$ in panels (d) and (h)--(j), respectively.
    (c), (d) Comparison between different spectral exponents with the fixed qubit--qubit coupling strength ($J / \omega_q = 1/8$).
    Three cases for the exponents, $s = 1$ (blue curves), $1/2$ (red curves), and $1/8$ (green curves), are depicted.
    Inset of (c): Time traces of $2|\rho_{23}(t)|$ (dashed curve) and $2\sqrt{\rho_{11}(t)\rho_{44}(t)}$ (solid curve) for $s = 1/8$, both of which constitute the concurrence.
    Inset of (d): Magnification of the main plot around the time in which the concurrence takes the maximum value (linear--linear plot).
    (e)--(j) Comparison between different strengths of the qubit--qubit coupling with the fixed spectral exponents: $s = 1$ in panels (e) and (h), $1/2$ in panels (f) and (i), and $1/8$ in panels (g) and (j).
    The stronger ($J/\omega_q = 1/8$, blue curves) and weaker ($J/\omega_q = 1/12$, red curves) couplings are considered.
    The filled circles indicate the maximum value of the concurrence.
    \label{fig:iSWAPHomo}}
\end{figure*}

\textcolor{black}{Interestingly, ESD is observed only for the cases without RWA; it is \emph{not} observed when the RWA is imposed [see the insets of Figs.~\ref{fig:RWA_subOhmic}(iv)--\ref{fig:RWA_subOhmic}(vi)]:}
The concurrence reaches exactly zero at a finite time for the exact results, while it approaches zero \emph{only asymptotically} for the RWA cases.
The time at which the ESD occurs is larger for smaller $s$, corresponding to the magnitude of the difference between the exact and approximate results.
Similar to the finite dark periods discussed above, the counter-rotating terms play a crucial role in the ESD.
We here emphasize that even if the initial state is prepared into the maximally entangled Bell states, and even if the temperature is zero, the ESD occurs by appropriately choosing the spectral noise power and by exactly including the counter-rotating terms.

Now, we discuss our results in light of the previous study~\cite{WangNJP2013}.
The ESD for the maximally entangled Bell initial states at zero temperature has already been suggested there, in which a Lorentzian noise spectrum with a much broader linewidth was considered with a fixed ratio $\kappa_j / \lambda_j$.
When the spectral density has narrow peaks, a few reservoir modes with frequencies located around the peaks are strongly coupled to the qubit system.
This leads to coherent dynamics of the composite system (qubits and those reservoir modes), and hence oscillatory behavior is observed.
\textcolor{black}{The frequency of this oscillation becomes higher for larger coupling strength $\kappa_j$ [cf.~Figs.~\ref{fig:RWA_Lorentz}(a)-(iii) and (vi)].
We remark that this oscillatory behavior must be distinguished from the Lamb shift~\cite{Breuer2002, NakamuraPRB2024}, which affects only the dynamics of off-diagonal elements~\cite{Breuer2002}, while here both diagonal and off-diagonal elements are influenced by the coherent dynamics mentioned above (cf.~Appendix~\ref{sec:appRWA}).}
In contrast, when the spectral density has broad peaks or a broadband structure, a larger number of reservoir modes with a wide range of frequencies interfere, and thus the oscillatory behavior is suppressed.
Note that too large values of the linewidth $\lambda_j$ in Eq.~\eqref{eq:sdLorentz} result in unphysical predictions, since the conditions $S_j^{\beta \to \infty}(\omega < 0) = 0$ at zero temperature and $S_j^\beta(-\omega) = e^{-\beta \hbar \omega} S_j^\beta(\omega)$ at finite temperature must be fulfilled (see also Appendix \ref{sec:appRWA}).

\section{Dynamics of entanglement during and after operation of a two-qubit gate} \label{sec:resultISWAP}
In the previous section, we focused on a fairly theoretical perspective and elucidated the role of the counter-rotating terms in the interaction between the qubit system and the respective noisy reservoirs.
For this purpose, the temperature of the reservoirs was set to zero, and the initial states were taken as the ideal Bell states.
Here, we move towards a more practical setting:
Entangled states of the two-qubit system are prepared by applying a \sqrtiswap{} gate, in the presence of noise sources, to a factorized state, $\hat{\rho}_\mathrm{tot}(t) = \ketbra{10}{10} \otimes \hat{\rho}_{R, 1}^{\eq} \otimes \hat{\rho}_{R, 2}^{\eq}$, and the dynamics of disentanglement after this operation (idling) is monitored.
Here, the \sqrtiswap{} gate is defined by
\begin{align}
    \begin{bmatrix}
        1 & & & \\
        & 1/\sqrt{2} & -i / \sqrt{2} & \\
        & -i / \sqrt{2} & 1 / \sqrt{2} & \\
        & & & 1
    \end{bmatrix}\, .
\end{align}

The schematics of this sequence are depicted in Fig.~\ref{fig:iSWAPHomo}(a):
A direct qubit--qubit coupling, i.e., a finite $J$ in Eq.~\eqref{eq:totH}, acts over a finite time span to create an entangled state, ideally $(\ket{10} - i\ket{01})/\sqrt{2}$ for an isolated system, and is switched off at time $t = \pi / (4J)$.
\textcolor{black}{Ideally, the superposition of those states $\ket{10}$ and $\ket{01}$ is conserved even after switching off the coupling [cf.~Fig.~\ref{fig:iSWAPHomo}(b)].}
We adopt a finite temperature of the reservoirs $\beta \hbar \omega_q = 5$ and consider broadband noise only, see Eq.~\eqref{eq:sdOhmic}.
Since no single-qubit gates are applied [$\Omega_1 (t) = \Omega_2(t) = 0$], the RDO remains in $X$ states during the time evolution, which is proven in Appendix~\ref{sec:appXState}.

We proceed in four steps: First, we explore the role of various spectral exponents (identical spectral exponents for both reservoirs) as well as the gate times; second, the strength of the qubit--reservoirs coupling is varied; third, heterogeneous reservoirs (different spectral exponents) are considered; and fourth, the focus is laid on memory effects.

\subsection{Dependence on spectral exponent and gate time}
Figure~\ref{fig:iSWAPHomo}(c) displays the concurrence for various spectral exponents with fixed system--reservoir coupling strength, $2 \pi \hbar \kappa = 0.04$.
\textcolor{black}{Comparing Fig.~\ref{fig:iSWAPHomo}(b), one finds a degraded concurrence and its decay in Fig.~\ref{fig:iSWAPHomo}(c).}
Obviously, the gate performance in terms of the concurrence is the best for the case with $s = 1/2$ at the end of the gate [vertical dashed line in Fig.~\ref{fig:iSWAPHomo}(c)].
This trend was already predicted for single-qubit gates in previous studies~\cite{NakamuraPRR2024}:
The relatively small portion of both low- and high-frequency modes of the spectral noise power is beneficial for generating entangled states.

For an Ohmic reservoir ($s = 1$) and in the short-time domain, $0 \leq \omega_q t / \pi \lesssim 0.1$, entanglement is never generated.
This is because the negative contribution of the diagonal terms $\sqrt{\rho_{11}(t)\rho_{44}(t)}$ prevails over the generation of the coherence $|\rho_{23}(t)|$.
According to Eq.~\eqref{eq:concXState}, the concurrence remains zero when $|\rho_{23}(t)| < \sqrt{\rho_{11}(t)\rho_{44}(t)}$ even if the coherence $|\rho_{23}(t)|$ takes a finite value.
Because of the fast increase of the populations $\rho_{11}(t)$ and $\rho_{44}(t)$ similar to the case in Sec.~\ref{sec:resultRWA}, the above inequality holds in the short-time period.

For the deep sub-Ohmic reservoir ($s = 1/8$), the concurrence exhibits interesting behavior during the idling after the gate operation:
Compared to the other two noise cases, faster disentanglement and more intense oscillations emerge.
Both components of the concurrence in Eq.~\eqref{eq:concXState} contribute to this behavior, i.e., the fast decoherence of $|\rho_{23}(t)|$ [dashed curve in the inset of Fig.~\ref{fig:iSWAPHomo}(c)] determines the envelope of the concurrence, while the oscillatory behavior is enhanced by the dynamics of $\sqrt{\rho_{11}(t)\rho_{44}(t)}$ (solid curve).
Similar to single-qubit dynamics, decoherence of the off-diagonal elements sets in faster as the spectral exponent gets smaller.
The oscillatory behavior of the population is caused by enhanced memory effects (retarded feedback) of the reservoir~\cite{NakamuraPRR2024}.
In this case, the oscillation of $\rho_{11}(t)$ is especially reflected in the concurrence.

Next, we explore effects of the qubit--qubit coupling strength (gate time).
Figures~\ref{fig:iSWAPHomo}(e)--\ref{fig:iSWAPHomo}(g) depict the concurrence for $J / \omega_q = 1/8$ and $1/12$, with various exponents, (e) $s = 1$, (f) $1/2$, and (g) $1/8$.
Note that the blue curves in Figs.~\ref{fig:iSWAPHomo}(e)--\ref{fig:iSWAPHomo}(g) correspond to each curve in Fig.~\ref{fig:iSWAPHomo}(c).
The blue and red vertical lines correspond to the final time of the gate operation with the coupling strength $J/\omega_q = 1/8$ and $1/12$, respectively.
One finds a similar behavior after the time indicated by the red line ($\omega_q t / \pi > 3$).
The states of the compound consisting of the two-qubit system correlated with their reservoirs at the time $\omega_q t / \pi = 3$ appear to be almost the same regardless of the different values for the qubit coupling parameter.
This is because of the relatively small coupling strengths, $\omega_q/8$ and $\omega_q/12$:
The Larmor precession of the qubits and the system--reservoir coupling dominate the time evolution of the correlated system.
This minor contribution of the qubit--qubit coupling can be verified through the peak position of the concurrence.
For the weaker coupling case, the maximum concurrence is located at a time much smaller than the time indicated by the red line.
For the deep sub-Ohmic case, both of the peaks with the stronger and weaker coupling are located at almost the same time.
Because of the relatively strong coupling to the reservoirs, the direct qubit--qubit coupling can no longer generate the entanglement during the time period $2 \leq \omega_q t / \pi \leq 3$ in this case.
It is worth noting that the peak position of the case $s = 1/8$ and $J / \omega_q = 1/12$ is highly affected by the oscillation of the population $\rho_{11}(t)$.

\subsection{Weaker system--reservoir coupling}
In this subsection, we focus on the dynamics with a very weak system--reservoir coupling and set $2 \pi \hbar \kappa=0.004$.
The time traces of the concurrence are displayed in Figs.~\ref{fig:iSWAPHomo}(d) and \ref{fig:iSWAPHomo}(h)--\ref{fig:iSWAPHomo}(j).
In Fig.~\ref{fig:iSWAPHomo}(d), one can see that the gate performance improves compared to the strong system--reservoir-coupling case [Fig.~\ref{fig:iSWAPHomo}(c)], as expected.
The difference of the concurrences between different spectral exponents is negligible during the gate application [inset of Fig.~\ref{fig:iSWAPHomo}(d)].
However, one finds an obvious difference during the idling:
Only the concurrence with the deep sub-Ohmic reservoir exhibits oscillations, a finding that requires to take 
the whole profile of the spectral noise power into account, even in this small system--reservoir-coupling regime.

Comparing the dynamics with different $J$ in Figs.~\ref{fig:iSWAPHomo}(h)-\ref{fig:iSWAPHomo}(j), we see that the concurrences after the time $\omega_q t / \pi = 3$ are almost the same, which is consistent with the case with the larger $\kappa$ in Figs.~\ref{fig:iSWAPHomo}(e)--\ref{fig:iSWAPHomo}(g).
Even in the case of the weaker system--reservoir coupling, the contribution of the reservoirs to the system dynamics is found to be dominant compared to the direct qubit--qubit coupling in our setting.
The main difference between the lager and smaller $\kappa$ is that the peak with the smaller $J$ is almost on the red vertical dashed line.
The concurrence with $J / \omega_q = 1/12$ manifestly increases during the period $2 \leq \omega_q t / \pi \leq 3$.

It is worth noting that the peak position in the case with  $J/\omega_q  = 1/12$ and $s = 1/8$ [red curve in Fig.~\ref{fig:iSWAPHomo}(j)] is located  after $\omega_q t / \pi = 3$.
The oscillation of $\sqrt{\rho_{11}(t)\rho_{44}(t)}$, which results from the memory effects of the reservoirs~\cite{NakamuraPRR2024}, can generate the entanglement even after switching off the direct qubit--qubit coupling.

\begin{table}[t]
    \caption{\textcolor{black}{Fidelity between the ideal states [cf.~Fig.~\ref{fig:iSWAPHomo}(b)] and dissipative states at the end of the gate operation [vertical lines in Figs.~\ref{fig:iSWAPHomo}(c) and \ref{fig:iSWAPHomo}(d)].
    The qubit--qubit coupling $J$, the spectral exponent $s$, and the qubit--reservoir coupling $\kappa$ are varied.}
    \label{tbl:fidelity_iSWAP}}
    \centering
    \begin{ruledtabular}
    \begin{tabular}{rcccccc}
        $J$ & \multicolumn{3}{c}{$\omega_q/8$} & \multicolumn{3}{c}{$\omega_q/12$}\\
        \cmidrule{2-4} \cmidrule{5-7}
        \diagbox{$2\pi\hbar\kappa$}{$s$} & $1$ & $1/2$ & $1/8$ & $1$ & $1/2$ & $1/8$ \\
        \midrule
        $0.04~~$ & $0.74$ & $0.76$ & $0.76$ & $0.66$ & $0.67$ & $0.63$ \\
        $0.004$ & $0.97$ & $0.97$ & $0.97$ & $0.96$ & $0.96$ & $0.95$
    \end{tabular}        
    \end{ruledtabular}
\end{table}

\subsection{\textcolor{black}{Gate fidelity}}
It is instructive to include at this point an analysis of the \sqrtiswap{}-gate fidelity.
Its dependence on various parameters is collected in  Table~\ref{tbl:fidelity_iSWAP}, where the fidelities at the time $t = \pi / (4 J)$ between the ideal isolated system [Fig.~\ref{fig:iSWAPHomo}(b)] and the dissipative systems in Figs.~\ref{fig:iSWAPHomo}(c)--\ref{fig:iSWAPHomo}(j) are listed.
One can see that the difference in the fidelity caused by the spectral exponent is larger for the weaker qubit--qubit coupling $J$ (longer gate time).
Non-Markovianity plays a more important role for the longer gate time.
As indicated in the previous subsection, the weaker system--reservoir coupling leads to the better fidelity.

Note that we only consider a single initial state, $\ket{10}$, in this analysis.
Usually, average gate fidelity of different initial states should be considered to measure the gate quality.
We address this quantity and discuss gate qualities in more detail in the next section.

\subsection{Heterogeneous environment}
Next, heterogeneous environments, i.e., $s_1 \neq s_2$,  are analyzed, which implies that each qubit is sufficiently separated in space.
For simplicity, we only focus on different values for $s_j$ while in principle other parameters can be considered as well (cf.~Ref.~\cite{WangPRA2019}).

\begin{figure}
    \centering
    \includegraphics[width=0.84\linewidth]{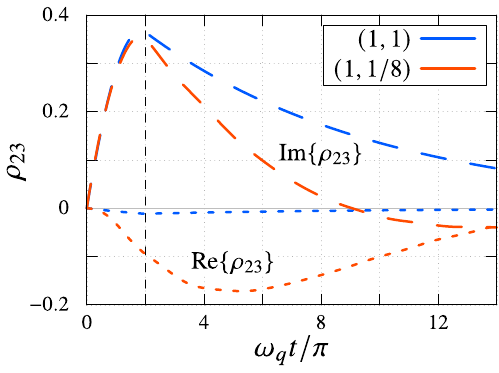}
    \caption{Dynamics of the off-diagonal element $\rho_{23}(t)$ in heterogeneous environments ($s_1 \neq s_2$) during and after application of a \sqrtiswap{} gate.
    The dotted and dashed curves represent the real and imaginary part of $\rho_{23}(t)$, respectively.
    The results with the exponents $(s_1, s_2) = (1, 1/8)$ (red curve) are depicted, and a homogeneous case $(s_1, s_2) = (1, 1)$ is also displayed as a reference (blue curve).
    The vertical dashed line indicates the end time of the gate operation.
    The coupling strengths between the qubits and the qubits and reservoirs are fixed to $J/\omega_q = 1/8$ and $2 \pi \hbar \kappa = 0.04$, respectively.
    \label{fig:iSWAPHetero}}
\end{figure}

\begin{figure*}
    \centering
    \includegraphics[width=0.99\linewidth]{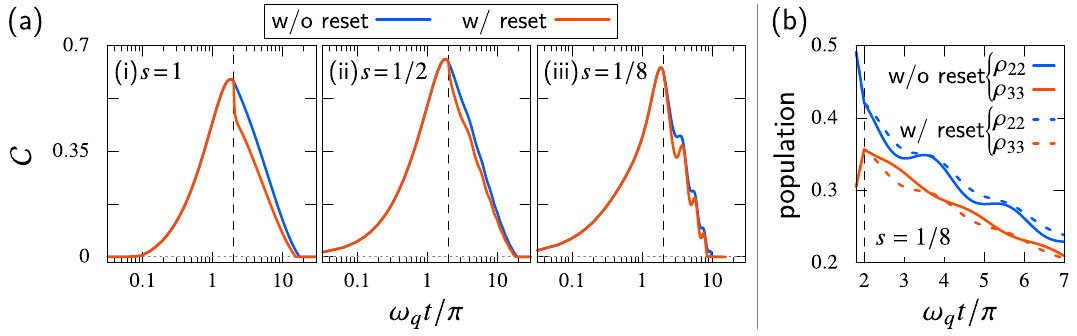}
    \caption{(a) Time traces of the concurrence $\mathcal{C}(t)$ during and after application of a \sqrtiswap{} gate (linear--log plots). The exact results (w/o reset, blue curves) and the results including an environmental reset (w/ reset, red curves) at the time $\omega_q t / \pi = 2$ (vertical dashed line) are depicted.
    For the details of the environmental reset, see the main text.
    The homogeneous environments are considered here ($s_1 = s_2 = s$), and the spectral exponent is varied in subpanels (i)--(iii).
    The coupling strengths between the qubits and the qubits and the reservoirs are fixed to $J/\omega_q = 1/8$ and $2 \pi \hbar \kappa = 0.04$, respectively.
    (b) Dynamics of the populations $\rho_{22} = \braket{10|\hat{\rho}_S(t)|10}$ (blue curves) and $\rho_{33} = \braket{01|\hat{\rho}_S(t)|01}$ (red curves) during the same gate sequence as (a) (linear--linear plot).
    Again, the exact results (w/o reset, solid curves) and those including the reset (w/ reset, dotted curves) are depicted.
    Here, only the case with the exponent $s = 1/8$ is considered, and the other parameters are the same as (a).
    \label{fig:iSWAPReset}}
\end{figure*}

Special interest is laid on the dynamics of the off-diagonal element $\rho_{23} (t)$ to illustrate the impact of the heterogeneity of the environment, and a more general discussion is given in Appendix~\ref{sec:appISWAPHetero}.
Figure~\ref{fig:iSWAPHetero} shows the dynamics of $\rho_{23}(t)$ during and after the application of a \sqrtiswap{} gate.
A homogeneous [$(s_1, s_2) = (1, 1)$] and a heterogeneous [$(s_1, s_2) = (1, 1/8)$] environment are considered.
In an ideal case with an isolated system, the imaginary part reaches $0.5$ at the end of the gate operation (vertical dashed line in Fig.~\ref{fig:iSWAPHetero}) and remains at this value during the idling, while the real part is always zero.
In the presence of noise, the imaginary part never reaches $0.5$ during the gate operation and begins to decay after switching off the qubit--qubit coupling.
A main difference in the imaginary part between the homogeneous and heterogeneous environment is that in the homogeneous case, it always takes positive values, and its decay is monotonic, while it takes both positive and negative values, and damped oscillation is observed in the heterogeneous case.

A clearer difference is seen in the real part of the matrix element:
For the homogeneous environment, the value remains almost zero, while a significant growth is found in the heterogeneous environment during the application of the gate.
Within the rotating frame with the effective Larmor frequencies $\tilde{\omega}_1^q$ and $\tilde{\omega}_2^{q}$, the Hamiltonian of the qubit--qubit coupling is expressed as
\begin{align}
    \hbar J (\hat{\sigma}_1^{+}\hat{\sigma}_2^{-}e^{i(\tilde{\omega}_1^q - \tilde{\omega}_2^q)t} + \hat{\sigma}_1^{-}\hat{\sigma}_2^{+} e^{-i(\tilde{\omega}_1^q - \tilde{\omega}_2^q)t})\, ,
    \label{eq:cplRotating}
\end{align}
with a time-independent coupling strength $J$.
When the effective frequencies of each qubit are the same, the time-dependent part of the above Hamiltonian vanishes, and the state $\ket{10}$ is ideally transformed into the state $(\ket{10} - i \ket{01})/ \sqrt{2}$ after a corresponding time period, which indicates the off-diagonal element $\rho_{23} = i /2$.
This is true for the isolated systems, because the effective Larmor frequencies are exactly the same in our study, $\tilde{\omega}_1^q = \tilde{\omega}_2^q = \omega_q$.
While the effective frequencies deviate from the bare frequency due to reservoir effects~\cite{NakamuraPRB2024} (cf.~Lamb shift), they take the same value in the homogeneous case since all the parameter values of each reservoir are the same.
This is the reason why the dynamics of the real part are suppressed, although the values are not exactly zero owing to reservoir-induced effects.
By contrast, the qubits rotate at different frequencies in the heterogeneous case ($\tilde{\omega}_1^q \neq \tilde{\omega}_2^q$).
Therefore, the exponential part in Eq.~\eqref{eq:cplRotating} plays a role, and both the real and imaginary part take nonzero values.
The real part in the heterogeneous case keeps increasing after the gate operation in a certain time period and then decreases.
This is related to the oscillation of the imaginary part:
Since the decay dynamics of the absolute value $|\rho_{23} (t)|$ was found to be similar in the homogeneous and heterogeneous case (results are not shown), it is concluded that the reservoirs only modulate the phase of the matrix element over time [note that the system Hamiltonian without the qubit--qubit coupling does not affect the off-diagonal element $\rho_{23} (t)$, so that the oscillation does not occur when the system is isolated].

\textcolor{black}{It is worth noting that this oscillatory behavior induced by the heterogeneous environment can be observed only by monitoring the single matrix element, $\rho_{23}(t)$.
Experimentally, readout of a certain matrix element of a system density operator is less expensive than procedures for obtaining the fidelity and the concurrence.
Namely, the heterogeneity can be detected in a relatively simple manner.}

\subsection{Remarks on memory effects}
Finally, we investigate memory effects of the reservoirs and for this, 
return to homogeneous environments $s_1 = s_2  =s$.
Because of a finite decay rate of the autocorrelation function $C_j (t)$, the dynamics of the RDO at a given time is affected by its properties in the past.
In this sense, the environment has ``memory effects'' that cannot be described within a standard Lindblad equation.
In particular, the dynamics after the application of the \sqrtiswap{} gate is affected by the one during the gate application.

Here, we elucidate this correlation between the gate-application segment and the idling segment through the following reset technique:
The contribution of the RDO during the gate operation $0 \leq t \leq t_g = \pi / (4J)$ to the subsequent dynamics can be disregarded by resetting the total density operator to a factorized state, i.e.,  $\hat{\rho}_\mathrm{tot} (t) \to \mathrm{tr}_R\{\hat{\rho}_\mathrm{tot}(t)\}\otimes \hat{\rho}_{R, 1}^{\eq} \otimes \hat{\rho}_{R, 2}^{\eq}$, at the time $t_g$.
This reset corresponds to setting the ADOs to zero, which is expressed as $\hat{\rho}_{(\vec{m}, \vec{n}) \neq (\vec{0}, \vec{0})}(t) \to 0$ within the framework of the HEOM approach.
In experiments, we can obtain results with the reset scheme by exploiting the Kraus operators for single-qubit systems, $\hat{\rho}_\mathrm{single}(t) = \sum_{i, j} \chi_{i, j} (t) \hat{A}_i \hat{\rho}_\mathrm{single}(0) \hat{A}_j$ (cf.~quantum process tomography~\cite{ChuangJMOpt1997}).
We assume factorizing initial conditions $\hat{\rho}_\mathrm{single}(0) \otimes \hat{\rho}_{R}^{\eq}$.
Since the reservoirs are independent of each other, the dynamics of the two-qubit systems after the gate operation is evaluated with the tensor products of the Kraus operators for the single-qubit systems as
\begin{align}
    \hat{\rho}_S(t+t_g) = \sum_{i, j} \sum_{k, l} \chi_{i, j}(t) \chi_{k, l}(t)
    \hat{A}_i \otimes \hat{A}_k \hat{\rho}_S(t_g) \hat{A}_j \otimes \hat{A}_l \, .
\end{align}

Figure~\ref{fig:iSWAPReset}(a) depicts  the concurrence of the system with and without the above reset scheme.
The coupling strengths between the qubits and the qubits and the reservoirs are fixed to $J / \omega_q = 1/8$ and $2 \pi \hbar \kappa = 0.04$, respectively.
Exact results without reset (blue curves) correspond to the curves in Fig.~\ref{fig:iSWAPHomo}(c).
One can see that the concurrence with the reset scheme is always worse than the result without.
Since the factorized state at time $t_g$ is far from the correlated state $\hat{\rho}_\mathrm{tot}(t_g)$, it gradually moves towards a correlated state at the expense of the qubit--qubit entanglement.
Especially in the Ohmic case ($s = 1$), one finds a sudden change of the state right after the time $t_g$, which is caused by the large portion of high-frequency modes of the spectral noise power [cf.~sudden decay of the concurrence in Fig.~\ref{fig:RWA_subOhmic}(i)].

\begin{figure*}
    \centering
    \includegraphics[width=1.0\linewidth]{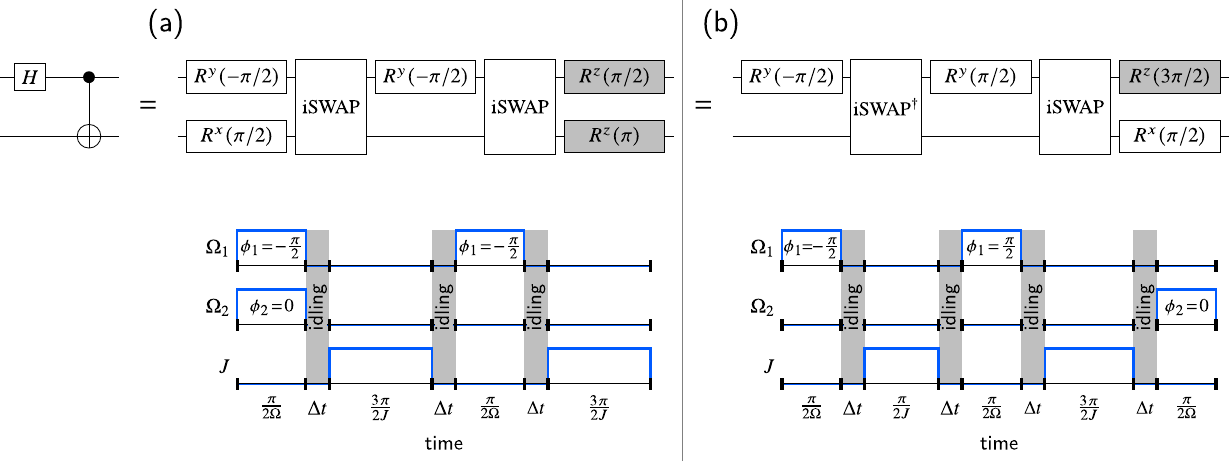}
    \caption{(a), (b) Decomposition schemes of the Hadamard ($H$) + CNOT sequence.
    Upper panels: Corresponding sequences consisting of the elemental gates.
    The $R^{\alpha} (\theta)$ gate is a single-qubit gate with rotation around an axis $\alpha$ by an angle $\theta$.
    The shaded boxes associated with $z$-axis rotation correspond to the virtual $Z$ gates, which are not applied in our simulations.
    The iSWAP$^\dagger$ gate is the complex conjugate of the iSWAP gate.
    Lower panels: Schematics of pulse sequences.
    The sequences of the pulse amplitude of each qubit, $\Omega_1(t)$ and $\Omega_2(t)$, and the strength of the qubit--qubit coupling $J(t)$ are depicted.
    The sequence is expressed as a step function.
    For the single-qubit gates, the phase of the pulse $\phi_j$ is indicated in the rectangular area where the gate is applied.
    Idlings during which no gates are applied are inserted between successive gates (gray boxes), and their durations are denoted by $\Delta t$.
    \label{fig:schematicH+CNOT}}
\end{figure*}

In previous studies~\cite{BellomoPRL2007}, it was proposed that once the above Kraus operators for the single-qubit systems are obtained, the dynamics of the two-qubit system can be predicted.
However, this is found \emph{not} to be true when memory effects are present, correlating the idling segment with the pulse-application segment.
\textcolor{black}{The initial state was assumed to be a factorized state in which two qubits are entangled, but in practice, this assumption does not hold because of the two-qubit-gate operation which gives also rise to qubit--reservoir correlations.}
Predictions with the Kraus operators will underestimate the concurrence, as discussed above, where, of course, the discrepancy depends on the strength of the system--reservoir coupling.
For example, we find that the maximum difference of the concurrences between cases with and without reset is approximately $0.01$ in a weaker coupling case ($2 \pi \hbar \kappa = 0.004$, results are not shown), which is approximately ten times as small as the difference in Fig.~\ref{fig:iSWAPReset}(a).
Nevertheless, for reliable predictions for high-perfomance qubits, it seems mandatory to include memory effects.

In Fig.~\ref{fig:iSWAPReset}(b), the dynamics of the populations $\rho_{22} (t)$ and $\rho_{33}(t)$ with and without the reset scheme is depicted.
In the deep sub-Ohmic case ($s = 1/8$), the populations exhibit oscillations~\cite{NakamuraPRR2024}, and one observes a qualitative difference of the phase of the oscillation between the two cases:
In the case without reset, the difference $\rho_{22}(t) - \rho_{33}(t)$ increases and decreases over time.
By contrast, the difference is almost time independent in the case with the reset. 
It is concluded that the phases of the oscillations are almost opposite in the former case, while the oscillations are almost in phase in the latter.

\section{Application to Hadamard + CNOT sequences} \label{sec:resultHCNOT}
As a final demonstration for a practical gate operations, we consider a sequence consisting of a single Hadamard ($H$) and a single CNOT gate.
Two decomposition schemes of the $H$ + CNOT sequence based on previous studies~\cite{SchuchPRA2003,ThorwartPRA2001} are displayed in Figs.~\ref{fig:schematicH+CNOT}(a) and \ref{fig:schematicH+CNOT}(b).
In the upper panels, the single-qubit gates consist of $R^{\alpha}(\theta)$, where the rotation axis and angle are denoted by $\alpha$ and $\theta$, respectively.
When the rotation axis sits in the $x$--$y$ plane, the rotation operator is expressed as $\exp[-i \theta_j (\hat{\sigma}_j^x \cos \phi_j + \hat{\sigma}_j^y \sin \phi_j)/2]$, corresponding to the time evolution operator with respect to the first term of Eq.~\eqref{eq:eachH} within the rotating frame under the condition $\omega_j^{\mathrm{ex}} = \omega_j^q$.
Here, we adopt this condition (see below).
The gate time is determined by the angle and the pulse amplitude given by $\theta_j / \Omega_j$.
For the $z$-axis rotation, we consider the virtual $Z$ gate~\cite{McKayPRA2017}, which, as a gate terminating the sequence, is not included in our simulation as it 
 only changes the phase of the RDO. Thus, it has no effect when only populations are monitored at the end of the sequence.
The iSWAP gate is expressed in the matrix form as
\begin{align}
    \begin{bmatrix}
        1 &&& \\
        &&i& \\
        &i&& \\
        &&&1
    \end{bmatrix}\, ,
\end{align}
with the iSWAP$^\dagger$ gate being the complex conjugate.
We can apply these gates by virtue of the direct qubit--qubit coupling, and the gate times of them are given by $3 \pi / (2J)$ and $\pi / (2J)$, respectively.

The lower panels of Figs.~\ref{fig:schematicH+CNOT}(a) and \ref{fig:schematicH+CNOT}(b) are schematics of the corresponding pulse sequences.
Here, we consider the limit of instantaneously
switching on and off the respective gates.
\textcolor{black}{The rectangular areas ($\Omega_j \neq 0$ or $J \neq 0$) correspond to the elemental gates appearing in the upper panels of Figs.~\ref{fig:schematicH+CNOT}(a) and \ref{fig:schematicH+CNOT}(b), and for the single-qubit gates, the phase $\phi_j$ is indicated in the corresponding area.}
Gate times are indicated below the baseline, and we adopt identical values for the amplitudes of each pulse, i.e., $\Omega_1 = \Omega_2 = \Omega$.
In addition, idlings are inserted between successive gates.
This is motivated by the actual experimental situation:
Since experimentally  step functions for pulses cannot be prepared, one has to add ``buffer time'' in order to make sure that application of a pulse is completely finished with no overlap with  the subsequent pulse.
Here, the duration of the idling is denoted by $\Delta t$.

For the simulations, the following parameter values have been used:
For the single-qubit gates, the drive frequency and the amplitude of the pulse is given by $\omega_1^{\mathrm{ex}} = \omega_2^{\mathrm{ex}} = \omega_q$ and $\Omega = \omega_q$, respectively; the strength of the qubit--qubit coupling is set to $J = \omega_q / 8$; we consider homogeneous environments, $s_1 = s_2 = s$, with
coupling strength to the qubit system given by $2 \pi \hbar \kappa = 0.004$ corresponding to the weaker coupling cases in Sec.~\ref{sec:resultISWAP}.
We vary the spectral exponent $s$, the initial state of the system $\rho_S(0) = \ketbra{\psi_0}{\psi_0}$, and the idling time $\Delta t$ as $s = 1$, $1/2$, and $1/8$, $\ket{\psi_0} = \ket{11}$, $\ket{10}$, $\ket{01}$, and $\ket{00}$, and $\omega_q \Delta t = 0$ and $\pi/2$, respectively.
The total time of the whole sequence is evaluated as $\pi / \Omega + 3 \pi / J + 3 \Delta t$ and $3\pi / (2\Omega) + 2\pi / J + 4 \Delta t$ for sequences (Seqs.)~(a) and (b).
With these parameter values, one obtains the total time $25 \pi / \omega_q$ and $26.5 \pi / \omega_q$ for Seq.~(a) without and with the idling, and $17.5 \pi / \omega_q$ and $19.5 \pi / \omega_q$ for Seq.~(b) without and with the idling, respectively.

\begin{table}[h]
    \caption{Fidelity between the ideal states and the final states obtained with the time evolution according to Seqs.~(a) and (b) in Fig.~\ref{fig:schematicH+CNOT}.
    The values are the average of the different spectral exponents ($s=1$, $1/2$, and $1/8$).
    Initial states $\ket{\psi_0}$ and idling times $\Delta t$ are varied.
    \label{tbl:fidelity_variousIdling}}
    \begin{ruledtabular}
    \begin{tabular}{cccccc}
        & \diagbox{$\omega_q \Delta t$}{$\ket{\psi_0}$} & $\ket{11}$ & $\ket{10}$ & $\ket{01}$ & $\ket{00}$ \\
        \cmidrule{1-2} \cmidrule{3-6}
        \multirow{2}{*}{Seq. (a)} & $0$ & $0.75$ & $0.75$ & $0.75$ & $0.76$ \\
        & $\pi / 2$ & $0.74$ & $0.74$ & $0.74$ & $0.74$ \\
        \addlinespace
        \multirow{2}{*}{Seq. (b)} & $0$ & $0.81$ & $0.86$ & $0.82$ & $0.86$ \\
        & $\pi / 2$ & $0.79$ & $0.85$ & $0.80$ & $0.85$ \\
    \end{tabular}
    \end{ruledtabular}

    \caption{Concurrence of the final states of Seqs.~(a) and (b) in Fig.~\ref{fig:schematicH+CNOT}.
    The values are the average of the different spectral exponents ($s=1$, $1/2$, and $1/8$).
    The same parameter values as Table~\ref{tbl:fidelity_variousIdling} are considered.
    \label{tbl:concurrence_variousIdling}}
    \begin{ruledtabular}
    \begin{tabular}{cccccc}
        & \diagbox{$\omega_q \Delta t$}{$\ket{\psi_0}$} & $\ket{11}$ & $\ket{10}$ & $\ket{01}$ & $\ket{00}$ \\
        \cmidrule{1-2} \cmidrule{3-6}
        \multirow{2}{*}{Seq. (a)} & $0$ & $0.55$ & $0.61$ & $0.55$ & $0.63$ \\
        & $\pi / 2$ & $0.53$ & $0.59$ & $0.53$ & $0.61$ \\
        \addlinespace
        \multirow{2}{*}{Seq. (b)} & $0$ & $0.65$ & $0.75$ & $0.67$ & $0.76$ \\
        & $\pi / 2$ & $0.62$ & $0.74$ & $0.63$ & $0.75$ \\
    \end{tabular}
    \end{ruledtabular}
\end{table}

\subsection{Analysis of performance}
The performance of the $H$ + CNOT sequence is investigated through the fidelity and the concurrence as specified in Eqs.~\eqref{eq:conc} and \eqref{eq:fidelity} taken at the end time.

We first turn to the effects of the initial states and the idling times on the performance.
Here, the initial state is expressed as $\hat{\rho}_S(0) = \ketbra{\psi_0}{\psi_0}$.
In Tables \ref{tbl:fidelity_variousIdling} and \ref{tbl:concurrence_variousIdling}, the average fidelities and concurrences of the different spectral exponents are listed, respectively.
The following trends can be identified:
(1) Sequence (b) performs better than Seq.~(a).
(2) The sequences without the idling ($\Delta t = 0$) are better than those with the idling ($\omega_q \Delta t = \pi / 2$).
(3) The initial states $\ket{10}$ and $\ket{00}$ indicate better performance compared to the states $\ket{11}$ and $\ket{01}$.
From the trends (1) and (2), it is concluded, as expected, that shorter sequences are preferable in the presence of noise.
Although partial recovery of the fidelity during transient times is observed when idlings are inserted, consistent with previous studies~\cite{NakamuraPRR2024} (see Appendix~\ref{sec:appHCNOT}), this recovery does not contribute to the final performance.
With the inclusion of the idling, results are even worse compared to the results without the idling.
We found that the partial recovery of the fidelity is caused by intense reconfiguration processes of the reservoirs that occur when the total system is far away from a stationary state~\cite{NakamuraPRR2024}.
During the long-time period of the sequence, the reservoirs approach the equilibrium state, and in such a long-time domain, significant feedback from the reservoirs to the system cannot be observed.
As a result, longer sequence times are detrimental.

As for the trend (1), the final gate plays a role:
For Seq.~(a), the final gate is the iSWAP gate, and the populations $\rho_{11}(t)$ and $\rho_{44}(t)$ are not affected by the gate.
In this sense, those populations are under the free relaxation process during the gate operation, and the performance deteriorates.
By contrast, the final gate $R^x(\pi/2)$ in Seq.~(b) applied to the second qubit affects all RDO elements, and hence the relaxation process of $\rho_{11}(t)$ and $\rho_{44}(t)$ is suppressed.
As a result, Seq.~(b) performs better than Seq.~(a).

The trend (3) is obvious when one looks at the concurrences (Table~\ref{tbl:concurrence_variousIdling}), while the fidelities (Table~\ref{tbl:fidelity_variousIdling}) for Seq.~(a) take almost the same value for the various initial states.
The reason for trend (3) is related to the final states of the two-qubit system:
With the initial states $\ket{11}$ and $\ket{01}$, the ideal final state is $(\ket{11} + e^{i\varphi}\ket{00})/ \sqrt{2}$, while it is $(\ket{10} + e^{i\varphi}\ket{01})/ \sqrt{2}$ with the initial states $\ket{10}$ and $\ket{00}$ with a certain phase $\varphi$.
Note that the CNOT gate is expressed in matrix form as $\ketbra{1}{1} \otimes I + \ketbra{0}{0} \otimes X$ in this study, where $I$ and $X$ are the identity and $X$ gate, respectively.
As discussed in Sec.~\ref{sec:resultRWA}, larger values of $\sqrt{\rho_{22}(t)\rho_{33}(t)}$ compared to $\sqrt{\rho_{11}(t)\rho_{44}(t)}$ lead to lower concurrences of states near $\ket{\Psi}$ compared to those near $\ket{\Phi}$.
Since the single-qubit gates are applied, the RDO at the end time is no longer in the $X$ states, and, strictly speaking, Eq.~\eqref{eq:concXState} cannot be used in this case.
\textcolor{black}{Thus, the eigenvalues $\{\Lambda_j\}$ in Eq.~\eqref{eq:conc} are numerically calculated in this case.}
However, the state is near the $X$ state (i.e., the values of the matrix elements that do not appear in the $X$ state are almost zero), and hence the above discussion is roughly valid.
In terms of the fidelity, we must maintain the population of the most excited state $\rho_{11}(t)$ at the value $0.5$ in the cases where the ideal state is $(\ket{11} + e^{i\varphi}\ket{00})/ \sqrt{2}$.
This is more difficult than the task to maintain the populations of $\rho_{22}(t)$ and $\rho_{33}(t)$ for the ideal state $(\ket{10} + e^{i\varphi}\ket{01})/ \sqrt{2}$ due to the relaxation process.

\begin{table}[t]
    \caption{Average fidelity $F$ and concurrence $\mathcal{C}$ of different initial states ($\ket{\psi_0} = \ket{11}$, $\ket{10}$, $\ket{01}$, and $\ket{00}$).
    Results with various spectral exponents $s$ for Seqs.~(a) and (b) are listed.
    The idling time $\Delta t$ is fixed to $0$.
    \label{tbl:performance_variousExponent}}
    \begin{ruledtabular}
    \begin{tabular}{ccccccc}
        & \multicolumn{3}{c}{Seq. (a)} & \multicolumn{3}{c}{Seq. (b)} \\
        \cmidrule{2-4}
        \cmidrule{5-7}
        & $s=1$ & $s=1/2$ & $s=1/8$ & $s=1$ & $s=1/2$ & $s=1/8$ \\
        \midrule
        $F$ & $0.74$ & $0.77$ & $0.75$ & $0.83$ & $0.85$ & $0.83$ \\
        $\mathcal{C}$ & $0.58$ & $0.60$ & $0.57$ & $0.71$ & $0.72$ & $0.70$ \\
    \end{tabular}
    \end{ruledtabular}
\end{table}

Next, we turn to the performance for various spectral exponents, i.e., different types of reservoirs.
The average fidelities and concurrences of the four initial states are listed in Table~\ref{tbl:performance_variousExponent}.
Again, two sequences (a) and (b) are considered, and the idling time is fixed to zero here.
One finds that in both sequences, and in terms of both fidelity and concurrence, the exponent $s = 1/2$ is the best among the three types of reservoirs.
As discussed in Sec.~\ref{sec:resultISWAP} and previous studies~\cite{NakamuraPRR2024}, the profile of the spectral noise power plays a key role in the sequence performance.
Interestingly, the fidelity is better in the deep sub-Ohmic case than in the Ohmic case while for the concurrence the opposite is true.
It seems that the large portion of the low-frequency modes of the spectral noise power slightly contributes to the deterioration of the entanglement.

\begin{figure}[t]
    \centering
    \includegraphics[width=0.84\linewidth]{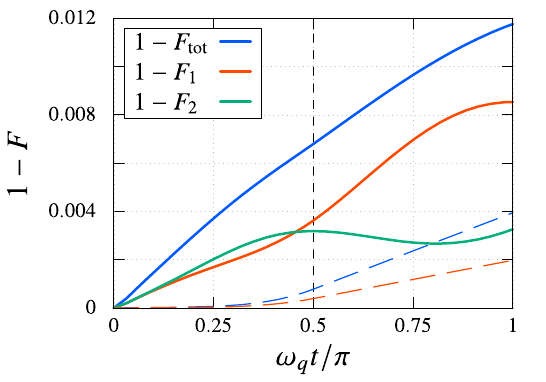}
    \caption{Time traces of the \textcolor{black}{infidelity $1-F(t)$} for Seq.~(a) with the initial state $\ketbra{00}{00}$ \textcolor{black}{(solid curves)}.
    The spectral exponent is $s = 1/8$.
    The dynamics during the first single-qubit gate $R^y(-\pi/2) \otimes R^x (\pi / 2)$ in the period $0 \leq \omega_q t / \pi \leq 0.5$ and the fist idling ($0.5 \leq \omega_q t / \pi \leq 1$) are depicted.
    The total fidelity is decomposed into the two parts of each qubit, $F_\mathrm{tot} (t) = F_1(t) F_2(t)$, and all the total and partial infidelities are displayed.
    \textcolor{black}{The infidelities obtained with the Lindblad equation are depicted as dashed curves, whose color scheme is same as that of the numerically rigorous ones.
    The partial fidelities $F_1$ and $F_2$ are identical for the Lindblad case, and therefore only one red dashed curve is displayed.}
    \label{fig:partialFidelity}}
\end{figure}

\subsection{Remarks on monotonic decay and non-Markovianity}
Finally, we discuss relations between monotonic decay and Markovianity.
The term ``Markovian'' process is sometimes used solely on the basis of monotonic decay of coherences, populations, or fidelities in open quantum systems.
\textcolor{black}{Here, we examine this assertion: When monotonic increase of infidelity is observed, then the process follows Markovian dynamics.}

Figure~\ref{fig:partialFidelity} displays the time traces of the infidelity under Seq.~(a) with the initial state $\ketbra{00}{00}$.
Here, those during the first single-qubit gate application [$R^y(-\pi / 2) \otimes R^x(\pi / 2)$] and the first idling (with the duration $\omega_q \Delta t = \pi / 2$) in the deep sub-Ohmic case $s = 1/8$ are  depicted.
Since no entanglement is generated until the end  in Fig.~\ref{fig:partialFidelity}, the total fidelity of the two-qubit system can be expressed as the product of the partial fidelities of each single qubit in the form of $F_\mathrm{tot}(t) = F_1(t) F_2(t)$.
\textcolor{black}{The time traces of the total infidelity, $1-F_\mathrm{tot}(t)$, and the partial infidelities of each qubit, $1-F_1(t)$ and $1-F_2(t)$ are shown in Fig.~\ref{fig:partialFidelity}.}

\textcolor{black}{One finds that the increase of the total infidelity is monotonic,} while in contrast the partial infidelities oscillate.
In fact, one even observes a partial recovery of $F_2(t)$ during the first idling.
The dynamics of single-qubit systems under pulse sequences in the presence of noise have been extensively studied in previous studies~\cite{NakamuraPRR2024}, where oscillatory patterns and the partial recovery of fidelities are attributed to retarded feedback (memory effects) of the reservoir.
As another figure of merit, we consider the Breuer--Laine--Piilo quantifiers~\cite{BreuerRMP2016} of the two-qubit system and each single-qubit subsystem in the time period $0 \leq \omega_q t / \pi \leq 1$.
We confirm that both are at least greater than zero, which leads to the conclusion that the reduced dynamics is non-Markovian on the basis of the definition of the complete positive indivisibility~\cite{RivasRPP2014}.
This suggests a scenario where a system evolves actually under a non-Markovian process even if \textcolor{black}{the total infidelity increases apparently monotonically.}
We emphasize that especially when multiple-qubit systems are considered and collective dynamics plays a role, the correspondence between the monotonic decay of fidelities and Markovianity no longer holds.
We comment in passing that when the system is coupled with heterogeneous environments ($s_1\neq s_2$), the total fidelity also oscillates.

\textcolor{black}{Next, we compare the above results with the dynamics of the infidelity obtained within the Lindblad approach.
We choose to take the ``worst'' Lindblad equation that has been widely used in various studies.
Then, the dissipator is obtained only on the basis of the bare Zeeman Hamiltonian, $\hbar \omega_q (\hat{\sigma}_1^z + \hat{\sigma}_2^z)/ 2$.
We stress that for systems subject to time-dependent Hamiltonians, for example, for single-qubit gates, the dissipator must be dynamically adapted~\cite{GulacsiPRR2025}.
The Lamb shift appears to be negligible since the system--reservoir coupling is very small ($2 \pi \hbar \kappa = 0.004$).}

\textcolor{black}{The partial fidelities $F_1(t)$ and $F_2(t)$ are found to be identical owing to the secular approximation imposed in the Lindblad equation~\cite{NakamuraPRB2024}.
Furthermore, the oscillatory behavior predicted by the numerically rigorous results (solid curves in Fig. \ref{fig:partialFidelity}) cannot be seen within Lindblad (dashed curves).
Even worse, the fidelities are overestimated when the Lindblad equation is adopted.
More accurate numerical methods must be used for better estimation of noise effects.}

\section{Concluding remarks} \label{sec:conclusion}
In this paper, we studied, based on a non-perturbative simulation method for open quantum dynamics (FP-HEOM in combination with tensor-train methods), the generation and destruction of the entanglement between two qubits and performances of single- and two-qubit gates.
While it is well known that the entanglement can be generated when both qubits are coupled to the same reservoir, we here assumed that each qubit is coupled to its individual noise source in order to focus on their detrimental effects on the entanglement. Various types of noise sources have been prepared and investigated.

The main findings are summarized as follows.
(I) We investigated effects on disentanglement dynamics of the counter-rotating terms that appear in the coupling Hamiltonian between the system and reservoir.
While these effects have been examined to some extent in previous studies within the framework of the HEOM~\cite{WangNJP2013}, a more extensive analysis has been conducted here by considering various types of noise spectra and sweeping through parameter space.
We found that in general, the RWA induces significant errors in terms of the entanglement (concurrence) in cases where the Bell state $(\ket{10} + \ket{01}) / \sqrt{2}$ should be maintained.
This is because the population of the state $\ket{11}$, which plays a crucial role in the concurrence, is incorrectly evaluated as being zero within the RWA.
When the other Bell state $(\ket{11} + \ket{00}) / \sqrt{2}$ is considered, the error caused by the RWA is weaker but not negligible.

Focusing on noise-specific effects, we have obtained the following main findings:
(1)~For both resonant and detuned Lorentzian noise spectra, the dark periods during which the concurrence remains zero have been identified.
In detuned cases, the dependence of the concurrence on the sign of the detuning is completely absent when the RWA is imposed.
(2)~For broadband noise spectra, the concurrence monotonically decays, and ESD is observed.
Both phenomena are induced by the counter-rotating terms.
In previous studies, the RWA has been introduced in order to obtain analytical expressions and to explore disentanglement dynamics with various initial states other than the Bell states discussed here.
Since the effects of the counter-rotating terms are obvious and substantial even for the limited initial states discussed in this study, it is interesting to investigate also other initial states. This is left for future work.

(II) Next, we turned to a detailed investigation of generation and destruction processes of bipartite entanglement.
In the above settings, the ideal Bell states have been initially prepared, but noise already plays a role during the preparation in practice.
Elevated temperatures for broadband noise spectra have been considered beyond the more theoretical limit of zero temperature.
We here focused on the transformation of the qubit states $\ket{10} \to (\ket{10} - i \ket{01})/\sqrt{2}$ through the use of the \sqrtiswap{} gate.

The main findings are related to (1) characteristic times of the reservoir dynamics and (2) the execution time of the \sqrtiswap{} gate as follows:
(1) When the reservoir dynamics is fast enough (Ohmic case with the spectral exponent $s = 1$), the entanglement cannot be generated in a short-time domain owing to fast excitation processes to $\ket{11}$.
By contrast, when the reservoir dynamics is sluggish (deep sub-Ohmic case, $s = 1/8$), disentanglement dynamics becomes faster after the gate operation.
The low-frequency modes of the noise spectral power mainly contribute to these dynamics.
The intermediate case, $s = 1/2$, has been found to indicate the best performance in terms of the entanglement in our setting.
(2) For a longer gate time (i.e., smaller coupling strength between the two qubits), the entanglement begins to decay even during the application of the gate when
the system--reservoir coupling prevails over the qubit--qubit coupling.
As other minor findings, we remark on (3) heterogeneous environments and (4) the construction of the Kraus operators.
(3) For heterogeneous reservoirs, the real part of the off-diagonal element $\rho_{23}(t)$ grows because of different effective Larmor frequencies between the two qubits.
\textcolor{black}{We here emphasize that this behavior can be detected using relatively simple experimental protocols.}
\textcolor{black}{(4) While the construction methods of the Kraus operators for the two-qubit systems out of those for the single-qubit systems have been proposed in previous studies~\cite{BellomoPRL2007}, those methods might not work well because the assumption of factorized initial states generally does not hold.}

(III) Finally, we have applied our method to a sequence consisting of a single $H$ and a single CNOT gate and measured the performance of the sequence by means of the fidelity and the concurrence.
We have compared two decomposition schemes of the $H$ + CNOT sequence and have found that the shorter time of the total sequence leads to the better performance.
Although by inserting idlings in which no gates are applied, the fidelity can be recovered transiently, the final performance worsens compared to the case without the idlings due to the longer total time of the sequence.
We have revealed the dependence of the sequence performance on (1) initial states and (2) noise source (here, we again consider broadband noise spectra only, and the temperature is set to a nonzero value).
(1) Starting with the initial states $\ket{10}$ and $\ket{00}$, a better performance is achieved compared to the initial states $\ket{11}$ and $\ket{01}$.
This is because of the ideal final states: $(\ket{10} + e^{i\varphi}\ket{01}) / \sqrt{2}$ for the former initial states and $(\ket{11} + e^{i\varphi}\ket{00}) / \sqrt{2}$ for the latter.
The preparation of the former final state has been found to be an easier task than the latter in the presence of noise mainly because of the relaxation process of the population.
(2) Similar to the result (II), the performance with the intermediate spectral exponent $s = 1/2$ has been found to be beneficial compared to the Ohmic and the deep sub-Ohmic case.
The same argument as the result (II) holds in this case.

In the simulations (I) and (II), we have varied the coupling strength between the system and reservoir and have found the somewhat expected result that the errors caused by the approximations, that is, the RWA in case (I) and the reset of the memory effects in case (II), are smaller when the coupling strength becomes weaker.
This implies that approximate methods might be sufficient to analyze performances of quantum computations qualitatively; however, minor errors will play a crucial role for high-precision predictions in further improved quantum devices.

In this study, we restricted ourselves to a simple $H$ + CNOT sequence.
The reservoirs have also been assumed to be independent of each other to focus on detrimental effects of noise.
It is interesting to investigate noise effects in other architectures; for example, spatially correlated noise, including noise-induced crosstalk, as well as reservoir-induced entanglement, see, e.g., Ref.~\cite{GulacsiPRR2025-2}.
Proposing better pulse shapes through, for example, optimal control techniques~\cite{WilhelmARXIV2020}, is one of the next steps to be incorporated in the FP-HEOM approach straightforwardly and without further approximations.
Those subjects are left for future work.
\textcolor{black}{We are currently developing a user-friendly interface for the simulation framework applied in this study that will help non-experts to directly apply it in order to learn about the dissipative dynamics of their specific qubit systems.}

\section*{Acknowledgement}
This work was supported by the BMFTR through QSolid and the Cluster4Future QSens (Project QComp) and the DFG through Grant No. AN336/17-1 (FOR2724).
K.N.~acknowledges support from the State of Baden-W{\"u}rttemberg through bwHPC (Justus II).

\appendix
\section{Time evolution method} \label{sec:appHEOMandRWA}
In this appendix, we present detailed form of the free-pole hierarchical equations of motion (FP-HEOM) and analytical expression of dynamics of the reduced density operator (RDO) under the rotating wave approximation (RWA) in Eq.~\eqref{eq:HI_RWA}.
Before we move to each concrete topic, we specify the reservoir operators $\hat{H}_{R, j}$ and $\hat{B}_j$ in Eqs.~\eqref{eq:eachH}, \eqref{eq:HI_RWA}, and \eqref{eq:HI_noRWA}.
Since Gaussian noise is considered in this study, the reservoir Hamiltonian is expressed as a collection of an infinite number of harmonic oscillators.
The Hamiltonian reads
\begin{align}
    \hat{H}_{R, j} = \sum_{k} \Biggl(\frac{\hat{p}_{j, k}^2}{2 m_{j, k}} + \frac{1}{2}m_{j, k} \omega_{j, k}^2 \hat{x}_{j, k}^2 \Biggr)\, ,
\end{align}
with $\hat{p}_{j, k}$, $\hat{x}_{j, k}$, $m_{j, k}$, and $\omega_{j, k}$ being the momentum, position, mass, and angular frequency of the $k$th mode of the $j$th reservoir, respectively.
The coupling term in Eq.~\eqref{eq:HI_noRWA}, $\hat{X}_j$, is the collective position of the reservoir given by $\hat{X}_j = \sum_{k} c_{j, k} \hat{x}_{j, k}$.
Using the standard transformation of $\hat{x}_{j, k}$ and $\hat{p}_{j, k}$ to the annihilation operator of the $k$th mode of the $j$th reservoir,
\begin{align}
    \hat{b}_{j, k} = \sqrt{\frac{m_{j, k} \omega_{j, k}}{2 \hbar}} \Biggl(\hat{x}_{j, k} + \frac{i}{m_{j, k} \omega_{j, k}}\hat{p}_{j, k}\Biggr)\, ,
\end{align}
the operator $\hat{B}_j$ in Eq.~\eqref{eq:HI_RWA} is rewritten as $\hat{B}_j = \sum_{k}g_{j, k}\hat{b}_{j, k}$, where $g_{j, k} = c_{j, k} \sqrt{\hbar/(2m_{j, k}\omega_{j, k})}$.
Accordingly, the spectral density is given by
\begin{align}
    J_j(\omega) = & \sum_{k} \frac{c_{j, k}^2}{2 m_{j, k}\omega_{j, k}} \delta(\omega - \omega_{j, k}) \\
    = & \sum_{k} \frac{g_{j, k}^2}{\hbar} \delta(\omega - \omega_{j, k})\, .
\end{align}

In this model, the relaxation and the decoherence time for single-qubit systems are evaluated within the Born--Markov approximation and the second-order perturbative treatment as 
\begin{align}
    T_1 = \frac{T_2}{2} = & \frac{1}{2 \pi [S^\beta(\omega_q) + S^\beta(-\omega_q)]} \\
    = & \frac{\tanh(\beta \hbar \omega_q / 2)}{2\pi \hbar J(\omega_q)}\, .
\end{align}
Here, we assume the condition $J(-\omega_q) = -J(\omega_q)$.
\textcolor{black}{Equation~\eqref{eq:T1Detuned} is obtained through this equation.}
When the temperature is sufficiently low ($\beta \to \infty$), and when the reservoir is resonant in the Lorentzian case [$\Delta = 0$ in Eq.~\eqref{eq:sdLorentz}] and the cutoff frequency is sufficiently large in the broadband case [$\omega_c \to \infty$ in Eq.~\eqref{eq:sdOhmic}], the relation between the coupling strength $\kappa$ and the relaxation time $T_1$ in Eq.~\eqref{eq:approxT1} is obtained.  

\subsection{Details of the free-pole hierarchical equations of motion (FP-HEOM)} \label{sec:appHEOM}

Even for multiple-reservoir systems, the form of the FP-HEOM is almost the same as the original version for single-reservoir systems~\cite{XuPRL2022}.
For the sake of simplicity, we refer interested readers to Refs.~\cite{Tanimura2014,Nakamura18PRA,NakamuraJCP2021} and here only display the detailed form of the equation:
\begin{align}
    \frac{\partial}{\partial t}\hat{\rho}_{\vec{m}, \vec{n}}(t) = & -\frac{i}{\hbar} [\hat{H}_S (t), \hat{\rho}_{\vec{m}, \vec{n}}(t)] \\
    & - \sum_{j= 1}^{2}\Biggl\{\sum_{k=1}^{K_j}[(m_{j, k}z_{j, k} + n_{j, k}z^*_{j, k}) \hat{\rho}_{\vec{m}, \vec{n}}(t) \\
    & + \sqrt{(m_{j, k}+1) d_{j, k}}[\hat{\sigma}_j^x, \hat{\rho}_{\vec{m}+\vec{e}_{j, k}, \vec{n}}(t)] \\
    & - \sqrt{(n_{j, k} + 1)d^*_{j, k}}[\hat{\sigma}_j^x, \hat{\rho}_{\vec{m}, \vec{n}+\vec{e}_{j, k}}(t)] \\
    & - \sqrt{m_{j, k} d_{j, k}} \hat{\sigma}_j^x \hat{\rho}_{\vec{m}-\vec{e}_{j, k}, \vec{n}}(t) \\
    & - \sqrt{n_{j, k} d^*_{j, k}} \hat{\rho}_{\vec{m}, \vec{n}-\vec{e}_{j, k}}(t) \hat{\sigma}_j^x] \Biggr\}\, .
    \label{eq:DetailedHEOM}
\end{align}
Here, we have introduced
\begin{align}
    \hat{H}_S(t) = \hat{H}_{S, 1}(t) + \hat{H}_{S, 2}(t) + \hbar J(t)(\hat{\sigma}_1^{+}\hat{\sigma}_2^{-} + \hat{\sigma}_1^{-}\hat{\sigma}_2^{+})\, .
\end{align}
As mentioned in the main text, the vector $(\vec{m}, \vec{n})$ distinguishes each auxiliary density operator (ADO).
The vector $\vec{e}_{j, k}$ is the unit vector of $(j, k)$th element.
The values $d_{j, k}$ and $z_{j, k}$ are given by Eq.~\eqref{eq:CF} in the main text.
The superoperator $\mathcal{L}_{j, k}^+$ ($\mathcal{L}_{j, k}^-$) in Eq.~\eqref{eq:HEOM} corresponds to the third and fourth lines (fifth and sixth lines) of Eq.~\eqref{eq:DetailedHEOM}.
To terminate the infinite-dimensional equation of motion [Eq.~\eqref{eq:DetailedHEOM}], we always set zero to the ADOs any $(j, k)$th element of which ($m_{j, k}$ or $n_{j, k}$) is greater than a certain value $\mathcal{N}_{j, \max}$.
We varied the values $\mathcal{N}_{j, \max}$ and confirmed that the results reported in this paper converge.

\subsection{Analytical expression of the dynamics under the rotating wave approximation} \label{sec:appRWA}
At zero temperature and without gate operations [$\Omega_1(t) = \Omega_2(t) = J(t) = 0$], the dynamics of the RDO with the Hamiltonian in Eqs.~\eqref{eq:totH} and ~\eqref{eq:HI_RWA} are expressed analytically as follows~\cite{Breuer2002,BellomoPRL2007,LiPRA2010}: for the diagonal elements,
\begin{align}
    \rho_{11}(t) = & \rho_{11}(0) |h_1(t)|^2|h_2(t)|^2\, , \\
    \rho_{22}(t) = & \rho_{22}(0) |h_1(t)|^2 + \rho_{11}(0) |h_1(t)|^2 [1 - |h_2(t)|^2]\, ,\\
    \rho_{33}(t) = & \rho_{33}(0) |h_2(t)|^2 + \rho_{11}(0) |h_2(t)|^2 [1 - |h_1(t)|^2]\, , \\
    \rho_{44}(t) = & 1 - [\rho_{11}(t) + \rho_{22}(t) + \rho_{33}(t)]\, ,
    \label{eq:rhoDiagRWA}
\end{align}
and for the off-diagonal elements,
\begin{align}
    \rho_{12}(t) = & \rho_{12}(0) |h_1(t)|^2 h_2(t)\, ,\\
    \rho_{13}(t) = & \rho_{13}(0) |h_2(t)|^2h_1(t)\, ,\\
    \rho_{14}(t) = & \rho_{14}(0) h_1(t) h_2(t)\, , \\
    \rho_{23}(t) = & \rho_{23}(0) h_1(t) h_2^*(t)\, , \\
    \rho_{24}(t) = & \rho_{24}(0) h_1(t) + \rho_{13}(0) h_1(t) [1 - |h_2(t)|^2]\, , \\
    \rho_{34}(t) = & \rho_{34}(0) h_2(t) + \rho_{12}(0) h_2(t) [1 - |h_1(t)|^2]\, ,
    \label{eq:rhoOffDiagRWA}
\end{align}
with the Hermitian condition $\rho_{ij}(t) = \rho_{ji}^*(t)$.
Here, the matrix element is expressed as $\rho_{ij}(t) = \braket{i|\rho_{S}(t)|j}$ with the standard basis $\mathcal{B} = \{\ket{1}, \ket{2}, \ket{3}, \ket{4}\} = \{\ket{11}, \ket{10}, \ket{01}, \ket{00}\}$ as mentioned in the main text.
The function $h_j(t) = e^{-i\omega_j^qt} \tilde{h}_j(t)$ ($j \in \{1, 2\}$) is evaluated through the solution of the following integro-differential equation:
\begin{align}
    \frac{d}{dt}\tilde{h}_j(t) = - \int_{0}^{t} dt' f_j(t-t') \tilde{h}_j(t')\, ,
    \label{eq:intDiffh}
\end{align}
where
\begin{align}
    f_j(t) = & \hbar \int_{0}^\infty d\omega J_j(\omega) e^{i(\omega_j^q-\omega)t} \\
    = & e^{i\omega_j^q t} C_j(t) \Bigr|_{\beta \to \infty}\, ,
    \label{eq:cfRWA}
\end{align}
with the initial condition $\tilde{h}_j(0) = 1$.
Note that in the limit $\beta \to \infty$ (zero-temperature limit), the spectral noise power is reduced to $S_j^{\beta \to \infty} (\omega) \to \hbar \theta(\omega) J_j(\omega)$, with $\theta(\omega)$ being the Heaviside step function.
The lower bound of the integral in Eq.~\eqref{eq:cfRWA} can be approximated with $-\infty$ as long as the condition $J_j(\omega < 0) \simeq 0$ is fulfilled.
For the Lorentzian noise spectrum [Eq.~\eqref{eq:sdLorentz}], this approximation holds when we set small values to $\lambda_j$ and $\Delta_j$, so that the following analytical expression is obtained:
\begin{align}
    C_j(t) \simeq & \hbar \int_{-\infty}^{\infty} d\omega \frac{\kappa_j \omega_j^q \lambda_j^2}{(\omega_j^q - \Delta_j - \omega)^2 + \lambda_j^2} e^{-i\omega t} \\
    = & \pi \hbar \kappa_j \omega_j^q \lambda_j \: \exp\bigl[-\lambda_j|t|-i(\omega_j^q-\Delta_j)t\bigr] \, .
    \label{eq:cfLorentz}
\end{align}
Note in passing that $2\pi \hbar \kappa_j \omega_j^q$ corresponds to the quantity $\gamma_{0}$ in other literature~\cite{Breuer2002,BellomoPRL2007}.
With this autocorrelation function $C_j(t)$ and $f_j(t) = e^{i \omega_j^q t} C_j(t)$, one can solve the integro-differential equation [Eq.~\eqref{eq:intDiffh}] analytically as
\begin{align}
    h_j(t; \Delta_j) = & \exp\biggl[-\biggl(\frac{\lambda_j - i \Delta_j}{2} + i \omega_j^q \biggr)t\biggr] \\
    & \times \biggl(\cosh \frac{d_j(\Delta _j) t}{2} + \frac{\lambda_j - i\Delta _j}{d_j(\Delta _j)} \sinh \frac{d_j(\Delta_j) t}{2}\biggr)\, ,
\end{align}
with the function $d_j(\Delta_j) = \sqrt{(\lambda_j- i \Delta_j)^2 - 4 \pi \hbar \kappa_j \omega_j^q \lambda_j}$.
Since the equation $d_j^*(\Delta_j) = d_j(-\Delta_j)$ holds, it follows that the absolute value $|h_j(t; \Delta_j)|$ is independent of the sign of $\Delta _j$, namely, $|h_j(t; \Delta_j)| = |h_j(t; -\Delta_j)|$.
Substituting Eqs.~\eqref{eq:rhoDiagRWA} and \eqref{eq:rhoOffDiagRWA} into Eq.~\eqref{eq:concXState} and considering the above property of $|h_j(t)|$, one finds that the concurrence of systems in the $X$ state with a Lorentzian noise spectrum under the RWA takes same values irrespective of the sign of $\Delta _j$.
As mentioned in the main text, this property does not hold without the RWA.

\textcolor{black}{When the reservoirs are identical, $h_1(t) = h_2(t) = h(t)$, and the initial state is the Bell state, $\hat{\rho}_S(0) = \ketbra{\Phi}{\Phi}$, the elements $\rho_{22}(t)$ and $\rho_{23}(t)$ exhibit the same behavior: $\rho_{22}(t) = \rho_{23}(t) = 0.5 |h(t)|^2$.
This implies that oscillation can be observed in both of the diagonal and off-diagonal elements, which contrasts with the Lamb shift as mentioned in the main text.}

Using the adaptive Antoulas--Anderson algorithm (AAA algorithm)~\cite{NakatsukasaSIAM2018}, we can expand the autocorrelation function in a series of exponential functions described in Eq.~\eqref{eq:CF} in both zero-temperature and finite-temperature settings.
To evaluate $f_j(t)$ in Eq.~\eqref{eq:cfRWA} with the expansion Eq.~\eqref{eq:CF}, one has to solve the following time-differential equation~\cite{IkedaJCP2020}:
\begin{align}
    \frac{d}{dt} \tilde{h}_j(t) = & - \sum_{k=1}^{K_j} \tilde{e}_{j, k}(t), \\
    \frac{d}{dt} \tilde{e}_{j, k}(t) = & -(z_{j, k} - i \omega_j^q) \tilde{e}_{j, k}(t) + d_{j, k} \tilde{h}_{j}(t)\, ,
\end{align}
with the definition
\begin{align}
    \tilde{e}_{j, k}(t) = \int_{0}^{t} dt' d_{j, k} e^{-(z_{j, k} - i \omega_j^q) (t-t')} \tilde{h}_j(t')\, .
\end{align}

\section{Dissipative dynamics of $X$ states} \label{sec:appXState}
Here, we discuss whether the RDO is in an $X$ state at any time under the condition that the initial RDO is in an $X$ state.
The discussion is simpler for the RWA case without pulse application [$\Omega_1(t) = \Omega_2(t) = J(t) = 0$]:
From Eq.~\eqref{eq:rhoOffDiagRWA}, one finds that the elements $\rho_{12}(t)$, $\rho_{13}(t)$, $\rho_{24}(t)$, and $\rho_{34}(t)$ are always zero when initial values of those elements are set to zero.
For the proof in the non-RWA case, we start to introduce the following ``$\bar{X}$ state'' for the sake of convenience:
\begin{align}
    \hat{\rho}_S =
    \begin{bmatrix}
        0 & \rho_{12} & \rho_{13} & 0\\
        \rho_{21} & 0 & 0 & \rho_{24} \\
        \rho_{31} & 0 & 0 & \rho_{34} \\
        0 & \rho_{42} & \rho_{43} & 0
    \end{bmatrix}\, ,
\end{align}
and investigate to which state $X$ states are mapped with the superoperators in Eq.~\eqref{eq:DetailedHEOM}.
Here, we first consider the case in which no single-qubit gates are applied [$\Omega_1(t) = \Omega_2(t) = 0$].
The superoperators $[\hat{\sigma}_j^z, \bullet]$ and $[\hat{\sigma}_1^{+}\hat{\sigma}_2^{-} + \hat{\sigma}_1^{-}\hat{\sigma}_2^{+}, \bullet]$ map the $X$ states to $X$ states (and the $\bar{X}$ states to $\bar{X}$ states), while the superoperators $\hat{\sigma}_j^x \bullet$ and $\bullet \hat{\sigma}_j^x$ map the $X$ states to $\bar{X}$ states and vice versa.
Since the former superoperators do not change the indices $(\vec{m}, \vec{n})$ in the FP-HEOM formalism while the latter change the depth $\mathcal{N} = \sum_{j, k} (m_{j, k} + n_{j, k})$ by $\pm 1$, it is concluded that the ADOs with even depths are in $X$ states, and those with odd depths are in $\bar{X}$ states provided that the initial RDO $\hat{\rho}_{\vec{0}, \vec{0}}(0)$ is in an $X$ state, and the other ADOs $\hat{\rho}_{(\vec{m}, \vec{n}) \neq (\vec{0}, \vec{0})}(0)$ are set to zero (factorized initial states).
When single-qubit gates are applied [$\Omega_j(t) \neq 0$], the system Hamiltonian $\hat{H}_{S, j}(t)$ includes the operator $\hat{\sigma}_j^x$, and hence the ADOs become mixture of $X$ and $\bar{X}$ states by just applying the system Hamiltonian.
In summary, an initial $X$ state of the RDO is conserved during the time evolution Eq.~\eqref{eq:HEOM} as long as no single gates are applied.

\section{Tensor-train (TT) representation of FP-HEOM} \label{sec:appTT}
Since the ADOs as well as the RDO must be taken into account, the computational cost of the FP-HEOM method can be quite large.
Especially, it becomes prohibitive when multiple reservoirs are considered.
To avoid this problem, we utilize the tensor-train (TT) representation.
Since this topic has not been carefully presented previously, we give a more detailed account in the sequel.

For this purpose, we first rewrite the FP-HEOM in Eq.~\eqref{eq:DetailedHEOM} in an extended Hilbert space.
Defining the extended density operator,
\begin{align}
    \hat{W}(t) = & \sum_{\vec{m}, \vec{n}} \hat{\rho}_{\vec{m}, \vec{n}}(t) \ket{\vec{m}, \vec{n}} \\
    = & \sum_{\substack{i, j, k, l \\ \vec{m}, \vec{n}}} \braket{ij | \hat{\rho}_{\vec{m}, \vec{n}}(t) |kl} |ij, kl\rangle\rangle  \otimes \ket{\vec{m}, \vec{n}}\, ,
    \label{eq:extendedRho}
\end{align}
and the annihilation operators acting on the $(j, k)$th element of $\vec{m}$ and $\vec{n}$,
\begin{align}
    \hat{a}_{j, k} \ket{\vec{m}, \vec{n}} = &\sqrt{m_{j, k}} \ket{\vec{m}-\vec{e}_{j, k}, \vec{n}}\, , \\
    \hat{b}_{j, k} \ket{\vec{m}, \vec{n}} = &\sqrt{n_{j, k}} \ket{\vec{m}, \vec{n}-\vec{e}_{j, k}}\, ,    
\end{align}
respectively, we obtain the following time-differential equation for the extended states $\hat{W}(t)$:
\begin{align}
    \frac{\partial}{\partial t} \hat{W}(t) = & \Biggl(-\frac{i}{\hbar}(\underrightarrow{\hat{H}_S(t)} - \underleftarrow{\hat{H}_S(t)})\otimes \hat{1}_{R}
    \\ &
    - \sum_{j=1}^{2} \Biggl\{\sum_{k=1}^{K_j} [\hat{1}_{S} \otimes (z_{j, k} \hat{m}_{j, k} + z_{j, k}^* \hat{n}_{j, k})
    \\ &
    + (\underrightarrow{\hat{\sigma}_j^x} - \underleftarrow{\hat{\sigma}_j^x}) \otimes \Bigl(\sqrt{d_{j, k}} \hat{a}_{j, k} - \sqrt{d_{j, k}^*}\hat{b}_{j, k}\Bigr)
    \\ &
    - \sqrt{d_{j, k}} \underrightarrow{\hat{\sigma}_j^x} \otimes \hat{a}_{j, k}^\dagger
    - \sqrt{d_{j, k}^*} \underleftarrow{\hat{\sigma}_j^x} \otimes \hat{b}_{j, k}^\dagger ] \Biggr\} \Biggr) \hat{W}(t)\, 
    \\
    = & \mathcal{L}_\mathrm{tot}(t) \hat{W}(t)\, .
    \label{eq:TTHEOM}
\end{align}
Here, the number operator for the $k$th mode of the $j$th reservoir is given by $\hat{m}_{j, k} = \hat{a}_{j, k}^\dagger \hat{a}_{j, k}$ and $\hat{n}_{j, k} = \hat{b}_{j, k}^\dagger \hat{b}_{j, k}$ (of course, complex conjugates $\hat{a}_{j, k}^\dagger$ and $\hat{b}_{j, k}^\dagger$ are the creation operators for the respective modes).
The operators $\hat{1}_S$ and $\hat{1}_R$ are the identity operator acting on the system and reservoir, respectively, and $\underrightarrow{\hat{O}}$ ($\underleftarrow{\hat{O}}$) is a system operator acting on the ket (bra) vector of the RDO.
The vectors $\bra{ij}$ and $\ket{kl}$ in Eq.~\eqref{eq:extendedRho} correspond to vectors in the standard product basis $\mathcal{B}$, and $|ij, kl\rangle\rangle$ is vectorization of $\ketbra{ij}{kl}$.
With this vectorization, the operator $\underrightarrow{\hat{O}}$ is expressed as the standard form, $\underrightarrow{\hat{O}} \equiv \hat{O}$, while the operator $\underleftarrow{\hat{O}}$ becomes the transpose of the original operator, $\underleftarrow{\hat{O}} \equiv \hat{O}^T$.

For the TT representation, we arrange the state vector in the order of
\begin{align}
    \ket{i} \otimes & \ket{m_{1, 1}} \otimes \ket{n_{1, 1}} \otimes \ket{m_{1, 2}} \otimes \cdots \otimes \ket{n_{1, K_1}} \otimes
    \ket{k}
    \otimes \ket{j} 
    \\
    \otimes & \ket{m_{2, 1}} \otimes \ket{n_{2, 1}} \otimes \ket{m_{2, 2}} \otimes \cdots \otimes \ket{n_{2, K_2}} \otimes \ket{l}
\end{align}
and expand the coefficient as
\begin{align}
    & \braket{ij|\hat{\rho}_{\vec{m}, \vec{n}}(t)|kl} \\
    & = 
    \begin{aligned}[t]
   &  \sum_{\substack{r_1, \ldots, \\ r_{\mathcal{M}-1}}} C_{1}^t(i, r_1) C_2^t(r_1, m_{1, 1}, r_2) \cdots 
    C_{2K_1+2}^t(r_{2K_1+1}, k, r_{2K_1+2}) \\
    & \times C_{2K_1+3}^t(r_{2K_1+2}, j, r_{2K_1+3}) \cdots
    C_{\mathcal{M}}^t(r_{\mathcal{M}-1}, l)\, ,
    \end{aligned}
\end{align}
where $\mathcal{M} = 2 (K_1 + K_2) + 4$ is the total number of TT cores, $C_k^t(:, :, :) \in \mathbb{C}^{R_{k-1} \times N_k \times R_k}$.
The integer $N_k$ corresponds to the dimension of the Hilbert space of the $k$th element [i.e., the number of system levels for the ket- and bra-vector parts ($k = 1$, $2K_1+2$, $2K_1+3$, and $\mathcal{M}$), $\mathcal{N}_{1, \max}+1$ for $2 \leq k \leq 2K_1+1$, and $\mathcal{N}_{2, \max}+1$ for $2K_1+4 \leq k \leq \mathcal{M}-1$], and $R_{k}$ is referred to as the ``TT rank.''
The ranks $R_{0}$ and $R_{\mathcal{M}}$, which are allocated at the edges of the TT, are fixed to $1$.
The ranks for the other cores $R_k$ ($1 \leq k \leq \mathcal{M}-1$) are bounded by a maximum value $R_{\max}$:
If a core can be analytically expressed with a smaller rank than $R_{\max}$, the rank is adopted.
Otherwise, approximate expression with the rank $R_{\max}$ is adopted.
We varied $R_{\max}$ and confirmed that the obtained results converge.

To utilize the TT representation, we decompose the operator $\mathcal{L}_\mathrm{tot}(t)$ in the TT operator form.
When we rewrite Eq.~\eqref{eq:TTHEOM} as
\begin{align}
    \frac{\partial}{\partial t } & \braket{ij|\hat{\rho}_{\vec{m}, \vec{n}}(t) | kl} \\
    & = \sum_{\substack{i', j', k', l' \\ \vec{m}', \vec{n}'}} \bigl[\mathcal{L}_\mathrm{tot}(t)\bigr]_{i, j, k, l, \vec{m}, \vec{n}}^{i', j', k', l', \vec{m}', \vec{n}'} \braket{i'j'|\hat{\rho}_{\vec{m}', \vec{n}'}(t) |k'l'}\, ,
\end{align}
the TT operator is given by
\begin{align}
    & \bigl[\mathcal{L}_\mathrm{tot}(t)\bigr]_{i, j, k, l, \vec{m}, \vec{n}}^{i', j', k', l', \vec{m}', \vec{n}'} \\
    & = 
    \begin{aligned}[t]
        & \sum_{\substack{\tilde{r}_1, \ldots, \\ \tilde{r}_{\mathcal{M}-1}}} D_{1}^t(i, i', \tilde{r}_1)D_2^t(\tilde{r}_1, m_{1, 1}, m'_{1, 1}, \tilde{r}_2) \\
        & \times \cdots D_{\mathcal{M}}^t(\tilde{r}_{\mathcal{M}-1}, l, l')\, .
    \end{aligned}    
\end{align}
Note that the ranks $\{\tilde{R}_k\}$ for the cores of the TT operators, $D_k^t(:, :, :, :) \in \mathbb{C}^{\tilde{R}_{k-1} \times N_k \times N_k \times \tilde{R}_{k}}$, are different from $\{R_k\}$.
Since the system Hamiltonian is time dependent, the TT operators are also time dependent, which is indicated by the superscript of $D_k^t$ (and accordingly, $C_k^t$).
The time-differential equations of the TT cores are given by
\begin{align}
    \frac{\partial}{\partial t} C_j^t(:, :, :)
    = & \mathcal{P}_j\bigl[\{D_k^t\}, \{C_k^t\}\bigr]\, .
\end{align}
Here, $\mathcal{P}_j[\{D_k^t\}, \{C_k^t\}]$ provides an update of the $j$th TT core obtained with the projection of $\mathcal{L}_{\mathrm{tot}}(t) \hat{W}(t)$.
Note that all the TT cores and TT operators are needed to obtain the $j$th update $\mathcal{P}_j[\{D_k^t\}, \{C_k^t\}]$.
Strictly, additional time-differential equations must be taken into account for the time integration of the whole TT cores in order to ensure the orthogonality of the cores:
For more details, see, for example, Ref.~\cite{LubichSIAMJNA2015}.

If the TT operators are time dependent, the computational cost to obtain $\mathcal{P}_j[\{D_k^t\}, \{C_k^t\}]$ is too high.
\textcolor{black}{To avoid this problem, we further decompose the TT operators $\{D_k^t\}$ into the five parts, $\{D_k^{(0)}\}$, $\{D_k^{(l, c)}\}$, and $\{D_k^{(l, s)}\}$ ($l = 1, 2$), and the differential equation reads}
\begin{align}
    &
    \frac{\partial}{\partial t} C_j^t(:, :, :)
    \\
    = & \mathcal{P}_j\bigl[\{D_k^{(0)}\}, \{C_k^t\}\bigr]
    \\ & 
    + \sum_{l = 1}^{2} \Omega_l \bigl\{\cos (\omega_l^{\mathrm{ex}}t + \phi_l) \mathcal{P}_j\bigl[\{D_k^{(l, c)}\}, \{C_k^t\}\bigr]
    \\ &
    + \sin (\omega_l^{\mathrm{ex}}t + \phi_l) \mathcal{P}_j\bigl[\{D_k^{(l, s)}\}, \{C_k^t\}\bigr]\bigr\}\, .
\end{align}
The matrix representation of the TT cores arranged with respect to the first and last indices is given as follows:
for the time-independent part,
\begin{gather}
    \bigl[D_1^{(0)} (u, :, :, v)\bigr] = 
    \begin{bmatrix}
        \displaystyle -\frac{i \omega_1^q}{2} \hat{\sigma}^z & \hat{1} & O & \hat{\sigma}^x & - i J \hat{\sigma}^+ & -i J \hat{\sigma}^-
    \end{bmatrix}\, ,
\end{gather}
\begin{gather}
    \bigl[D_{2K_1+2}^{(0)} (u, :, :, v)\bigr] \!=\!
    \begin{bmatrix}
        \hat{1} & O & O & O & O & O \\[0.5em]
        \displaystyle \frac{i \omega_1^q}{2} \bigl(\hat{\sigma}^z\bigr)^T & \hat{1} & O & O & \bigl(\hat{\sigma}^+\bigr)^T & \bigl(\hat{\sigma}^-\bigr)^T \\[1em]
        \bigl(\hat{\sigma}^x\bigr)^T & O & O & O & O & O \\
        O & O & O & O & O & O \\
        O & O & \hat{1} & O & O & O \\
        O & O & O & \hat{1} & O & O
    \end{bmatrix}\, ,
\end{gather}
\begin{gather}
    \bigl[D_{2K_1+3}^{(0)} (u, :, :, v)\bigr] =
    \begin{bmatrix}
        \hat{1} & O & O & O & O & O \\[0.5em]
        \displaystyle -\frac{i\omega_2^q}{2} \hat{\sigma}^z & \hat{1} & O & \hat{\sigma}^x & O & O \\[1em]
        \hat{\sigma}^- & O & O & O & O & O \\
        \hat{\sigma}^+ & O & O & O & O & O \\
        O & O & O & O & \hat{1} & O \\
        O & O & O & O & O & \hat{1}
    \end{bmatrix}\, ,
\end{gather}
\begin{gather}
    \bigl[D_{\mathcal{M}}^{(0)}(u, :, :, v)\bigr] = 
    \begin{bmatrix}
        \hat{1} \\[0.5em] \displaystyle \frac{i\omega_2^q}{2} \bigl(\hat{\sigma}^z\bigr)^T \\[1em] \bigl(\hat{\sigma}^x\bigr)^T \\ O \\ iJ \bigl(\hat{\sigma}^-\bigr)^T \\ iJ \bigl(\hat{\sigma}^+\bigr)^T
    \end{bmatrix}\, ,
\end{gather}
\begin{align}
    & 
    \bigl[D_{p_j + 2k}^{(0)}(u, :, :, v)\bigr]
    \\ = &  
    \begin{bmatrix}
        \hat{1} & O & O & O & O & O \\
        -z_{j, k}\hat{m}_j & \hat{1} & \sqrt{d_{j, k}} \hat{a}_j & O & O & O \\
        O & O & \hat{1} & O & O & O \\
        \sqrt{d_{j, k}}(\hat{a}_j^\dagger - \hat{a}_j) & O & O & \hat{1} & O & O \\
        O & O & O & O & \hat{1} & O \\
        O & O & O & O & O & \hat{1}
    \end{bmatrix} \\
    & \hspace*{7em} (1 \leq k \leq K_j, \quad j = 1, 2)\, ,
    \label{eq:TTO_FPHEOM1}
\end{align}
\begin{align}
    &
    \bigl[\!D_{p_j + 2k+1}^{(0)}\!(u, :, :, v)\!\bigr]
    \\ = &
    \begin{bmatrix}
        \hat{1} & O & O & O & O & O \\
        -z_{j, k}^*\hat{m}_j & \hat{1} & \sqrt{d_{j, k}^*} (\hat{a}_j^\dagger \!-\! \hat{a}_j) & O & O & O \\
        O & O & \hat{1} & O & O & O \\
        \sqrt{d_{j, k}^*}\hat{a}_j & O & O & \hat{1} & O & O \\
        O & O & O & O & \hat{1} & O \\
        O & O & O & O & O & \hat{1}
    \end{bmatrix} \\
    &\hspace*{7em}(1 \leq k \leq K_j, \quad j = 1, 2)\, ,
    \label{eq:TTO_FPHEOM2}
\end{align}
and for the time-dependent parts for the single-qubit gates,
\begin{align}
    \bigl[D_1^{(1, c)} (u, :, :, v)\bigr] = &
    \begin{bmatrix}
        - i\hat{\sigma}^x / 2 & \hat{1}
    \end{bmatrix}\, ,
    \\
    \bigl[D_1^{(1, s)} (u, :, :, v)\bigr] = &
    \begin{bmatrix}
        - i \hat{\sigma}^y / 2 & \hat{1}
    \end{bmatrix}\, ,
\end{align}
\begin{align}
    \bigl[D_{2K_1+2}^{(1, c)} (u, :, :, v)\bigr] = &
    \begin{bmatrix}
        \hat{1} \\ i \bigl(\hat{\sigma}^x\bigr)^T / 2
    \end{bmatrix}\, , \\
    \bigl[D_{2K_1+2}^{(1, s)} (u, :, :, v)\bigr] = &
    \begin{bmatrix}
        \hat{1} \\ i \bigl(\hat{\sigma}^y\bigr)^T / 2
    \end{bmatrix}\, ,
\end{align}
\begin{align}    
    \bigl[D_{k}^{(1, c/s)} (u, :, :, v)\bigr] = &
    \begin{bmatrix}
        \hat{1} & O \\
        O & \hat{1}
    \end{bmatrix} & (2 \leq k \leq 2K_1+1)\, ,\\
    \bigl[D_{k}^{(1, c/s)} (u, :, :, v)\bigr] = &
    \begin{bmatrix}
        \hat{1}
    \end{bmatrix} & (2K_1+3 \leq k \leq \mathcal{M})\, ,
\end{align}
and
\begin{align}
    \bigl[D_{2K_1+3}^{(2, c)} (u, :, :, v)\bigr] = &
    \begin{bmatrix}
        - i\hat{\sigma}^x / 2 & \hat{1}
    \end{bmatrix}\, ,
    \\
    \bigl[D_{2K_1+3}^{(2, s)} (u, :, :, v)\bigr] = &
    \begin{bmatrix}
        - i \hat{\sigma}^y / 2 & \hat{1}
    \end{bmatrix}\, ,
\end{align}
\begin{align}
    \bigl[D_{\mathcal{M}}^{(2, c)} (u, :, :, v)\bigr] = &
    \begin{bmatrix}
        \hat{1} \\ i \bigl(\hat{\sigma}^x\bigr)^T / 2
    \end{bmatrix}\, , \\
    \bigl[D_{\mathcal{M}}^{(2, s)} (u, :, :, v)\bigr] = &
    \begin{bmatrix}
        \hat{1} \\ i \bigl(\hat{\sigma}^y\bigr)^T / 2
    \end{bmatrix}\, ,
\end{align}
\begin{align}    
    \bigl[D_{k}^{(2, c/s)} (u, :, :, v)\bigr] = &
    \begin{bmatrix}
        \hat{1}
    \end{bmatrix} & (1 \leq k \leq 2K_1+2)\, ,\\
    \bigl[D_{k}^{(2, c/s)} (u, :, :, v)\bigr] = &
    \begin{bmatrix}
        \hat{1} & O \\
        O & \hat{1}
    \end{bmatrix} & (2K_1+4 \leq k \leq \mathcal{M}-1)\, .
\end{align}
Since it is no longer necessary to distinguish each qubit and each reservoir mode here, we omit the subscript of the Pauli matrices, and $\hat{m}_j$ and $\hat{a}_j$ are the representative expressions for $\hat{m}_{j, k}$ and $\hat{n}_{j, k}$, and $\hat{a}_{j, k}$ and $\hat{b}_{j, k}$, respectively.
The dimension of $\hat{m}_j$ and $\hat{a}_j$ is truncated to $\mathcal{N}_{j, \max}+1$.
The operators $\hat{1}$ and $O$ are the identity and null operator with the size $N_k \times N_k$ for the $k$th TT operator.
The quantity $p_j$ in Eqs.~\eqref{eq:TTO_FPHEOM1} and \eqref{eq:TTO_FPHEOM2} indicates the starting point of the $j$th reservoir part, namely, $p_1 = 0$ and $p_2 = 2K_1+2$.
Here, we explicitly describe the transpose $(\bullet)^T$ even for the symmetric matrices in order to indicate the application to the bra vectors.
One can recover the original expression of $\mathcal{L}_{\mathrm{tot}}(t)$ in Eq.~\eqref{eq:TTHEOM} through the matrix multiplication with respect to the first and last indices $u$ and $v$.

\begin{table}[t]
    \caption{Maximum depth of the hierarchy $\mathcal{N}_{\max}$ for the Lorentzian noise spectra.
    The initial state is expressed as $\hat{\rho}_0 = \ketbra{\psi_0}{\psi_0}$.
    The number of modes $K$ is always $1$ in these spectra [see Eq.~\eqref{eq:cfLorentz}].
    Because of the small number of $K$, we do not use the TT representation and directly calculate Eq.~\eqref{eq:DetailedHEOM}, and therefore the TT ranks are not listed.
    \label{tbl:paramsLorentz}}
    \begin{ruledtabular}
    \begin{tabular}{ccccccc}
        $2 \pi \hbar \kappa $ & \multicolumn{3}{c}{$0.002$} & \multicolumn{2}{c}{$0.1$} & $5$\\
        $\Delta / \omega_q$ & \multicolumn{2}{c}{$0$} & $\pm 0.05$ & $0$ & $\pm 0.05$ & $0$, $\pm0.05$ \\
        \cmidrule{2-3}
        \cmidrule{4-4}
        \cmidrule{5-5}
        \cmidrule{6-6}
        \cmidrule{7-7}
        $\ket{\psi_0}$ & $\ket{\Phi}$ & $\ket{\Psi}$ & $\ket{\Phi}$, $\ket{\Psi}$ & \multicolumn{2}{c}{$\ket{\Phi}$, $\ket{\Psi}$} & $\ket{\Phi}$, $\ket{\Psi}$ \\
        \midrule
        $\mathcal{N}_{\max}$ & $2$ & $4$ & $2$ & $6$ & $4$ & $9$ \\
    \end{tabular}
    \end{ruledtabular}

    \caption{Number of modes $K$, maximum depth of the hierarchy $\mathcal{N}_{\max}$, and maximum TT rank $R_{\max}$ for the broadband noise spectra.
    The parameter values are independent of the initial states $\hat{\rho}_0$.
    The values of $\mathcal{N}_{\max}$ and $R_{\max}$ in the left column in the case with $s = 1/8$, $\beta \hbar \omega_q = 5$, and $2 \pi \hbar \kappa = 0.004$ are used for the results in Sec.~\ref{sec:resultISWAP}, while those in the right column are used for the results in Sec.~\ref{sec:resultHCNOT}.
    \label{tbl:paramsSubOhmic}}
    \begin{ruledtabular}
    \begin{tabular}{ccccccccccc}
        $s$ & \multicolumn{3}{c}{$1$} & \multicolumn{3}{c}{$1/2$} & \multicolumn{3}{c}{$1/8$} & $1/14$ \\
        $\beta \hbar \omega_q$ & $\infty$ & \multicolumn{2}{c}{$5$} & $\infty$ & \multicolumn{2}{c}{$5$} & \multicolumn{3}{c}{$5$} & $\infty$ \\
        \cmidrule(lr){2-2}
        \cmidrule(lr){3-4}
        \cmidrule(lr){5-5}
        \cmidrule(lr){6-7}
        \cmidrule(lr){8-10}
        \cmidrule(lr){11-11}
        $2 \pi \hbar \kappa$ & $0.04$ & $0.04$ & $0.004$ & $0.04$ & $0.04$ & $0.004$ & $0.04$ & \multicolumn{2}{c}{$0.004$} & $0.04$ \\
        \midrule
        $K$ & $15$ & $10$ & $10$ & $28$ & $24$ & $31$ & $8$ & \multicolumn{2}{c}{$8$} & $35$ \\
        $\mathcal{N}_{\max}$ & $3$ & $3$ & $2$ & $3$ & $3$ & $2$ & $8$ & $7$ & $5$ & $3$ \\
        $R_{\max}$ & $30$ & $30$ & $20$ & $30$ & $30$ & $20$ & $200$ & $50$ & $80$ & $30$
    \end{tabular}
    \end{ruledtabular}
\end{table}

\section{Parameter values used for FP-HEOM} \label{sec:appHEOMParams}
The parameter values for the FP-HEOM method and TT representation are listed in Tables \ref{tbl:paramsLorentz} and \ref{tbl:paramsSubOhmic}.
Those used in Sec.~\ref{sec:resultRWA} are in Table~\ref{tbl:paramsLorentz} for the Lorentzian noise spectra and in the columns with zero temperature ($\beta \hbar \omega_q \to \infty$) in Table~\ref{tbl:paramsSubOhmic} for the broadband noise spectra.
The results in Sec.~\ref{sec:resultISWAP} were obtained with the parameter values in the columns with $\beta \hbar \omega_q = 5$ in Table~\ref{tbl:paramsSubOhmic}, while those in Sec.~\ref{sec:resultHCNOT} were obtained with the parameter values in the columns with $2 \pi \hbar \kappa = 0.004$ in the same table.
For the heterogeneous environments discussed in Sec.~\ref{sec:resultISWAP}, the larger $R_{\max}$ was adopted.
We can set $\mathcal{N}_{\max}$ independently for each reservoir within our termination scheme.

As shown in Eq.~\eqref{eq:cfLorentz}, since the autocorrelation function for the Lorentzian spectra can be always expressed with a single damped mode ($K = 1$), we did not need to use the TT method.
Accordingly, we chose a different termination scheme for the FP-HEOM method:
When the depth of an ADO, $\mathcal{N} = \sum_{j, k}(m_{j, k} + n_{j, k})$, is greater than $\mathcal{N}_{\max}$, the ADO is always set to zero.
One finds peculiar increase of $\mathcal{N}_{\max}$ in the case with $2 \pi \hbar \kappa = 0.002$, $\Delta / \omega_q = 0$, and $\hat{\rho}_0 = \ketbra{\Psi}{\Psi}$.
The larger depth of the hierarchy was needed to accurately describe the relaxation process of the most excited state $\rho_{11}(t)$.
In the detuned case for the intermediate coupling $2 \pi \hbar \kappa = 0.1$, the smaller value of $\mathcal{N}_{\max}$ compared to the resonant case results from the effective weaker system--reservoir coupling [Eq.~\eqref{eq:T1Detuned}].
In principle, FP-HEOM simulation with maximum depth $\mathcal{N}_{\max}$ includes the system--reservoir interaction up to the $2\mathcal{N}_{\max}$th order, and hence the weaker the coupling becomes, the fewer hierarchy is needed.
The same trend can be seen in the case with the broadband spectra for the different $\kappa$.

One can see contribution of high-order terms with respect to the system--reservoir coupling through the comparison between the zero- and finite-temperature cases with deep sub-Ohmic reservoirs in Table~\ref{tbl:paramsSubOhmic}.
In the deep sub-Ohmic case ($s = 1/14$) at zero temperature, the maximum depth of the hierarchy is the same as the other two cases, $s = 1$, $1/2$, while it increases at finite temperature ($s = 1/8$).
When the temperature is zero, the spectral noise power reaches zero at the origin of the frequency, which is indicated by the approximate expression $S^{\beta \to \infty} (\omega \simeq 0) \simeq \omega^{s}$.
By contrast, when the temperature is nonzero and the exponent $s$ is less than $1$, it diverges at the origin, which is expressed as $S^\beta (\omega \simeq 0) \simeq 1/\omega^{1-s}$.
It is concluded that in order to accurately describe the effects caused by the slow dynamics of the reservoir, which are enhanced in nonzero-temperature cases, we must include higher-order terms of the system--reservoir coupling.

Finally, we mention the parameter values $\mathcal{N}_{\max}$ and $R_{\max}$ in the cases with $s = 1/8$.
We found that required values for both $\mathcal{N}_{\max}$ and $R_{\max}$ become larger as the total simulation time $\omega_q t$ grows.
Further, it was found that with two different values of $R_{\max}$, the system dynamics are almost the same up to a certain time and then deviate from each other, instead of exhibiting different behavior throughout the entire period.
For example, we confirmed that the dynamics in Fig.~\ref{fig:iSWAPHomo}(c) with $s = 1/8$ converge up to the time $\omega_q t / \pi \simeq 14.3$ on the basis of this finding.
The values in the left column for the case with $s = 1/8$, $\beta \hbar \omega_q = 5$, and $2 \pi \hbar \kappa = 0.004$ were used in Sec.~\ref{sec:resultISWAP}, while those in the right column were used in Sec.~\ref{sec:resultHCNOT}.
The larger value $\mathcal{N}_{\max}$ in the left column is required for the longer simulation time, while the large value $R_{\max}$ in the right column is required for the more precise description of the dynamics with the finite values of $\Omega_1(t)$ and $\Omega_2(t)$ in Sec.~\ref{sec:resultHCNOT}.

\begin{figure}[t]
    \centering
    \includegraphics[width=0.95\linewidth]{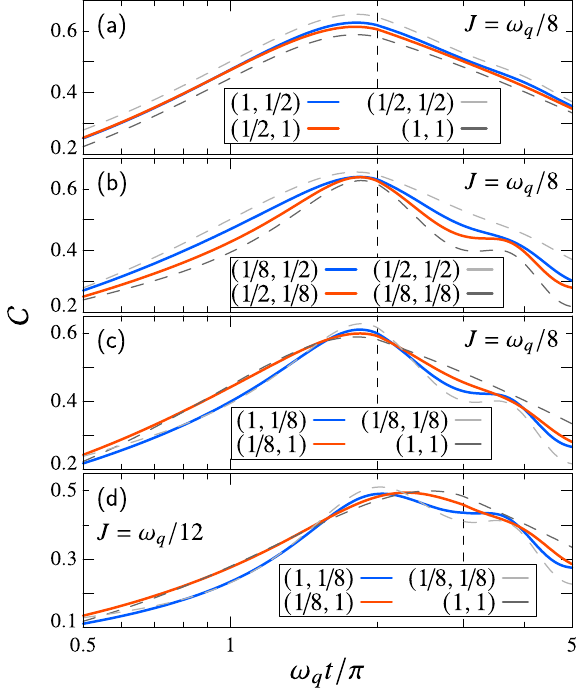}
    \caption{Time traces of the concurrence $\mathcal{C}(t)$ for various reservoirs during and after application of a \sqrtiswap{} gate (linear--log plots).
    Each qubit couples with a different reservoir (different spectral exponent, $s_1 \neq s_2$). The set of the two numbers in the legend indicates the spectral exponent of each reservoir, $(s_1, s_2)$.
    The dashed gray curves correspond to cases with the homogeneous environments and are depicted for reference purposes.
    The strength of the system--reservoir coupling is fixed to $2 \pi \hbar \kappa = 0.04$.
    Different sets of the spectral exponents are considered in (a)--(c) with the same coupling strength between the two qubits, $J / \omega_q = 1/8$.
    In (d), the set of the spectral exponents is the same as that in (c), but the qubit--qubit coupling is weaker ($J / \omega_q = 1/12$).
    The vertical dashed lines indicate the end time of the gate operation.
    \label{fig:iSWAPApp}}
\end{figure}

\begin{figure}[t]
    \centering
    \includegraphics[width=0.95\linewidth]{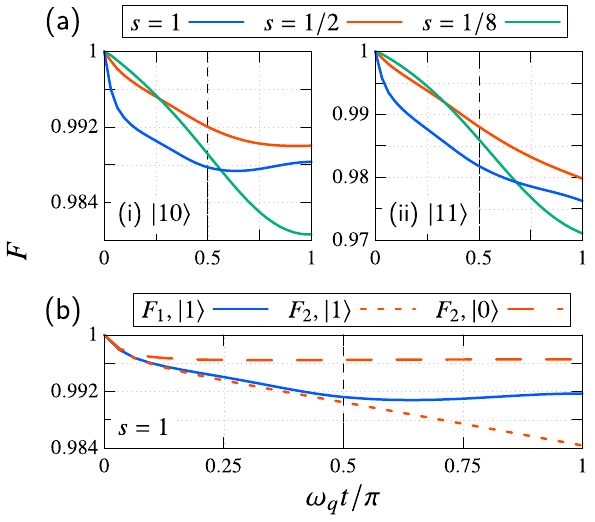}
    \caption{Time traces of the fidelity during the first gate operation [$R^{y}(-\pi/2) \otimes \hat{1}$] and the first idling for Seq.~(b) in Fig.~\ref{fig:schematicH+CNOT}.
    The vertical dashed lines indicate the end of the first gate application.
    (a) The results with the initial states (i) $\ket{10}$ and (ii) $\ket{11}$ are depicted.
    The spectral exponent is varied as $s = 1$ (blue curves), $1/2$ (red curves), and $1/8$ (green curves).
    (b) Partial fidelities in the Ohmic case, $s=1$.
    The fidelity of the first qubit to which the single-qubit gate is applied is depicted as a blue curve, while that of the second qubit without application of any gates is depicted as red curves, with the initial state being $\ket{1}$ (dotted curve) and $\ket{0}$ (dashed curve).
    \label{fig:partialRecovery}}
\end{figure}

\section{Dynamics of concurrence in heterogeneous environment} \label{sec:appISWAPHetero}
In this appendix, we investigate how the concurrence is affected by heterogeneous environments.
Following the scheme in Sec.~\ref{sec:resultISWAP}, a \sqrtiswap{} gate is applied with the initial state $\ketbra{10}{10}$, and idling dynamics after switching off the gate are monitored [cf.~schematic of Fig.~\ref{fig:iSWAPHomo}(a)].
The coupling strength between the system and reservoir is fixed to $2 \pi \hbar \kappa  = 0.04$.

Figure~\ref{fig:iSWAPApp} displays the time traces of the concurrence of systems in heterogeneous environments.
All the combinations of the spectral exponent, $s_j \in \{1, 1/2, 1/8\}$, are depicted.
Since the initial state is asymmetric with respect to each qubit, the dynamics with $(s_1, s_2) = (a, b)$ and $(b, a)$ are different.
\textcolor{black}{At the end time of the gate, the concurrences for the heterogeneous environments (blue and red curves) take the values between those for the homogeneous environments (dashed gray curves).}
The performance is determined by the characteristics of both of the reservoirs.

In Figs.~\ref{fig:iSWAPApp}(b) and \ref{fig:iSWAPApp}(c), one can see enhanced oscillation during the idling in the setting where the second qubit is coupled with the deep sub-Ohmic reservoir, $s_2 = 1/8$.
As discussed in the main text, the oscillatory behavior is mainly attributed to the behavior of the most excited state $\rho_{11}(t)$.
Since the initial state is $\ketbra{10}{10}$, the growth of the population $\rho_{11}(t)$ is mainly caused by the second reservoir under the excitation process from $\ket{10}$ to $\ket{11}$.
Hence, the second reservoir mainly contributes to the oscillatory behavior of the concurrence.
\textcolor{black}{We here again emphasize that the excitation process to the state $\ket{11}$ cannot be described within the RWA.}

Another interesting phenomenon relating to this oscillatory behavior is found in a weaker qubit--qubit coupling case [$J/\omega_q = 1/12$, Fig.~\ref{fig:iSWAPApp}(d)].
At the end of the gate operation, the concurrence with the reservoirs $(s_1, s_2) = (1/8, 1)$ is larger than that with $(1, 1/8)$, which is the opposite result in the stronger coupling case in Fig.~\ref{fig:iSWAPApp}(c).
In Fig.~\ref{fig:iSWAPApp}(d), the concurrence with $(1, 1/8)$ at the time $\omega_q t / \pi = 2$ is larger than that with $(1/8, 1)$, but owing to the oscillation, the concurrence in the former case gets worse as time goes on.

Noting that switching the reservoirs $s_1 \leftrightarrow s_2$ corresponds to switching the initial state $\ket{10} \leftrightarrow \ket{01}$, it is implied that we can avoid certain sub-Ohmic noise effects by switching the roles of the qubit, for example, a target and a control qubit.\\

\section{Partial recovery of the fidelity during the first Hadamard gate} \label{sec:appHCNOT}
Here, we illustrate partial recovery of the fidelity occurring during an idling phase, which is discussed in the main text (Sec.~\ref{sec:resultHCNOT}).
One can see this phenomenon during the first idling in the Ohmic case with the initial state $\ket{10}$ for Seq.~(b) in Fig.~\ref{fig:partialRecovery}(a)-(i):
The relation $F(t = 0.5\pi/\omega_q) < F(t = \pi / \omega_q)$ is observed.
We also found this recovery in the Ohmic case with the initial state $\ket{00}$ for Seq.~(b) (results are not shown).
For the other parameter values, the fidelity after the idling phases cannot exceed the values before those phases [cf.~Figs.~\ref{fig:partialRecovery}(a)-(i) and (ii)].
For deeper analysis, we again decompose the total fidelity into the components for each qubit, $F_\mathrm{tot}(t) = F_1(t) F_2(t)$.
As indicated in Fig.~\ref{fig:partialRecovery}(b), the partial fidelity of the first qubit is recovered after the application of the single-qubit gate, $R^y(-\pi /2)$.
\textcolor{black}{When the second qubit, which is not subject to any gate operations until the end of the first idling phase, is in the ground state, the fidelity is almost time stationary.}
As a result, the total fidelity for the Ohmic reservoir in Fig.~\ref{fig:partialRecovery}(a)-(i) exhibits the recovery behavior.
By contrast, the second qubit experiences the intense decay of the fidelity when the initial state is prepared into $\ket{1}$ [red dotted curve in Fig.~\ref{fig:partialRecovery}(b)], and the recovery behavior of the total fidelity cannot be observed in Fig.~\ref{fig:partialRecovery}(a)-(ii) (blue curve).
Despite this partial recovery, the fidelity at the end of the whole sequence is worse compared to that without the idling, as discussed in the main text.

\bibliography{reference,qubit,tt}

\begin{thebibliography}{89}%
\makeatletter
\providecommand \@ifxundefined [1]{%
 \@ifx{#1\undefined}
}%
\providecommand \@ifnum [1]{%
 \ifnum #1\expandafter \@firstoftwo
 \else \expandafter \@secondoftwo
 \fi
}%
\providecommand \@ifx [1]{%
 \ifx #1\expandafter \@firstoftwo
 \else \expandafter \@secondoftwo
 \fi
}%
\providecommand \natexlab [1]{#1}%
\providecommand \enquote  [1]{``#1''}%
\providecommand \bibnamefont  [1]{#1}%
\providecommand \bibfnamefont [1]{#1}%
\providecommand \citenamefont [1]{#1}%
\providecommand \href@noop [0]{\@secondoftwo}%
\providecommand \href [0]{\begingroup \@sanitize@url \@href}%
\providecommand \@href[1]{\@@startlink{#1}\@@href}%
\providecommand \@@href[1]{\endgroup#1\@@endlink}%
\providecommand \@sanitize@url [0]{\catcode `\\12\catcode `\$12\catcode
  `\&12\catcode `\#12\catcode `\^12\catcode `\_12\catcode `\%12\relax}%
\providecommand \@@startlink[1]{}%
\providecommand \@@endlink[0]{}%
\providecommand \url  [0]{\begingroup\@sanitize@url \@url }%
\providecommand \@url [1]{\endgroup\@href {#1}{\urlprefix }}%
\providecommand \urlprefix  [0]{URL }%
\providecommand \Eprint [0]{\href }%
\providecommand \doibase [0]{https://doi.org/}%
\providecommand \selectlanguage [0]{\@gobble}%
\providecommand \bibinfo  [0]{\@secondoftwo}%
\providecommand \bibfield  [0]{\@secondoftwo}%
\providecommand \translation [1]{[#1]}%
\providecommand \BibitemOpen [0]{}%
\providecommand \bibitemStop [0]{}%
\providecommand \bibitemNoStop [0]{.\EOS\space}%
\providecommand \EOS [0]{\spacefactor3000\relax}%
\providecommand \BibitemShut  [1]{\csname bibitem#1\endcsname}%
\let\auto@bib@innerbib\@empty
\bibitem [{\citenamefont {Place}\ \emph {et~al.}(2021)\citenamefont {Place},
  \citenamefont {Rodgers}, \citenamefont {Mundada}, \citenamefont {Smitham},
  \citenamefont {Fitzpatrick}, \citenamefont {Leng}, \citenamefont {Premkumar},
  \citenamefont {Bryon}, \citenamefont {Vrajitoarea}, \citenamefont {Sussman}
  \emph {et~al.}}]{PlaceNATCOMMUN2021}%
  \BibitemOpen
  \bibfield  {author} {\bibinfo {author} {\bibfnamefont {A.~P.~M.}\
  \bibnamefont {Place}}, \bibinfo {author} {\bibfnamefont {L.~V.~H.}\
  \bibnamefont {Rodgers}}, \bibinfo {author} {\bibfnamefont {P.}~\bibnamefont
  {Mundada}}, \bibinfo {author} {\bibfnamefont {B.~M.}\ \bibnamefont
  {Smitham}}, \bibinfo {author} {\bibfnamefont {M.}~\bibnamefont
  {Fitzpatrick}}, \bibinfo {author} {\bibfnamefont {Z.}~\bibnamefont {Leng}},
  \bibinfo {author} {\bibfnamefont {A.}~\bibnamefont {Premkumar}}, \bibinfo
  {author} {\bibfnamefont {J.}~\bibnamefont {Bryon}}, \bibinfo {author}
  {\bibfnamefont {A.}~\bibnamefont {Vrajitoarea}}, \bibinfo {author}
  {\bibfnamefont {S.}~\bibnamefont {Sussman}}, \emph {et~al.},\ }\bibfield
  {title} {\bibinfo {title} {{New material platform for superconducting
  transmon qubits with coherence times exceeding 0.3 milliseconds}},\ }\href
  {https://doi.org/10.1038/s41467-021-22030-5} {\bibfield  {journal} {\bibinfo
  {journal} {Nat. Commun.}\ }\textbf {\bibinfo {volume} {12}},\ \bibinfo
  {pages} {1779} (\bibinfo {year} {2021})}\BibitemShut {NoStop}%
\bibitem [{\citenamefont {Wang}\ \emph {et~al.}(2022)\citenamefont {Wang},
  \citenamefont {Li}, \citenamefont {Xu}, \citenamefont {Li}, \citenamefont
  {Wang}, \citenamefont {Yang}, \citenamefont {Mi}, \citenamefont {Liang},
  \citenamefont {Su}, \citenamefont {Yang} \emph {et~al.}}]{WangNPJQI2022}%
  \BibitemOpen
  \bibfield  {author} {\bibinfo {author} {\bibfnamefont {C.}~\bibnamefont
  {Wang}}, \bibinfo {author} {\bibfnamefont {X.}~\bibnamefont {Li}}, \bibinfo
  {author} {\bibfnamefont {H.}~\bibnamefont {Xu}}, \bibinfo {author}
  {\bibfnamefont {Z.}~\bibnamefont {Li}}, \bibinfo {author} {\bibfnamefont
  {J.}~\bibnamefont {Wang}}, \bibinfo {author} {\bibfnamefont {Z.}~\bibnamefont
  {Yang}}, \bibinfo {author} {\bibfnamefont {Z.}~\bibnamefont {Mi}}, \bibinfo
  {author} {\bibfnamefont {X.}~\bibnamefont {Liang}}, \bibinfo {author}
  {\bibfnamefont {T.}~\bibnamefont {Su}}, \bibinfo {author} {\bibfnamefont
  {C.}~\bibnamefont {Yang}}, \emph {et~al.},\ }\bibfield  {title} {\bibinfo
  {title} {{Towards practical quantum computers: transmon qubit with a lifetime
  approaching 0.5 milliseconds}},\ }\href
  {https://doi.org/10.1038/s41534-021-00510-2} {\bibfield  {journal} {\bibinfo
  {journal} {npj Quantum Inf.}\ }\textbf {\bibinfo {volume} {8}},\ \bibinfo
  {pages} {3} (\bibinfo {year} {2022})}\BibitemShut {NoStop}%
\bibitem [{\citenamefont {Wang}\ \emph
  {et~al.}(2025{\natexlab{a}})\citenamefont {Wang}, \citenamefont {Lu},
  \citenamefont {Zhan}, \citenamefont {Ma}, \citenamefont {Wu}, \citenamefont
  {Sun}, \citenamefont {Deng}, \citenamefont {Bai}, \citenamefont {Bao} \emph
  {et~al.}}]{WangPhysRevAppl2025}%
  \BibitemOpen
  \bibfield  {author} {\bibinfo {author} {\bibfnamefont {F.}~\bibnamefont
  {Wang}}, \bibinfo {author} {\bibfnamefont {K.}~\bibnamefont {Lu}}, \bibinfo
  {author} {\bibfnamefont {H.}~\bibnamefont {Zhan}}, \bibinfo {author}
  {\bibfnamefont {L.}~\bibnamefont {Ma}}, \bibinfo {author} {\bibfnamefont
  {F.}~\bibnamefont {Wu}}, \bibinfo {author} {\bibfnamefont {H.}~\bibnamefont
  {Sun}}, \bibinfo {author} {\bibfnamefont {H.}~\bibnamefont {Deng}}, \bibinfo
  {author} {\bibfnamefont {Y.}~\bibnamefont {Bai}}, \bibinfo {author}
  {\bibfnamefont {F.}~\bibnamefont {Bao}}, \emph {et~al.},\ }\bibfield  {title}
  {\bibinfo {title} {{High-coherence fluxonium qubits manufactured with a
  wafer-scale-uniformity process}},\ }\href
  {https://doi.org/10.1103/PhysRevApplied.23.044064} {\bibfield  {journal}
  {\bibinfo  {journal} {Phys. Rev. Applied}\ }\textbf {\bibinfo {volume}
  {23}},\ \bibinfo {pages} {044064} (\bibinfo {year}
  {2025}{\natexlab{a}})}\BibitemShut {NoStop}%
\bibitem [{\citenamefont {Tuokkola}\ \emph {et~al.}(2025)\citenamefont
  {Tuokkola}, \citenamefont {Sunada}, \citenamefont {Kivij{\"a}rvi},
  \citenamefont {Albanese}, \citenamefont {Gr{\"o}nberg}, \citenamefont
  {Kaikkonen}, \citenamefont {Vesterinen}, \citenamefont {Govenius},\ and\
  \citenamefont {M{\"o}tt{\"o}nen}}]{TuokkolaNATCOMMUN2025}%
  \BibitemOpen
  \bibfield  {author} {\bibinfo {author} {\bibfnamefont {M.}~\bibnamefont
  {Tuokkola}}, \bibinfo {author} {\bibfnamefont {Y.}~\bibnamefont {Sunada}},
  \bibinfo {author} {\bibfnamefont {H.}~\bibnamefont {Kivij{\"a}rvi}}, \bibinfo
  {author} {\bibfnamefont {J.}~\bibnamefont {Albanese}}, \bibinfo {author}
  {\bibfnamefont {L.}~\bibnamefont {Gr{\"o}nberg}}, \bibinfo {author}
  {\bibfnamefont {J.-P.}\ \bibnamefont {Kaikkonen}}, \bibinfo {author}
  {\bibfnamefont {V.}~\bibnamefont {Vesterinen}}, \bibinfo {author}
  {\bibfnamefont {J.}~\bibnamefont {Govenius}},\ and\ \bibinfo {author}
  {\bibfnamefont {M.}~\bibnamefont {M{\"o}tt{\"o}nen}},\ }\bibfield  {title}
  {\bibinfo {title} {{Methods to achieve near-millisecond energy relaxation and
  dephasing times for a superconducting transmon qubit}},\ }\href
  {https://doi.org/10.1038/s41467-025-61126-0} {\bibfield  {journal} {\bibinfo
  {journal} {Nat. Commun.}\ }\textbf {\bibinfo {volume} {16}},\ \bibinfo
  {pages} {5421} (\bibinfo {year} {2025})}\BibitemShut {NoStop}%
\bibitem [{\citenamefont {Neg\^{\i}rneac}\ \emph {et~al.}(2021)\citenamefont
  {Neg\^{\i}rneac}, \citenamefont {Ali}, \citenamefont {Muthusubramanian},
  \citenamefont {Battistel}, \citenamefont {Sagastizabal}, \citenamefont
  {Moreira}, \citenamefont {Marques}, \citenamefont {Vlothuizen}, \citenamefont
  {Beekman}, \citenamefont {Zachariadis} \emph {et~al.}}]{NegirneacPRL2021}%
  \BibitemOpen
  \bibfield  {author} {\bibinfo {author} {\bibfnamefont {V.}~\bibnamefont
  {Neg\^{\i}rneac}}, \bibinfo {author} {\bibfnamefont {H.}~\bibnamefont {Ali}},
  \bibinfo {author} {\bibfnamefont {N.}~\bibnamefont {Muthusubramanian}},
  \bibinfo {author} {\bibfnamefont {F.}~\bibnamefont {Battistel}}, \bibinfo
  {author} {\bibfnamefont {R.}~\bibnamefont {Sagastizabal}}, \bibinfo {author}
  {\bibfnamefont {M.~S.}\ \bibnamefont {Moreira}}, \bibinfo {author}
  {\bibfnamefont {J.~F.}\ \bibnamefont {Marques}}, \bibinfo {author}
  {\bibfnamefont {W.~J.}\ \bibnamefont {Vlothuizen}}, \bibinfo {author}
  {\bibfnamefont {M.}~\bibnamefont {Beekman}}, \bibinfo {author} {\bibfnamefont
  {C.}~\bibnamefont {Zachariadis}}, \emph {et~al.},\ }\bibfield  {title}
  {\bibinfo {title} {{High-Fidelity Controlled-$Z$ Gate with Maximal
  Intermediate Leakage Operating at the Speed Limit in a Superconducting
  Quantum Processor}},\ }\href {https://doi.org/10.1103/PhysRevLett.126.220502}
  {\bibfield  {journal} {\bibinfo  {journal} {Phys. Rev. Lett.}\ }\textbf
  {\bibinfo {volume} {126}},\ \bibinfo {pages} {220502} (\bibinfo {year}
  {2021})}\BibitemShut {NoStop}%
\bibitem [{\citenamefont {Sung}\ \emph {et~al.}(2021)\citenamefont {Sung},
  \citenamefont {Ding}, \citenamefont {Braum\"uller}, \citenamefont
  {Veps\"al\"ainen}, \citenamefont {Kannan}, \citenamefont {Kjaergaard},
  \citenamefont {Greene}, \citenamefont {Samach}, \citenamefont {McNally},
  \citenamefont {Kim} \emph {et~al.}}]{SungPRX2021}%
  \BibitemOpen
  \bibfield  {author} {\bibinfo {author} {\bibfnamefont {Y.}~\bibnamefont
  {Sung}}, \bibinfo {author} {\bibfnamefont {L.}~\bibnamefont {Ding}}, \bibinfo
  {author} {\bibfnamefont {J.}~\bibnamefont {Braum\"uller}}, \bibinfo {author}
  {\bibfnamefont {A.}~\bibnamefont {Veps\"al\"ainen}}, \bibinfo {author}
  {\bibfnamefont {B.}~\bibnamefont {Kannan}}, \bibinfo {author} {\bibfnamefont
  {M.}~\bibnamefont {Kjaergaard}}, \bibinfo {author} {\bibfnamefont
  {A.}~\bibnamefont {Greene}}, \bibinfo {author} {\bibfnamefont {G.~O.}\
  \bibnamefont {Samach}}, \bibinfo {author} {\bibfnamefont {C.}~\bibnamefont
  {McNally}}, \bibinfo {author} {\bibfnamefont {D.}~\bibnamefont {Kim}}, \emph
  {et~al.},\ }\bibfield  {title} {\bibinfo {title} {{Realization of
  High-Fidelity CZ and $ZZ$-Free iSWAP Gates with a Tunable Coupler}},\ }\href
  {https://doi.org/10.1103/PhysRevX.11.021058} {\bibfield  {journal} {\bibinfo
  {journal} {Phys. Rev. X}\ }\textbf {\bibinfo {volume} {11}},\ \bibinfo
  {pages} {021058} (\bibinfo {year} {2021})}\BibitemShut {NoStop}%
\bibitem [{\citenamefont {Kandala}\ \emph {et~al.}(2021)\citenamefont
  {Kandala}, \citenamefont {Wei}, \citenamefont {Srinivasan}, \citenamefont
  {Magesan}, \citenamefont {Carnevale}, \citenamefont {Keefe}, \citenamefont
  {Klaus}, \citenamefont {Dial},\ and\ \citenamefont {McKay}}]{KandalaPRL2021}%
  \BibitemOpen
  \bibfield  {author} {\bibinfo {author} {\bibfnamefont {A.}~\bibnamefont
  {Kandala}}, \bibinfo {author} {\bibfnamefont {K.~X.}\ \bibnamefont {Wei}},
  \bibinfo {author} {\bibfnamefont {S.}~\bibnamefont {Srinivasan}}, \bibinfo
  {author} {\bibfnamefont {E.}~\bibnamefont {Magesan}}, \bibinfo {author}
  {\bibfnamefont {S.}~\bibnamefont {Carnevale}}, \bibinfo {author}
  {\bibfnamefont {G.~A.}\ \bibnamefont {Keefe}}, \bibinfo {author}
  {\bibfnamefont {D.}~\bibnamefont {Klaus}}, \bibinfo {author} {\bibfnamefont
  {O.}~\bibnamefont {Dial}},\ and\ \bibinfo {author} {\bibfnamefont {D.~C.}\
  \bibnamefont {McKay}},\ }\bibfield  {title} {\bibinfo {title} {{Demonstration
  of a High-Fidelity \textsc{cnot} Gate for Fixed-Frequency Transmons with
  Engineered $ZZ$ Suppression}},\ }\href
  {https://doi.org/10.1103/PhysRevLett.127.130501} {\bibfield  {journal}
  {\bibinfo  {journal} {Phys. Rev. Lett.}\ }\textbf {\bibinfo {volume} {127}},\
  \bibinfo {pages} {130501} (\bibinfo {year} {2021})}\BibitemShut {NoStop}%
\bibitem [{\citenamefont {Li}\ \emph {et~al.}(2024)\citenamefont {Li},
  \citenamefont {Kubo}, \citenamefont {Ho}, \citenamefont {Yan}, \citenamefont
  {Nakamura},\ and\ \citenamefont {Goto}}]{LiPRX2024}%
  \BibitemOpen
  \bibfield  {author} {\bibinfo {author} {\bibfnamefont {R.}~\bibnamefont
  {Li}}, \bibinfo {author} {\bibfnamefont {K.}~\bibnamefont {Kubo}}, \bibinfo
  {author} {\bibfnamefont {Y.}~\bibnamefont {Ho}}, \bibinfo {author}
  {\bibfnamefont {Z.}~\bibnamefont {Yan}}, \bibinfo {author} {\bibfnamefont
  {Y.}~\bibnamefont {Nakamura}},\ and\ \bibinfo {author} {\bibfnamefont
  {H.}~\bibnamefont {Goto}},\ }\bibfield  {title} {\bibinfo {title}
  {{Realization of High-Fidelity CZ Gate Based on a Double-Transmon Coupler}},\
  }\href {https://doi.org/10.1103/PhysRevX.14.041050} {\bibfield  {journal}
  {\bibinfo  {journal} {Phys. Rev. X}\ }\textbf {\bibinfo {volume} {14}},\
  \bibinfo {pages} {041050} (\bibinfo {year} {2024})}\BibitemShut {NoStop}%
\bibitem [{\citenamefont {Wang}\ \emph
  {et~al.}(2025{\natexlab{b}})\citenamefont {Wang}, \citenamefont {Liu},
  \citenamefont {Chen}, \citenamefont {Du}, \citenamefont {Ying}, \citenamefont
  {Wang}, \citenamefont {Huo}, \citenamefont {Peng}, \citenamefont {Zhu},
  \citenamefont {Chen}, \citenamefont {Lu},\ and\ \citenamefont
  {Pan}}]{WangPRL2025}%
  \BibitemOpen
  \bibfield  {author} {\bibinfo {author} {\bibfnamefont {C.}~\bibnamefont
  {Wang}}, \bibinfo {author} {\bibfnamefont {F.-M.}\ \bibnamefont {Liu}},
  \bibinfo {author} {\bibfnamefont {H.}~\bibnamefont {Chen}}, \bibinfo {author}
  {\bibfnamefont {Y.-F.}\ \bibnamefont {Du}}, \bibinfo {author} {\bibfnamefont
  {C.}~\bibnamefont {Ying}}, \bibinfo {author} {\bibfnamefont {J.-W.}\
  \bibnamefont {Wang}}, \bibinfo {author} {\bibfnamefont {Y.-H.}\ \bibnamefont
  {Huo}}, \bibinfo {author} {\bibfnamefont {C.-Z.}\ \bibnamefont {Peng}},
  \bibinfo {author} {\bibfnamefont {X.}~\bibnamefont {Zhu}}, \bibinfo {author}
  {\bibfnamefont {M.-C.}\ \bibnamefont {Chen}}, \bibinfo {author}
  {\bibfnamefont {C.-Y.}\ \bibnamefont {Lu}},\ and\ \bibinfo {author}
  {\bibfnamefont {J.-W.}\ \bibnamefont {Pan}},\ }\bibfield  {title} {\bibinfo
  {title} {{Longitudinal and Nonlinear Coupling for High-Fidelity Readout of a
  Superconducting Qubit}},\ }\href {https://doi.org/10.1103/98n9-13y4}
  {\bibfield  {journal} {\bibinfo  {journal} {Phys. Rev. Lett.}\ }\textbf
  {\bibinfo {volume} {135}},\ \bibinfo {pages} {060803} (\bibinfo {year}
  {2025}{\natexlab{b}})}\BibitemShut {NoStop}%
\bibitem [{\citenamefont {Arute}\ \emph {et~al.}(2019)\citenamefont {Arute},
  \citenamefont {Arya}, \citenamefont {Babbush}, \citenamefont {Bacon},
  \citenamefont {Bardin}, \citenamefont {Barends}, \citenamefont {Biswas},
  \citenamefont {Boixo}, \citenamefont {Brandao}, \citenamefont {Buell} \emph
  {et~al.}}]{GoogleNATURE2019}%
  \BibitemOpen
  \bibfield  {author} {\bibinfo {author} {\bibfnamefont {F.}~\bibnamefont
  {Arute}}, \bibinfo {author} {\bibfnamefont {K.}~\bibnamefont {Arya}},
  \bibinfo {author} {\bibfnamefont {R.}~\bibnamefont {Babbush}}, \bibinfo
  {author} {\bibfnamefont {D.}~\bibnamefont {Bacon}}, \bibinfo {author}
  {\bibfnamefont {J.~C.}\ \bibnamefont {Bardin}}, \bibinfo {author}
  {\bibfnamefont {R.}~\bibnamefont {Barends}}, \bibinfo {author} {\bibfnamefont
  {R.}~\bibnamefont {Biswas}}, \bibinfo {author} {\bibfnamefont
  {S.}~\bibnamefont {Boixo}}, \bibinfo {author} {\bibfnamefont {F.~G. S.~L.}\
  \bibnamefont {Brandao}}, \bibinfo {author} {\bibfnamefont {D.~A.}\
  \bibnamefont {Buell}}, \emph {et~al.},\ }\bibfield  {title} {\bibinfo {title}
  {{Quantum supremacy using a programmable superconducting processor}},\ }\href
  {https://doi.org/10.1038/s41586-019-1666-5} {\bibfield  {journal} {\bibinfo
  {journal} {{Nature (London)}}\ }\textbf {\bibinfo {volume} {574}},\ \bibinfo
  {pages} {505} (\bibinfo {year} {2019})}\BibitemShut {NoStop}%
\bibitem [{\citenamefont {{Google Quantum AI}}(2023)}]{GoogleNATURE2023}%
  \BibitemOpen
  \bibfield  {author} {\bibinfo {author} {\bibnamefont {{Google Quantum AI}}},\
  }\bibfield  {title} {\bibinfo {title} {{Suppressing quantum errors by scaling
  a surface code logical qubit}},\ }\href
  {https://doi.org/10.1038/s41586-022-05434-1} {\bibfield  {journal} {\bibinfo
  {journal} {{Nature (London)}}\ }\textbf {\bibinfo {volume} {614}},\ \bibinfo
  {pages} {676} (\bibinfo {year} {2023})}\BibitemShut {NoStop}%
\bibitem [{\citenamefont {Kim}\ \emph {et~al.}(2023)\citenamefont {Kim},
  \citenamefont {Eddins}, \citenamefont {Anand}, \citenamefont {Wei},
  \citenamefont {van~den Berg}, \citenamefont {Rosenblatt}, \citenamefont
  {Nayfeh}, \citenamefont {Wu}, \citenamefont {Zaletel}, \citenamefont {Temme}
  \emph {et~al.}}]{IBMNATURE2023}%
  \BibitemOpen
  \bibfield  {author} {\bibinfo {author} {\bibfnamefont {Y.}~\bibnamefont
  {Kim}}, \bibinfo {author} {\bibfnamefont {A.}~\bibnamefont {Eddins}},
  \bibinfo {author} {\bibfnamefont {S.}~\bibnamefont {Anand}}, \bibinfo
  {author} {\bibfnamefont {K.~X.}\ \bibnamefont {Wei}}, \bibinfo {author}
  {\bibfnamefont {E.}~\bibnamefont {van~den Berg}}, \bibinfo {author}
  {\bibfnamefont {S.}~\bibnamefont {Rosenblatt}}, \bibinfo {author}
  {\bibfnamefont {H.}~\bibnamefont {Nayfeh}}, \bibinfo {author} {\bibfnamefont
  {Y.}~\bibnamefont {Wu}}, \bibinfo {author} {\bibfnamefont {M.}~\bibnamefont
  {Zaletel}}, \bibinfo {author} {\bibfnamefont {K.}~\bibnamefont {Temme}},
  \emph {et~al.},\ }\bibfield  {title} {\bibinfo {title} {{Evidence for the
  utility of quantum computing before fault tolerance}},\ }\href
  {https://doi.org/10.1038/s41586-023-06096-3} {\bibfield  {journal} {\bibinfo
  {journal} {{Nature (London)}}\ }\textbf {\bibinfo {volume} {618}},\ \bibinfo
  {pages} {500} (\bibinfo {year} {2023})}\BibitemShut {NoStop}%
\bibitem [{\citenamefont {Havl{\'i}{\v{c}}ek}\ \emph
  {et~al.}(2019)\citenamefont {Havl{\'i}{\v{c}}ek}, \citenamefont
  {C{\'o}rcoles}, \citenamefont {Temme}, \citenamefont {Harrow}, \citenamefont
  {Kandala}, \citenamefont {Chow},\ and\ \citenamefont
  {Gambetta}}]{HavlicekNATURE2019}%
  \BibitemOpen
  \bibfield  {author} {\bibinfo {author} {\bibfnamefont {V.}~\bibnamefont
  {Havl{\'i}{\v{c}}ek}}, \bibinfo {author} {\bibfnamefont {A.~D.}\ \bibnamefont
  {C{\'o}rcoles}}, \bibinfo {author} {\bibfnamefont {K.}~\bibnamefont {Temme}},
  \bibinfo {author} {\bibfnamefont {A.~W.}\ \bibnamefont {Harrow}}, \bibinfo
  {author} {\bibfnamefont {A.}~\bibnamefont {Kandala}}, \bibinfo {author}
  {\bibfnamefont {J.~M.}\ \bibnamefont {Chow}},\ and\ \bibinfo {author}
  {\bibfnamefont {J.~M.}\ \bibnamefont {Gambetta}},\ }\bibfield  {title}
  {\bibinfo {title} {{Supervised learning with quantum-enhanced feature
  spaces}},\ }\href {https://doi.org/10.1038/s41586-019-0980-2} {\bibfield
  {journal} {\bibinfo  {journal} {{Nature (London)}}\ }\textbf {\bibinfo
  {volume} {567}},\ \bibinfo {pages} {209} (\bibinfo {year}
  {2019})}\BibitemShut {NoStop}%
\bibitem [{\citenamefont {Kandala}\ \emph {et~al.}(2017)\citenamefont
  {Kandala}, \citenamefont {Mezzacapo}, \citenamefont {Temme}, \citenamefont
  {Takita}, \citenamefont {Brink}, \citenamefont {Chow},\ and\ \citenamefont
  {Gambetta}}]{KandalaNATURE2017}%
  \BibitemOpen
  \bibfield  {author} {\bibinfo {author} {\bibfnamefont {A.}~\bibnamefont
  {Kandala}}, \bibinfo {author} {\bibfnamefont {A.}~\bibnamefont {Mezzacapo}},
  \bibinfo {author} {\bibfnamefont {K.}~\bibnamefont {Temme}}, \bibinfo
  {author} {\bibfnamefont {M.}~\bibnamefont {Takita}}, \bibinfo {author}
  {\bibfnamefont {M.}~\bibnamefont {Brink}}, \bibinfo {author} {\bibfnamefont
  {J.~M.}\ \bibnamefont {Chow}},\ and\ \bibinfo {author} {\bibfnamefont
  {J.~M.}\ \bibnamefont {Gambetta}},\ }\bibfield  {title} {\bibinfo {title}
  {{Hardware-efficient variational quantum eigensolver for small molecules and
  quantum magnets}},\ }\href {https://doi.org/10.1038/nature23879} {\bibfield
  {journal} {\bibinfo  {journal} {{Nature (London)}}\ }\textbf {\bibinfo
  {volume} {549}},\ \bibinfo {pages} {242} (\bibinfo {year}
  {2017})}\BibitemShut {NoStop}%
\bibitem [{\citenamefont {Rossmannek}\ \emph {et~al.}(2023)\citenamefont
  {Rossmannek}, \citenamefont {Pavo\v{s}evi\'{c}}, \citenamefont {Rubio},\ and\
  \citenamefont {Tavernelli}}]{RossmannekJPCL2023}%
  \BibitemOpen
  \bibfield  {author} {\bibinfo {author} {\bibfnamefont {M.}~\bibnamefont
  {Rossmannek}}, \bibinfo {author} {\bibfnamefont {F.}~\bibnamefont
  {Pavo\v{s}evi\'{c}}}, \bibinfo {author} {\bibfnamefont {A.}~\bibnamefont
  {Rubio}},\ and\ \bibinfo {author} {\bibfnamefont {I.}~\bibnamefont
  {Tavernelli}},\ }\bibfield  {title} {\bibinfo {title} {{Quantum Embedding
  Method for the Simulation of Strongly Correlated Systems on Quantum
  Computers}},\ }\href {https://doi.org/10.1021/acs.jpclett.3c00330} {\bibfield
   {journal} {\bibinfo  {journal} {J. Phys. Chem. Lett.}\ }\textbf {\bibinfo
  {volume} {14}},\ \bibinfo {pages} {3491} (\bibinfo {year}
  {2023})}\BibitemShut {NoStop}%
\bibitem [{\citenamefont {Willsch}\ \emph {et~al.}(2020)\citenamefont
  {Willsch}, \citenamefont {Willsch}, \citenamefont {Jin}, \citenamefont
  {De~Raedt},\ and\ \citenamefont {Michielsen}}]{WillschQIP2020}%
  \BibitemOpen
  \bibfield  {author} {\bibinfo {author} {\bibfnamefont {M.}~\bibnamefont
  {Willsch}}, \bibinfo {author} {\bibfnamefont {D.}~\bibnamefont {Willsch}},
  \bibinfo {author} {\bibfnamefont {F.}~\bibnamefont {Jin}}, \bibinfo {author}
  {\bibfnamefont {H.}~\bibnamefont {De~Raedt}},\ and\ \bibinfo {author}
  {\bibfnamefont {K.}~\bibnamefont {Michielsen}},\ }\bibfield  {title}
  {\bibinfo {title} {{Benchmarking the quantum approximate optimization
  algorithm}},\ }\href {https://doi.org/10.1007/s11128-020-02692-8} {\bibfield
  {journal} {\bibinfo  {journal} {Quantum Inf. Process.}\ }\textbf {\bibinfo
  {volume} {19}},\ \bibinfo {pages} {197} (\bibinfo {year} {2020})}\BibitemShut
  {NoStop}%
\bibitem [{\citenamefont {Garc{\'i}a-P{\'e}rez}\ \emph
  {et~al.}(2020)\citenamefont {Garc{\'i}a-P{\'e}rez}, \citenamefont {Rossi},\
  and\ \citenamefont {Maniscalco}}]{PerezNPJQI2020}%
  \BibitemOpen
  \bibfield  {author} {\bibinfo {author} {\bibfnamefont {G.}~\bibnamefont
  {Garc{\'i}a-P{\'e}rez}}, \bibinfo {author} {\bibfnamefont {M.~A.~C.}\
  \bibnamefont {Rossi}},\ and\ \bibinfo {author} {\bibfnamefont
  {S.}~\bibnamefont {Maniscalco}},\ }\bibfield  {title} {\bibinfo {title} {{IBM
  Q Experience as a versatile experimental testbed for simulating open quantum
  systems}},\ }\href {https://doi.org/10.1038/s41534-019-0235-y} {\bibfield
  {journal} {\bibinfo  {journal} {npj Quantum Inf.}\ }\textbf {\bibinfo
  {volume} {6}},\ \bibinfo {pages} {1} (\bibinfo {year} {2020})}\BibitemShut
  {NoStop}%
\bibitem [{\citenamefont {Dan}\ \emph {et~al.}(2025)\citenamefont {Dan},
  \citenamefont {Geva},\ and\ \citenamefont {Batista}}]{DanJCTC2025}%
  \BibitemOpen
  \bibfield  {author} {\bibinfo {author} {\bibfnamefont {X.}~\bibnamefont
  {Dan}}, \bibinfo {author} {\bibfnamefont {E.}~\bibnamefont {Geva}},\ and\
  \bibinfo {author} {\bibfnamefont {V.~S.}\ \bibnamefont {Batista}},\
  }\bibfield  {title} {\bibinfo {title} {{Simulating Non-Markovian Quantum
  Dynamics on NISQ Computers Using the Hierarchical Equations of Motion}},\
  }\href {https://doi.org/10.1021/acs.jctc.4c01565} {\bibfield  {journal}
  {\bibinfo  {journal} {J. Chem. Theory Comput.}\ }\textbf {\bibinfo {volume}
  {21}},\ \bibinfo {pages} {1530} (\bibinfo {year} {2025})}\BibitemShut
  {NoStop}%
\bibitem [{\citenamefont {Ithier}\ \emph {et~al.}(2005)\citenamefont {Ithier},
  \citenamefont {Collin}, \citenamefont {Joyez}, \citenamefont {Meeson},
  \citenamefont {Vion}, \citenamefont {Esteve}, \citenamefont {Chiarello},
  \citenamefont {Shnirman}, \citenamefont {Makhlin}, \citenamefont {Schriefl}
  \emph {et~al.}}]{IthierPRB2005}%
  \BibitemOpen
  \bibfield  {author} {\bibinfo {author} {\bibfnamefont {G.}~\bibnamefont
  {Ithier}}, \bibinfo {author} {\bibfnamefont {E.}~\bibnamefont {Collin}},
  \bibinfo {author} {\bibfnamefont {P.}~\bibnamefont {Joyez}}, \bibinfo
  {author} {\bibfnamefont {P.~J.}\ \bibnamefont {Meeson}}, \bibinfo {author}
  {\bibfnamefont {D.}~\bibnamefont {Vion}}, \bibinfo {author} {\bibfnamefont
  {D.}~\bibnamefont {Esteve}}, \bibinfo {author} {\bibfnamefont
  {F.}~\bibnamefont {Chiarello}}, \bibinfo {author} {\bibfnamefont
  {A.}~\bibnamefont {Shnirman}}, \bibinfo {author} {\bibfnamefont
  {Y.}~\bibnamefont {Makhlin}}, \bibinfo {author} {\bibfnamefont
  {J.}~\bibnamefont {Schriefl}}, \emph {et~al.},\ }\bibfield  {title} {\bibinfo
  {title} {{Decoherence in a superconducting quantum bit circuit}},\ }\href
  {https://doi.org/10.1103/PhysRevB.72.134519} {\bibfield  {journal} {\bibinfo
  {journal} {Phys. Rev. B}\ }\textbf {\bibinfo {volume} {72}},\ \bibinfo
  {pages} {134519} (\bibinfo {year} {2005})}\BibitemShut {NoStop}%
\bibitem [{\citenamefont {Rist{\`e}}\ \emph {et~al.}(2013)\citenamefont
  {Rist{\`e}}, \citenamefont {Bultink}, \citenamefont {Tiggelman},
  \citenamefont {Schouten}, \citenamefont {Lehnert},\ and\ \citenamefont
  {DiCarlo}}]{RisteNATCOMMUN2013}%
  \BibitemOpen
  \bibfield  {author} {\bibinfo {author} {\bibfnamefont {D.}~\bibnamefont
  {Rist{\`e}}}, \bibinfo {author} {\bibfnamefont {C.~C.}\ \bibnamefont
  {Bultink}}, \bibinfo {author} {\bibfnamefont {M.~J.}\ \bibnamefont
  {Tiggelman}}, \bibinfo {author} {\bibfnamefont {R.~N.}\ \bibnamefont
  {Schouten}}, \bibinfo {author} {\bibfnamefont {K.~W.}\ \bibnamefont
  {Lehnert}},\ and\ \bibinfo {author} {\bibfnamefont {L.}~\bibnamefont
  {DiCarlo}},\ }\bibfield  {title} {\bibinfo {title} {{Millisecond
  charge-parity fluctuations and induced decoherence in a superconducting
  transmon qubit}},\ }\href {https://doi.org/10.1038/ncomms2936} {\bibfield
  {journal} {\bibinfo  {journal} {Nat. Commun.}\ }\textbf {\bibinfo {volume}
  {4}},\ \bibinfo {pages} {1913} (\bibinfo {year} {2013})}\BibitemShut
  {NoStop}%
\bibitem [{\citenamefont {Cardani}\ \emph {et~al.}(2021)\citenamefont
  {Cardani}, \citenamefont {Valenti}, \citenamefont {Casali}, \citenamefont
  {Catelani}, \citenamefont {Charpentier}, \citenamefont {Clemenza},
  \citenamefont {Colantoni}, \citenamefont {Cruciani}, \citenamefont
  {D'Imperio}, \citenamefont {Gironi} \emph {et~al.}}]{CardaniNATCOMMUN2021}%
  \BibitemOpen
  \bibfield  {author} {\bibinfo {author} {\bibfnamefont {L.}~\bibnamefont
  {Cardani}}, \bibinfo {author} {\bibfnamefont {F.}~\bibnamefont {Valenti}},
  \bibinfo {author} {\bibfnamefont {N.}~\bibnamefont {Casali}}, \bibinfo
  {author} {\bibfnamefont {G.}~\bibnamefont {Catelani}}, \bibinfo {author}
  {\bibfnamefont {T.}~\bibnamefont {Charpentier}}, \bibinfo {author}
  {\bibfnamefont {M.}~\bibnamefont {Clemenza}}, \bibinfo {author}
  {\bibfnamefont {I.}~\bibnamefont {Colantoni}}, \bibinfo {author}
  {\bibfnamefont {A.}~\bibnamefont {Cruciani}}, \bibinfo {author}
  {\bibfnamefont {G.}~\bibnamefont {D'Imperio}}, \bibinfo {author}
  {\bibfnamefont {L.}~\bibnamefont {Gironi}}, \emph {et~al.},\ }\bibfield
  {title} {\bibinfo {title} {{Reducing the impact of radioactivity on quantum
  circuits in a deep-underground facility}},\ }\href
  {https://doi.org/10.1038/s41467-021-23032-z} {\bibfield  {journal} {\bibinfo
  {journal} {Nat. Commun.}\ }\textbf {\bibinfo {volume} {12}},\ \bibinfo
  {pages} {2733} (\bibinfo {year} {2021})}\BibitemShut {NoStop}%
\bibitem [{\citenamefont {Pan}\ \emph {et~al.}(2022)\citenamefont {Pan},
  \citenamefont {Zhou}, \citenamefont {Yuan}, \citenamefont {Nie},
  \citenamefont {Wei}, \citenamefont {Zhang}, \citenamefont {Li}, \citenamefont
  {Liu}, \citenamefont {Jiang}, \citenamefont {Catelani} \emph
  {et~al.}}]{PanNATCOMMUN2022}%
  \BibitemOpen
  \bibfield  {author} {\bibinfo {author} {\bibfnamefont {X.}~\bibnamefont
  {Pan}}, \bibinfo {author} {\bibfnamefont {Y.}~\bibnamefont {Zhou}}, \bibinfo
  {author} {\bibfnamefont {H.}~\bibnamefont {Yuan}}, \bibinfo {author}
  {\bibfnamefont {L.}~\bibnamefont {Nie}}, \bibinfo {author} {\bibfnamefont
  {W.}~\bibnamefont {Wei}}, \bibinfo {author} {\bibfnamefont {L.}~\bibnamefont
  {Zhang}}, \bibinfo {author} {\bibfnamefont {J.}~\bibnamefont {Li}}, \bibinfo
  {author} {\bibfnamefont {S.}~\bibnamefont {Liu}}, \bibinfo {author}
  {\bibfnamefont {Z.~H.}\ \bibnamefont {Jiang}}, \bibinfo {author}
  {\bibfnamefont {G.}~\bibnamefont {Catelani}}, \emph {et~al.},\ }\bibfield
  {title} {\bibinfo {title} {{Engineering superconducting qubits to reduce
  quasiparticles and charge noise}},\ }\href
  {https://doi.org/10.1038/s41467-022-34727-2} {\bibfield  {journal} {\bibinfo
  {journal} {Nat. Commun.}\ }\textbf {\bibinfo {volume} {13}},\ \bibinfo
  {pages} {7196} (\bibinfo {year} {2022})}\BibitemShut {NoStop}%
\bibitem [{\citenamefont {Tuorila}\ \emph {et~al.}(2019)\citenamefont
  {Tuorila}, \citenamefont {Stockburger}, \citenamefont {Ala-Nissila},
  \citenamefont {Ankerhold},\ and\ \citenamefont
  {M\"ott\"onen}}]{TuorilaPRR2019}%
  \BibitemOpen
  \bibfield  {author} {\bibinfo {author} {\bibfnamefont {J.}~\bibnamefont
  {Tuorila}}, \bibinfo {author} {\bibfnamefont {J.}~\bibnamefont
  {Stockburger}}, \bibinfo {author} {\bibfnamefont {T.}~\bibnamefont
  {Ala-Nissila}}, \bibinfo {author} {\bibfnamefont {J.}~\bibnamefont
  {Ankerhold}},\ and\ \bibinfo {author} {\bibfnamefont {M.}~\bibnamefont
  {M\"ott\"onen}},\ }\bibfield  {title} {\bibinfo {title} {{System-environment
  correlations in qubit initialization and control}},\ }\href
  {https://doi.org/10.1103/PhysRevResearch.1.013004} {\bibfield  {journal}
  {\bibinfo  {journal} {Phys. Rev. Research}\ }\textbf {\bibinfo {volume}
  {1}},\ \bibinfo {pages} {013004} (\bibinfo {year} {2019})}\BibitemShut
  {NoStop}%
\bibitem [{\citenamefont {Babu}\ \emph {et~al.}(2021)\citenamefont {Babu},
  \citenamefont {Tuorila},\ and\ \citenamefont {Ala-Nissila}}]{BabuNPJQI2021}%
  \BibitemOpen
  \bibfield  {author} {\bibinfo {author} {\bibfnamefont {A.~P.}\ \bibnamefont
  {Babu}}, \bibinfo {author} {\bibfnamefont {J.}~\bibnamefont {Tuorila}},\ and\
  \bibinfo {author} {\bibfnamefont {T.}~\bibnamefont {Ala-Nissila}},\
  }\bibfield  {title} {\bibinfo {title} {{State leakage during fast decay and
  control of a superconducting transmon qubit}},\ }\href
  {https://doi.org/10.1038/s41534-020-00357-z} {\bibfield  {journal} {\bibinfo
  {journal} {npj Quantum Inf.}\ }\textbf {\bibinfo {volume} {7}},\ \bibinfo
  {pages} {30} (\bibinfo {year} {2021})}\BibitemShut {NoStop}%
\bibitem [{\citenamefont {Papi\v{c}}\ \emph {et~al.}()\citenamefont
  {Papi\v{c}}, \citenamefont {Auer},\ and\ \citenamefont
  {de~Vega}}]{PapicARXIV2023}%
  \BibitemOpen
  \bibfield  {author} {\bibinfo {author} {\bibfnamefont {M.}~\bibnamefont
  {Papi\v{c}}}, \bibinfo {author} {\bibfnamefont {A.}~\bibnamefont {Auer}},\
  and\ \bibinfo {author} {\bibfnamefont {I.}~\bibnamefont {de~Vega}},\
  }\href@noop {} {\bibinfo {title} {{Fast Estimation of Physical Error
  Contributions of Quantum Gates}}},\ \Eprint
  {https://arxiv.org/abs/2305.08916} {arXiv:2305.08916} \BibitemShut {NoStop}%
\bibitem [{\citenamefont {Nakamura}\ and\ \citenamefont
  {Ankerhold}(2024{\natexlab{a}})}]{NakamuraPRR2024}%
  \BibitemOpen
  \bibfield  {author} {\bibinfo {author} {\bibfnamefont {K.}~\bibnamefont
  {Nakamura}}\ and\ \bibinfo {author} {\bibfnamefont {J.}~\bibnamefont
  {Ankerhold}},\ }\bibfield  {title} {\bibinfo {title} {{Gate operations for
  superconducting qubits and non-Markovianity}},\ }\href
  {https://doi.org/10.1103/PhysRevResearch.6.033215} {\bibfield  {journal}
  {\bibinfo  {journal} {Phys. Rev. Research}\ }\textbf {\bibinfo {volume}
  {6}},\ \bibinfo {pages} {033215} (\bibinfo {year}
  {2024}{\natexlab{a}})}\BibitemShut {NoStop}%
\bibitem [{\citenamefont {Gul\'acsi}\ and\ \citenamefont
  {Burkard}(2025)}]{GulacsiPRR2025}%
  \BibitemOpen
  \bibfield  {author} {\bibinfo {author} {\bibfnamefont {B.}~\bibnamefont
  {Gul\'acsi}}\ and\ \bibinfo {author} {\bibfnamefont {G.}~\bibnamefont
  {Burkard}},\ }\bibfield  {title} {\bibinfo {title} {{Temporally correlated
  quantum noise in driven quantum systems with applications to quantum gate
  operations}},\ }\href {https://doi.org/10.1103/PhysRevResearch.7.023073}
  {\bibfield  {journal} {\bibinfo  {journal} {Phys. Rev. Research}\ }\textbf
  {\bibinfo {volume} {7}},\ \bibinfo {pages} {023073} (\bibinfo {year}
  {2025})}\BibitemShut {NoStop}%
\bibitem [{\citenamefont {Gul\'acsi}\ and\ \citenamefont
  {Burkard}(2023)}]{GulasciPRB2022}%
  \BibitemOpen
  \bibfield  {author} {\bibinfo {author} {\bibfnamefont {B.}~\bibnamefont
  {Gul\'acsi}}\ and\ \bibinfo {author} {\bibfnamefont {G.}~\bibnamefont
  {Burkard}},\ }\bibfield  {title} {\bibinfo {title} {{Signatures of
  non-Markovianity of a superconducting qubit}},\ }\href
  {https://doi.org/10.1103/PhysRevB.107.174511} {\bibfield  {journal} {\bibinfo
   {journal} {Phys. Rev. B}\ }\textbf {\bibinfo {volume} {107}},\ \bibinfo
  {pages} {174511} (\bibinfo {year} {2023})}\BibitemShut {NoStop}%
\bibitem [{\citenamefont {Nakamura}\ and\ \citenamefont
  {Ankerhold}(2024{\natexlab{b}})}]{NakamuraPRB2024}%
  \BibitemOpen
  \bibfield  {author} {\bibinfo {author} {\bibfnamefont {K.}~\bibnamefont
  {Nakamura}}\ and\ \bibinfo {author} {\bibfnamefont {J.}~\bibnamefont
  {Ankerhold}},\ }\bibfield  {title} {\bibinfo {title} {{Qubit dynamics beyond
  Lindblad: Non-Markovianity versus rotating wave approximation}},\ }\href
  {https://doi.org/10.1103/PhysRevB.109.014315} {\bibfield  {journal} {\bibinfo
   {journal} {Phys. Rev. B}\ }\textbf {\bibinfo {volume} {109}},\ \bibinfo
  {pages} {014315} (\bibinfo {year} {2024}{\natexlab{b}})}\BibitemShut
  {NoStop}%
\bibitem [{\citenamefont {Nakamura}\ and\ \citenamefont
  {Ankerhold}(2025)}]{NakamuraPRB2025}%
  \BibitemOpen
  \bibfield  {author} {\bibinfo {author} {\bibfnamefont {K.}~\bibnamefont
  {Nakamura}}\ and\ \bibinfo {author} {\bibfnamefont {J.}~\bibnamefont
  {Ankerhold}},\ }\bibfield  {title} {\bibinfo {title} {{Impact of
  time-retarded noise on dynamical decoupling schemes for qubits}},\ }\href
  {https://doi.org/10.1103/PhysRevB.111.064503} {\bibfield  {journal} {\bibinfo
   {journal} {Phys. Rev. B}\ }\textbf {\bibinfo {volume} {111}},\ \bibinfo
  {pages} {064503} (\bibinfo {year} {2025})}\BibitemShut {NoStop}%
\bibitem [{\citenamefont {Babu}\ \emph {et~al.}(2023)\citenamefont {Babu},
  \citenamefont {Orell}, \citenamefont {Vadimov}, \citenamefont {Teixeira},
  \citenamefont {M\"ott\"onen},\ and\ \citenamefont {Silveri}}]{BabuPRR2023}%
  \BibitemOpen
  \bibfield  {author} {\bibinfo {author} {\bibfnamefont {A.~P.}\ \bibnamefont
  {Babu}}, \bibinfo {author} {\bibfnamefont {T.}~\bibnamefont {Orell}},
  \bibinfo {author} {\bibfnamefont {V.}~\bibnamefont {Vadimov}}, \bibinfo
  {author} {\bibfnamefont {W.}~\bibnamefont {Teixeira}}, \bibinfo {author}
  {\bibfnamefont {M.}~\bibnamefont {M\"ott\"onen}},\ and\ \bibinfo {author}
  {\bibfnamefont {M.}~\bibnamefont {Silveri}},\ }\bibfield  {title} {\bibinfo
  {title} {{Quantum error correction under numerically exact
  open-quantum-system dynamics}},\ }\href
  {https://doi.org/10.1103/PhysRevResearch.5.043161} {\bibfield  {journal}
  {\bibinfo  {journal} {Phys. Rev. Research}\ }\textbf {\bibinfo {volume}
  {5}},\ \bibinfo {pages} {043161} (\bibinfo {year} {2023})}\BibitemShut
  {NoStop}%
\bibitem [{\citenamefont {Gul\'acsi}\ \emph {et~al.}(2025)\citenamefont
  {Gul\'acsi}, \citenamefont {Kattem\"olle},\ and\ \citenamefont
  {Burkard}}]{GulacsiPRR2025-2}%
  \BibitemOpen
  \bibfield  {author} {\bibinfo {author} {\bibfnamefont {B.}~\bibnamefont
  {Gul\'acsi}}, \bibinfo {author} {\bibfnamefont {J.}~\bibnamefont
  {Kattem\"olle}},\ and\ \bibinfo {author} {\bibfnamefont {G.}~\bibnamefont
  {Burkard}},\ }\bibfield  {title} {\bibinfo {title} {Characterizing correlated
  noise with single-qubit operations},\ }\href
  {https://doi.org/10.1103/k2lx-vfqp} {\bibfield  {journal} {\bibinfo
  {journal} {Phys. Rev. Res.}\ }\textbf {\bibinfo {volume} {7}},\ \bibinfo
  {pages} {L042067} (\bibinfo {year} {2025})}\BibitemShut {NoStop}%
\bibitem [{\citenamefont {Zhang}\ \emph {et~al.}(2024)\citenamefont {Zhang},
  \citenamefont {Chen},\ and\ \citenamefont {Shi}}]{ZhangJCP2024}%
  \BibitemOpen
  \bibfield  {author} {\bibinfo {author} {\bibfnamefont {S.}~\bibnamefont
  {Zhang}}, \bibinfo {author} {\bibfnamefont {Y.}~\bibnamefont {Chen}},\ and\
  \bibinfo {author} {\bibfnamefont {Q.}~\bibnamefont {Shi}},\ }\bibfield
  {title} {\bibinfo {title} {{Simulating the operation of a quantum computer in
  a dissipative environment}},\ }\href {https://doi.org/10.1063/5.0185263}
  {\bibfield  {journal} {\bibinfo  {journal} {J. Chem. Phys.}\ }\textbf
  {\bibinfo {volume} {160}},\ \bibinfo {pages} {054101} (\bibinfo {year}
  {2024})}\BibitemShut {NoStop}%
\bibitem [{\citenamefont {Yu}\ and\ \citenamefont {Eberly}(2004)}]{YuPRL2004}%
  \BibitemOpen
  \bibfield  {author} {\bibinfo {author} {\bibfnamefont {T.}~\bibnamefont
  {Yu}}\ and\ \bibinfo {author} {\bibfnamefont {J.~H.}\ \bibnamefont
  {Eberly}},\ }\bibfield  {title} {\bibinfo {title} {{Finite-Time
  Disentanglement Via Spontaneous Emission}},\ }\href
  {https://doi.org/10.1103/PhysRevLett.93.140404} {\bibfield  {journal}
  {\bibinfo  {journal} {Phys. Rev. Lett.}\ }\textbf {\bibinfo {volume} {93}},\
  \bibinfo {pages} {140404} (\bibinfo {year} {2004})}\BibitemShut {NoStop}%
\bibitem [{\citenamefont {Yu}\ and\ \citenamefont
  {Eberly}(2006{\natexlab{a}})}]{YuOPTCOMMUN2006}%
  \BibitemOpen
  \bibfield  {author} {\bibinfo {author} {\bibfnamefont {T.}~\bibnamefont
  {Yu}}\ and\ \bibinfo {author} {\bibfnamefont {J.~H.}\ \bibnamefont
  {Eberly}},\ }\bibfield  {title} {\bibinfo {title} {{Sudden death of
  entanglement: Classical noise effects}},\ }\href
  {https://doi.org/https://doi.org/10.1016/j.optcom.2006.01.061} {\bibfield
  {journal} {\bibinfo  {journal} {Opt. Commun.}\ }\textbf {\bibinfo {volume}
  {264}},\ \bibinfo {pages} {393} (\bibinfo {year}
  {2006}{\natexlab{a}})}\BibitemShut {NoStop}%
\bibitem [{\citenamefont {Yu}\ and\ \citenamefont
  {Eberly}(2006{\natexlab{b}})}]{YuPRL2006}%
  \BibitemOpen
  \bibfield  {author} {\bibinfo {author} {\bibfnamefont {T.}~\bibnamefont
  {Yu}}\ and\ \bibinfo {author} {\bibfnamefont {J.~H.}\ \bibnamefont
  {Eberly}},\ }\bibfield  {title} {\bibinfo {title} {{Quantum Open System
  Theory: Bipartite Aspects}},\ }\href
  {https://doi.org/10.1103/PhysRevLett.97.140403} {\bibfield  {journal}
  {\bibinfo  {journal} {Phys. Rev. Lett.}\ }\textbf {\bibinfo {volume} {97}},\
  \bibinfo {pages} {140403} (\bibinfo {year} {2006}{\natexlab{b}})}\BibitemShut
  {NoStop}%
\bibitem [{\citenamefont {L\'opez}\ \emph {et~al.}(2008)\citenamefont
  {L\'opez}, \citenamefont {Romero}, \citenamefont {Lastra}, \citenamefont
  {Solano},\ and\ \citenamefont {Retamal}}]{LopezPRL2008}%
  \BibitemOpen
  \bibfield  {author} {\bibinfo {author} {\bibfnamefont {C.~E.}\ \bibnamefont
  {L\'opez}}, \bibinfo {author} {\bibfnamefont {G.}~\bibnamefont {Romero}},
  \bibinfo {author} {\bibfnamefont {F.}~\bibnamefont {Lastra}}, \bibinfo
  {author} {\bibfnamefont {E.}~\bibnamefont {Solano}},\ and\ \bibinfo {author}
  {\bibfnamefont {J.~C.}\ \bibnamefont {Retamal}},\ }\bibfield  {title}
  {\bibinfo {title} {{Sudden Birth versus Sudden Death of Entanglement in
  Multipartite Systems}},\ }\href
  {https://doi.org/10.1103/PhysRevLett.101.080503} {\bibfield  {journal}
  {\bibinfo  {journal} {Phys. Rev. Lett.}\ }\textbf {\bibinfo {volume} {101}},\
  \bibinfo {pages} {080503} (\bibinfo {year} {2008})}\BibitemShut {NoStop}%
\bibitem [{\citenamefont {Bellomo}\ \emph {et~al.}(2007)\citenamefont
  {Bellomo}, \citenamefont {Lo~Franco},\ and\ \citenamefont
  {Compagno}}]{BellomoPRL2007}%
  \BibitemOpen
  \bibfield  {author} {\bibinfo {author} {\bibfnamefont {B.}~\bibnamefont
  {Bellomo}}, \bibinfo {author} {\bibfnamefont {R.}~\bibnamefont {Lo~Franco}},\
  and\ \bibinfo {author} {\bibfnamefont {G.}~\bibnamefont {Compagno}},\
  }\bibfield  {title} {\bibinfo {title} {{Non-Markovian Effects on the Dynamics
  of Entanglement}},\ }\href {https://doi.org/10.1103/PhysRevLett.99.160502}
  {\bibfield  {journal} {\bibinfo  {journal} {Phys. Rev. Lett.}\ }\textbf
  {\bibinfo {volume} {99}},\ \bibinfo {pages} {160502} (\bibinfo {year}
  {2007})}\BibitemShut {NoStop}%
\bibitem [{\citenamefont {Bellomo}\ \emph {et~al.}(2008)\citenamefont
  {Bellomo}, \citenamefont {Lo~Franco},\ and\ \citenamefont
  {Compagno}}]{BellomoPRA2008}%
  \BibitemOpen
  \bibfield  {author} {\bibinfo {author} {\bibfnamefont {B.}~\bibnamefont
  {Bellomo}}, \bibinfo {author} {\bibfnamefont {R.}~\bibnamefont {Lo~Franco}},\
  and\ \bibinfo {author} {\bibfnamefont {G.}~\bibnamefont {Compagno}},\
  }\bibfield  {title} {\bibinfo {title} {Entanglement dynamics of two
  independent qubits in environments with and without memory},\ }\href
  {https://doi.org/10.1103/PhysRevA.77.032342} {\bibfield  {journal} {\bibinfo
  {journal} {Phys. Rev. A}\ }\textbf {\bibinfo {volume} {77}},\ \bibinfo
  {pages} {032342} (\bibinfo {year} {2008})}\BibitemShut {NoStop}%
\bibitem [{\citenamefont {Bellomo}\ \emph {et~al.}(2010)\citenamefont
  {Bellomo}, \citenamefont {Franco}, \citenamefont {Maniscalco},\ and\
  \citenamefont {Compagno}}]{BellomoPHYSSCR2010}%
  \BibitemOpen
  \bibfield  {author} {\bibinfo {author} {\bibfnamefont {B.}~\bibnamefont
  {Bellomo}}, \bibinfo {author} {\bibfnamefont {R.~L.}\ \bibnamefont {Franco}},
  \bibinfo {author} {\bibfnamefont {S.}~\bibnamefont {Maniscalco}},\ and\
  \bibinfo {author} {\bibfnamefont {G.}~\bibnamefont {Compagno}},\ }\bibfield
  {title} {\bibinfo {title} {{Two-qubit entanglement dynamics for two different
  non-Markovian environments}},\ }\href
  {https://doi.org/10.1088/0031-8949/2010/T140/014014} {\bibfield  {journal}
  {\bibinfo  {journal} {Phys. Scr.}\ }\textbf {\bibinfo {volume} {2010}},\
  \bibinfo {pages} {014014} (\bibinfo {year} {2010})}\BibitemShut {NoStop}%
\bibitem [{\citenamefont {Cao}\ and\ \citenamefont {Zheng}(2008)}]{CaoPRA2008}%
  \BibitemOpen
  \bibfield  {author} {\bibinfo {author} {\bibfnamefont {X.}~\bibnamefont
  {Cao}}\ and\ \bibinfo {author} {\bibfnamefont {H.}~\bibnamefont {Zheng}},\
  }\bibfield  {title} {\bibinfo {title} {{Non-Markovian disentanglement
  dynamics of a two-qubit system}},\ }\href
  {https://doi.org/10.1103/PhysRevA.77.022320} {\bibfield  {journal} {\bibinfo
  {journal} {Phys. Rev. A}\ }\textbf {\bibinfo {volume} {77}},\ \bibinfo
  {pages} {022320} (\bibinfo {year} {2008})}\BibitemShut {NoStop}%
\bibitem [{\citenamefont {Ma}\ \emph {et~al.}(2012)\citenamefont {Ma},
  \citenamefont {Sun}, \citenamefont {Wang},\ and\ \citenamefont
  {Nori}}]{MaPRA2012}%
  \BibitemOpen
  \bibfield  {author} {\bibinfo {author} {\bibfnamefont {J.}~\bibnamefont
  {Ma}}, \bibinfo {author} {\bibfnamefont {Z.}~\bibnamefont {Sun}}, \bibinfo
  {author} {\bibfnamefont {X.}~\bibnamefont {Wang}},\ and\ \bibinfo {author}
  {\bibfnamefont {F.}~\bibnamefont {Nori}},\ }\bibfield  {title} {\bibinfo
  {title} {{Entanglement dynamics of two qubits in a common bath}},\ }\href
  {https://doi.org/10.1103/PhysRevA.85.062323} {\bibfield  {journal} {\bibinfo
  {journal} {Phys. Rev. A}\ }\textbf {\bibinfo {volume} {85}},\ \bibinfo
  {pages} {062323} (\bibinfo {year} {2012})}\BibitemShut {NoStop}%
\bibitem [{\citenamefont {Wu}\ \emph {et~al.}(2013)\citenamefont {Wu},
  \citenamefont {Yu},\ and\ \citenamefont {Segal}}]{WuNJP2013}%
  \BibitemOpen
  \bibfield  {author} {\bibinfo {author} {\bibfnamefont {L.-A.}\ \bibnamefont
  {Wu}}, \bibinfo {author} {\bibfnamefont {C.~X.}\ \bibnamefont {Yu}},\ and\
  \bibinfo {author} {\bibfnamefont {D.}~\bibnamefont {Segal}},\ }\bibfield
  {title} {\bibinfo {title} {{Exact dynamics of interacting qubits in a thermal
  environment: results beyond the weak coupling limit}},\ }\href
  {https://doi.org/10.1088/1367-2630/15/2/023044} {\bibfield  {journal}
  {\bibinfo  {journal} {New J. Phys.}\ }\textbf {\bibinfo {volume} {15}},\
  \bibinfo {pages} {023044} (\bibinfo {year} {2013})}\BibitemShut {NoStop}%
\bibitem [{\citenamefont {Duan}\ \emph {et~al.}(2013)\citenamefont {Duan},
  \citenamefont {Wang}, \citenamefont {Chen},\ and\ \citenamefont
  {Zhao}}]{DuanJCP2013}%
  \BibitemOpen
  \bibfield  {author} {\bibinfo {author} {\bibfnamefont {L.}~\bibnamefont
  {Duan}}, \bibinfo {author} {\bibfnamefont {H.}~\bibnamefont {Wang}}, \bibinfo
  {author} {\bibfnamefont {Q.-H.}\ \bibnamefont {Chen}},\ and\ \bibinfo
  {author} {\bibfnamefont {Y.}~\bibnamefont {Zhao}},\ }\bibfield  {title}
  {\bibinfo {title} {{Entanglement dynamics of two qubits coupled individually
  to Ohmic baths}},\ }\href {https://doi.org/10.1063/1.4816122} {\bibfield
  {journal} {\bibinfo  {journal} {J. Chem. Phys.}\ }\textbf {\bibinfo {volume}
  {139}},\ \bibinfo {pages} {044115} (\bibinfo {year} {2013})}\BibitemShut
  {NoStop}%
\bibitem [{\citenamefont {Wang}\ and\ \citenamefont
  {Chen}(2013)}]{WangNJP2013}%
  \BibitemOpen
  \bibfield  {author} {\bibinfo {author} {\bibfnamefont {C.}~\bibnamefont
  {Wang}}\ and\ \bibinfo {author} {\bibfnamefont {Q.-H.}\ \bibnamefont
  {Chen}},\ }\bibfield  {title} {\bibinfo {title} {{Exact dynamics of quantum
  correlations of two qubits coupled to bosonic baths}},\ }\href
  {https://doi.org/10.1088/1367-2630/15/10/103020} {\bibfield  {journal}
  {\bibinfo  {journal} {New J. Phys.}\ }\textbf {\bibinfo {volume} {15}},\
  \bibinfo {pages} {103020} (\bibinfo {year} {2013})}\BibitemShut {NoStop}%
\bibitem [{\citenamefont {Wu}\ and\ \citenamefont {Liu}(2017)}]{WuPRA2017}%
  \BibitemOpen
  \bibfield  {author} {\bibinfo {author} {\bibfnamefont {W.}~\bibnamefont
  {Wu}}\ and\ \bibinfo {author} {\bibfnamefont {M.}~\bibnamefont {Liu}},\
  }\bibfield  {title} {\bibinfo {title} {{Effects of counter-rotating-wave
  terms on the non-Markovianity in quantum open systems}},\ }\href
  {https://doi.org/10.1103/PhysRevA.96.032125} {\bibfield  {journal} {\bibinfo
  {journal} {Phys. Rev. A}\ }\textbf {\bibinfo {volume} {96}},\ \bibinfo
  {pages} {032125} (\bibinfo {year} {2017})}\BibitemShut {NoStop}%
\bibitem [{\citenamefont {Sun}\ \emph {et~al.}(2018)\citenamefont {Sun},
  \citenamefont {Xu},\ and\ \citenamefont {Liu}}]{SunPRA2018}%
  \BibitemOpen
  \bibfield  {author} {\bibinfo {author} {\bibfnamefont {Z.}~\bibnamefont
  {Sun}}, \bibinfo {author} {\bibfnamefont {X.-Q.}\ \bibnamefont {Xu}},\ and\
  \bibinfo {author} {\bibfnamefont {B.}~\bibnamefont {Liu}},\ }\bibfield
  {title} {\bibinfo {title} {{Creation of quantum steering by interaction with
  a common bath}},\ }\href {https://doi.org/10.1103/PhysRevA.97.052309}
  {\bibfield  {journal} {\bibinfo  {journal} {Phys. Rev. A}\ }\textbf {\bibinfo
  {volume} {97}},\ \bibinfo {pages} {052309} (\bibinfo {year}
  {2018})}\BibitemShut {NoStop}%
\bibitem [{\citenamefont {Dijkstra}\ and\ \citenamefont
  {Tanimura}(2010)}]{DijkstraPRL2010}%
  \BibitemOpen
  \bibfield  {author} {\bibinfo {author} {\bibfnamefont {A.~G.}\ \bibnamefont
  {Dijkstra}}\ and\ \bibinfo {author} {\bibfnamefont {Y.}~\bibnamefont
  {Tanimura}},\ }\bibfield  {title} {\bibinfo {title} {{Non-Markovian
  Entanglement Dynamics in the Presence of System-Bath Coherence}},\ }\href
  {https://doi.org/10.1103/PhysRevLett.104.250401} {\bibfield  {journal}
  {\bibinfo  {journal} {Phys. Rev. Lett.}\ }\textbf {\bibinfo {volume} {104}},\
  \bibinfo {pages} {250401} (\bibinfo {year} {2010})}\BibitemShut {NoStop}%
\bibitem [{\citenamefont {Wang}\ \emph {et~al.}(2011)\citenamefont {Wang},
  \citenamefont {Li}, \citenamefont {Zou}, \citenamefont {Ge},\ and\
  \citenamefont {Guo}}]{WangPHYSICAA2011}%
  \BibitemOpen
  \bibfield  {author} {\bibinfo {author} {\bibfnamefont {H.-T.}\ \bibnamefont
  {Wang}}, \bibinfo {author} {\bibfnamefont {C.-F.}\ \bibnamefont {Li}},
  \bibinfo {author} {\bibfnamefont {Y.}~\bibnamefont {Zou}}, \bibinfo {author}
  {\bibfnamefont {R.-C.}\ \bibnamefont {Ge}},\ and\ \bibinfo {author}
  {\bibfnamefont {G.-C.}\ \bibnamefont {Guo}},\ }\bibfield  {title} {\bibinfo
  {title} {{Non-Markovian entanglement sudden death and rebirth of a two-qubit
  system in the presence of system-bath coherence}},\ }\href
  {https://doi.org/https://doi.org/10.1016/j.physa.2011.04.029} {\bibfield
  {journal} {\bibinfo  {journal} {Physica A}\ }\textbf {\bibinfo {volume}
  {390}},\ \bibinfo {pages} {3183} (\bibinfo {year} {2011})}\BibitemShut
  {NoStop}%
\bibitem [{\citenamefont {Paladino}\ \emph {et~al.}(2010)\citenamefont
  {Paladino}, \citenamefont {D’Arrigo}, \citenamefont {Mastellone},\ and\
  \citenamefont {Falci}}]{PaladinoPHYSICAE2010}%
  \BibitemOpen
  \bibfield  {author} {\bibinfo {author} {\bibfnamefont {E.}~\bibnamefont
  {Paladino}}, \bibinfo {author} {\bibfnamefont {A.}~\bibnamefont
  {D’Arrigo}}, \bibinfo {author} {\bibfnamefont {A.}~\bibnamefont
  {Mastellone}},\ and\ \bibinfo {author} {\bibfnamefont {G.}~\bibnamefont
  {Falci}},\ }\bibfield  {title} {\bibinfo {title} {{Relaxation processes in
  solid-state two-qubit gates}},\ }\href
  {https://doi.org/https://doi.org/10.1016/j.physe.2009.06.042} {\bibfield
  {journal} {\bibinfo  {journal} {Physica E}\ }\textbf {\bibinfo {volume}
  {42}},\ \bibinfo {pages} {439} (\bibinfo {year} {2010})}\BibitemShut
  {NoStop}%
\bibitem [{\citenamefont {Paladino}\ \emph {et~al.}(2011)\citenamefont
  {Paladino}, \citenamefont {D'Arrigo}, \citenamefont {Mastellone},\ and\
  \citenamefont {Falci}}]{PaladinoNJP2011}%
  \BibitemOpen
  \bibfield  {author} {\bibinfo {author} {\bibfnamefont {E.}~\bibnamefont
  {Paladino}}, \bibinfo {author} {\bibfnamefont {A.}~\bibnamefont {D'Arrigo}},
  \bibinfo {author} {\bibfnamefont {A.}~\bibnamefont {Mastellone}},\ and\
  \bibinfo {author} {\bibfnamefont {G.}~\bibnamefont {Falci}},\ }\bibfield
  {title} {\bibinfo {title} {{Decoherence times of universal two-qubit gates in
  the presence of broad-band noise}},\ }\href
  {https://doi.org/10.1088/1367-2630/13/9/093037} {\bibfield  {journal}
  {\bibinfo  {journal} {New J. Phys.}\ }\textbf {\bibinfo {volume} {13}},\
  \bibinfo {pages} {093037} (\bibinfo {year} {2011})}\BibitemShut {NoStop}%
\bibitem [{\citenamefont {D'Arrigo}\ and\ \citenamefont
  {Paladino}(2012)}]{D'ArrigoNJP2012}%
  \BibitemOpen
  \bibfield  {author} {\bibinfo {author} {\bibfnamefont {A.}~\bibnamefont
  {D'Arrigo}}\ and\ \bibinfo {author} {\bibfnamefont {E.}~\bibnamefont
  {Paladino}},\ }\bibfield  {title} {\bibinfo {title} {{Optimal operating
  conditions of an entangling two-transmon gate}},\ }\href
  {https://doi.org/10.1088/1367-2630/14/5/053035} {\bibfield  {journal}
  {\bibinfo  {journal} {New J. Phys.}\ }\textbf {\bibinfo {volume} {14}},\
  \bibinfo {pages} {053035} (\bibinfo {year} {2012})}\BibitemShut {NoStop}%
\bibitem [{\citenamefont {Loss}\ and\ \citenamefont
  {DiVincenzo}(1998)}]{LossPRA1998}%
  \BibitemOpen
  \bibfield  {author} {\bibinfo {author} {\bibfnamefont {D.}~\bibnamefont
  {Loss}}\ and\ \bibinfo {author} {\bibfnamefont {D.~P.}\ \bibnamefont
  {DiVincenzo}},\ }\bibfield  {title} {\bibinfo {title} {{Quantum computation
  with quantum dots}},\ }\href {https://doi.org/10.1103/PhysRevA.57.120}
  {\bibfield  {journal} {\bibinfo  {journal} {Phys. Rev. A}\ }\textbf {\bibinfo
  {volume} {57}},\ \bibinfo {pages} {120} (\bibinfo {year} {1998})}\BibitemShut
  {NoStop}%
\bibitem [{\citenamefont {Thorwart}\ and\ \citenamefont
  {H\"anggi}(2001)}]{ThorwartPRA2001}%
  \BibitemOpen
  \bibfield  {author} {\bibinfo {author} {\bibfnamefont {M.}~\bibnamefont
  {Thorwart}}\ and\ \bibinfo {author} {\bibfnamefont {P.}~\bibnamefont
  {H\"anggi}},\ }\bibfield  {title} {\bibinfo {title} {{Decoherence and
  dissipation during a quantum XOR gate operation}},\ }\href
  {https://doi.org/10.1103/PhysRevA.65.012309} {\bibfield  {journal} {\bibinfo
  {journal} {Phys. Rev. A}\ }\textbf {\bibinfo {volume} {65}},\ \bibinfo
  {pages} {012309} (\bibinfo {year} {2001})}\BibitemShut {NoStop}%
\bibitem [{\citenamefont {Cheng}\ and\ \citenamefont
  {Silbey}(2004)}]{ChengPRA2004}%
  \BibitemOpen
  \bibfield  {author} {\bibinfo {author} {\bibfnamefont {Y.~C.}\ \bibnamefont
  {Cheng}}\ and\ \bibinfo {author} {\bibfnamefont {R.~J.}\ \bibnamefont
  {Silbey}},\ }\bibfield  {title} {\bibinfo {title} {{Stochastic Liouville
  equation approach for the effect of noise in quantum computations}},\ }\href
  {https://doi.org/10.1103/PhysRevA.69.052325} {\bibfield  {journal} {\bibinfo
  {journal} {Phys. Rev. A}\ }\textbf {\bibinfo {volume} {69}},\ \bibinfo
  {pages} {052325} (\bibinfo {year} {2004})}\BibitemShut {NoStop}%
\bibitem [{\citenamefont {Storcz}\ \emph {et~al.}(2005)\citenamefont {Storcz},
  \citenamefont {Hellmann}, \citenamefont {Hrelescu},\ and\ \citenamefont
  {Wilhelm}}]{StorczPRA2005}%
  \BibitemOpen
  \bibfield  {author} {\bibinfo {author} {\bibfnamefont {M.~J.}\ \bibnamefont
  {Storcz}}, \bibinfo {author} {\bibfnamefont {F.}~\bibnamefont {Hellmann}},
  \bibinfo {author} {\bibfnamefont {C.}~\bibnamefont {Hrelescu}},\ and\
  \bibinfo {author} {\bibfnamefont {F.~K.}\ \bibnamefont {Wilhelm}},\
  }\bibfield  {title} {\bibinfo {title} {{Decoherence of a two-qubit system
  away from perfect symmetry}},\ }\href
  {https://doi.org/10.1103/PhysRevA.72.052314} {\bibfield  {journal} {\bibinfo
  {journal} {Phys. Rev. A}\ }\textbf {\bibinfo {volume} {72}},\ \bibinfo
  {pages} {052314} (\bibinfo {year} {2005})}\BibitemShut {NoStop}%
\bibitem [{\citenamefont {Xu}\ \emph {et~al.}(2022)\citenamefont {Xu},
  \citenamefont {Yan}, \citenamefont {Shi}, \citenamefont {Ankerhold},\ and\
  \citenamefont {Stockburger}}]{XuPRL2022}%
  \BibitemOpen
  \bibfield  {author} {\bibinfo {author} {\bibfnamefont {M.}~\bibnamefont
  {Xu}}, \bibinfo {author} {\bibfnamefont {Y.}~\bibnamefont {Yan}}, \bibinfo
  {author} {\bibfnamefont {Q.}~\bibnamefont {Shi}}, \bibinfo {author}
  {\bibfnamefont {J.}~\bibnamefont {Ankerhold}},\ and\ \bibinfo {author}
  {\bibfnamefont {J.~T.}\ \bibnamefont {Stockburger}},\ }\bibfield  {title}
  {\bibinfo {title} {{Taming Quantum Noise for Efficient Low Temperature
  Simulations of Open Quantum Systems}},\ }\href
  {https://doi.org/10.1103/PhysRevLett.129.230601} {\bibfield  {journal}
  {\bibinfo  {journal} {Phys. Rev. Lett.}\ }\textbf {\bibinfo {volume} {129}},\
  \bibinfo {pages} {230601} (\bibinfo {year} {2022})}\BibitemShut {NoStop}%
\bibitem [{\citenamefont {Xu}\ \emph {et~al.}(2026)\citenamefont {Xu},
  \citenamefont {Vadimov}, \citenamefont {Stockburger},\ and\ \citenamefont
  {Ankerhold}}]{Xu2026}%
  \BibitemOpen
  \bibfield  {author} {\bibinfo {author} {\bibfnamefont {M.}~\bibnamefont
  {Xu}}, \bibinfo {author} {\bibfnamefont {V.}~\bibnamefont {Vadimov}},
  \bibinfo {author} {\bibfnamefont {J.~T.}\ \bibnamefont {Stockburger}},\ and\
  \bibinfo {author} {\bibfnamefont {J.}~\bibnamefont {Ankerhold}},\ }\bibfield
  {title} {\bibinfo {title} {{Simulating Non-Markovian Dynamnics in Open
  Quantum Systems}},\ }\href {https://doi.org/10.1103/w3nw-hbjc} {\bibfield
  {journal} {\bibinfo  {journal} {Rev. Mod. Phys. (in press)}\ } (\bibinfo
  {year} {2026})}\BibitemShut {NoStop}%
\bibitem [{\citenamefont {Maniscalco}\ \emph {et~al.}(2008)\citenamefont
  {Maniscalco}, \citenamefont {Francica}, \citenamefont {Zaffino},
  \citenamefont {Lo~Gullo},\ and\ \citenamefont
  {Plastina}}]{ManiscalcoPRL2008}%
  \BibitemOpen
  \bibfield  {author} {\bibinfo {author} {\bibfnamefont {S.}~\bibnamefont
  {Maniscalco}}, \bibinfo {author} {\bibfnamefont {F.}~\bibnamefont
  {Francica}}, \bibinfo {author} {\bibfnamefont {R.~L.}\ \bibnamefont
  {Zaffino}}, \bibinfo {author} {\bibfnamefont {N.}~\bibnamefont {Lo~Gullo}},\
  and\ \bibinfo {author} {\bibfnamefont {F.}~\bibnamefont {Plastina}},\
  }\bibfield  {title} {\bibinfo {title} {{Protecting Entanglement via the
  Quantum Zeno Effect}},\ }\href
  {https://doi.org/10.1103/PhysRevLett.100.090503} {\bibfield  {journal}
  {\bibinfo  {journal} {Phys. Rev. Lett.}\ }\textbf {\bibinfo {volume} {100}},\
  \bibinfo {pages} {090503} (\bibinfo {year} {2008})}\BibitemShut {NoStop}%
\bibitem [{\citenamefont {D'Arrigo}\ \emph {et~al.}(2008)\citenamefont
  {D'Arrigo}, \citenamefont {Mastellone}, \citenamefont {Paladino},\ and\
  \citenamefont {Falci}}]{D'ArrigoNJP2008}%
  \BibitemOpen
  \bibfield  {author} {\bibinfo {author} {\bibfnamefont {A.}~\bibnamefont
  {D'Arrigo}}, \bibinfo {author} {\bibfnamefont {A.}~\bibnamefont
  {Mastellone}}, \bibinfo {author} {\bibfnamefont {E.}~\bibnamefont
  {Paladino}},\ and\ \bibinfo {author} {\bibfnamefont {G.}~\bibnamefont
  {Falci}},\ }\bibfield  {title} {\bibinfo {title} {{Effects of low-frequency
  noise cross-correlations in coupled superconducting qubits}},\ }\href
  {https://doi.org/10.1088/1367-2630/10/11/115006} {\bibfield  {journal}
  {\bibinfo  {journal} {New J. Phys.}\ }\textbf {\bibinfo {volume} {10}},\
  \bibinfo {pages} {115006} (\bibinfo {year} {2008})}\BibitemShut {NoStop}%
\bibitem [{\citenamefont {Mazzola}\ \emph {et~al.}(2009)\citenamefont
  {Mazzola}, \citenamefont {Maniscalco}, \citenamefont {Piilo}, \citenamefont
  {Suominen},\ and\ \citenamefont {Garraway}}]{MazzolaPRA2009}%
  \BibitemOpen
  \bibfield  {author} {\bibinfo {author} {\bibfnamefont {L.}~\bibnamefont
  {Mazzola}}, \bibinfo {author} {\bibfnamefont {S.}~\bibnamefont {Maniscalco}},
  \bibinfo {author} {\bibfnamefont {J.}~\bibnamefont {Piilo}}, \bibinfo
  {author} {\bibfnamefont {K.-A.}\ \bibnamefont {Suominen}},\ and\ \bibinfo
  {author} {\bibfnamefont {B.~M.}\ \bibnamefont {Garraway}},\ }\bibfield
  {title} {\bibinfo {title} {{Sudden death and sudden birth of entanglement in
  common structured reservoirs}},\ }\href
  {https://doi.org/10.1103/PhysRevA.79.042302} {\bibfield  {journal} {\bibinfo
  {journal} {Phys. Rev. A}\ }\textbf {\bibinfo {volume} {79}},\ \bibinfo
  {pages} {042302} (\bibinfo {year} {2009})}\BibitemShut {NoStop}%
\bibitem [{\citenamefont {Kast}\ and\ \citenamefont
  {Ankerhold}(2014)}]{KastPRB2014}%
  \BibitemOpen
  \bibfield  {author} {\bibinfo {author} {\bibfnamefont {D.}~\bibnamefont
  {Kast}}\ and\ \bibinfo {author} {\bibfnamefont {J.}~\bibnamefont
  {Ankerhold}},\ }\bibfield  {title} {\bibinfo {title} {{Bipartite entanglement
  dynamics of two-level systems in sub-Ohmic reservoirs}},\ }\href
  {https://doi.org/10.1103/PhysRevB.90.100301} {\bibfield  {journal} {\bibinfo
  {journal} {Phys. Rev. B}\ }\textbf {\bibinfo {volume} {90}},\ \bibinfo
  {pages} {100301} (\bibinfo {year} {2014})}\BibitemShut {NoStop}%
\bibitem [{\citenamefont {Hartmann}\ and\ \citenamefont
  {Strunz}(2020)}]{HartmannQUANTUM2020}%
  \BibitemOpen
  \bibfield  {author} {\bibinfo {author} {\bibfnamefont {R.}~\bibnamefont
  {Hartmann}}\ and\ \bibinfo {author} {\bibfnamefont {W.~T.}\ \bibnamefont
  {Strunz}},\ }\bibfield  {title} {\bibinfo {title} {Environmentally {I}nduced
  {E}ntanglement --- {A}nomalous {B}ehavior in the {A}diabatic {R}egime},\
  }\href {https://doi.org/10.22331/q-2020-10-22-347} {\bibfield  {journal}
  {\bibinfo  {journal} {{Quantum}}\ }\textbf {\bibinfo {volume} {4}},\ \bibinfo
  {pages} {347} (\bibinfo {year} {2020})}\BibitemShut {NoStop}%
\bibitem [{\citenamefont {Caldeira}\ and\ \citenamefont
  {Leggett}(1983)}]{CLModel}%
  \BibitemOpen
  \bibfield  {author} {\bibinfo {author} {\bibfnamefont {A.}~\bibnamefont
  {Caldeira}}\ and\ \bibinfo {author} {\bibfnamefont {A.}~\bibnamefont
  {Leggett}},\ }\bibfield  {title} {\bibinfo {title} {{Quantum tunnelling in a
  dissipative system}},\ }\href
  {https://doi.org/https://doi.org/10.1016/0003-4916(83)90202-6} {\bibfield
  {journal} {\bibinfo  {journal} {Ann. Phys.}\ }\textbf {\bibinfo {volume}
  {149}},\ \bibinfo {pages} {374} (\bibinfo {year} {1983})}\BibitemShut
  {NoStop}%
\bibitem [{\citenamefont {Breuer}\ and\ \citenamefont
  {Petruccione}(2002)}]{Breuer2002}%
  \BibitemOpen
  \bibfield  {author} {\bibinfo {author} {\bibfnamefont {H.-P.}\ \bibnamefont
  {Breuer}}\ and\ \bibinfo {author} {\bibfnamefont {F.}~\bibnamefont
  {Petruccione}},\ }\href@noop {} {\emph {\bibinfo {title} {{The Theory of Open
  Quantum Systems}}}}\ (\bibinfo  {publisher} {Oxford University Press},\
  \bibinfo {address} {Oxford},\ \bibinfo {year} {2002})\BibitemShut {NoStop}%
\bibitem [{\citenamefont {Weiss}(2012)}]{Weiss2012}%
  \BibitemOpen
  \bibfield  {author} {\bibinfo {author} {\bibfnamefont {U.}~\bibnamefont
  {Weiss}},\ }\href@noop {} {\emph {\bibinfo {title} {{Quantum Dissipative
  Systems}}}},\ \bibinfo {edition} {4th}\ ed.\ (\bibinfo  {publisher} {World
  Scientific},\ \bibinfo {address} {Singapore},\ \bibinfo {year}
  {2012})\BibitemShut {NoStop}%
\bibitem [{\citenamefont {Barone}\ and\ \citenamefont
  {Patern\`{o}}(1982)}]{Barone1982}%
  \BibitemOpen
  \bibfield  {author} {\bibinfo {author} {\bibfnamefont {A.}~\bibnamefont
  {Barone}}\ and\ \bibinfo {author} {\bibfnamefont {G.}~\bibnamefont
  {Patern\`{o}}},\ }\href@noop {} {\emph {\bibinfo {title} {{Physics and
  Applications of the Josephson Effect}}}}\ (\bibinfo  {publisher} {John Wiley
  \& Sons},\ \bibinfo {address} {New York},\ \bibinfo {year}
  {1982})\BibitemShut {NoStop}%
\bibitem [{\citenamefont {Wendin}\ and\ \citenamefont
  {Shumeiko}()}]{WendinARXIV2005}%
  \BibitemOpen
  \bibfield  {author} {\bibinfo {author} {\bibfnamefont {G.}~\bibnamefont
  {Wendin}}\ and\ \bibinfo {author} {\bibfnamefont {V.~S.}\ \bibnamefont
  {Shumeiko}},\ }\href@noop {} {\bibinfo {title} {{Superconducting Quantum
  Circuits, Qubits and Computing}}},\ \Eprint
  {https://arxiv.org/abs/cond-mat/0508729} {arXiv:cond-mat/0508729}
  \BibitemShut {NoStop}%
\bibitem [{\citenamefont {Glazman}\ and\ \citenamefont
  {Catelani}(2021)}]{GlazmanSPPLN2021}%
  \BibitemOpen
  \bibfield  {author} {\bibinfo {author} {\bibfnamefont {L.~I.}\ \bibnamefont
  {Glazman}}\ and\ \bibinfo {author} {\bibfnamefont {G.}~\bibnamefont
  {Catelani}},\ }\bibfield  {title} {\bibinfo {title} {{Bogoliubov
  quasiparticles in superconducting qubits}},\ }\href
  {https://doi.org/10.21468/SciPostPhysLectNotes.31} {\bibfield  {journal}
  {\bibinfo  {journal} {SciPost Phys. Lect. Notes}\ }\textbf {\bibinfo {volume}
  {31}} (\bibinfo {year} {2021})}\BibitemShut {NoStop}%
\bibitem [{\citenamefont {Machlup}(1954)}]{MachlupJAP1954}%
  \BibitemOpen
  \bibfield  {author} {\bibinfo {author} {\bibfnamefont {S.}~\bibnamefont
  {Machlup}},\ }\bibfield  {title} {\bibinfo {title} {{Noise in Semiconductors:
  Spectrum of a Two‐Parameter Random Signal}},\ }\href
  {https://doi.org/10.1063/1.1721637} {\bibfield  {journal} {\bibinfo
  {journal} {J. Appl. Phys.}\ }\textbf {\bibinfo {volume} {25}},\ \bibinfo
  {pages} {341} (\bibinfo {year} {1954})}\BibitemShut {NoStop}%
\bibitem [{\citenamefont {Paladino}\ \emph {et~al.}(2014)\citenamefont
  {Paladino}, \citenamefont {Galperin}, \citenamefont {Falci},\ and\
  \citenamefont {Altshuler}}]{PaladinoRMP2014}%
  \BibitemOpen
  \bibfield  {author} {\bibinfo {author} {\bibfnamefont {E.}~\bibnamefont
  {Paladino}}, \bibinfo {author} {\bibfnamefont {Y.~M.}\ \bibnamefont
  {Galperin}}, \bibinfo {author} {\bibfnamefont {G.}~\bibnamefont {Falci}},\
  and\ \bibinfo {author} {\bibfnamefont {B.~L.}\ \bibnamefont {Altshuler}},\
  }\bibfield  {title} {\bibinfo {title} {{$1/f$ noise: Implications for
  solid-state quantum information}},\ }\href
  {https://doi.org/10.1103/RevModPhys.86.361} {\bibfield  {journal} {\bibinfo
  {journal} {Rev. Mod. Phys.}\ }\textbf {\bibinfo {volume} {86}},\ \bibinfo
  {pages} {361} (\bibinfo {year} {2014})}\BibitemShut {NoStop}%
\bibitem [{\citenamefont {M\"uller}\ \emph {et~al.}(2019)\citenamefont
  {M\"uller}, \citenamefont {Cole},\ and\ \citenamefont
  {Lisenfeld}}]{MullerRPP2019}%
  \BibitemOpen
  \bibfield  {author} {\bibinfo {author} {\bibfnamefont {C.}~\bibnamefont
  {M\"uller}}, \bibinfo {author} {\bibfnamefont {J.~H.}\ \bibnamefont {Cole}},\
  and\ \bibinfo {author} {\bibfnamefont {J.}~\bibnamefont {Lisenfeld}},\
  }\bibfield  {title} {\bibinfo {title} {{Towards understanding
  two-level-systems in amorphous solids: insights from quantum circuits}},\
  }\href {https://doi.org/10.1088/1361-6633/ab3a7e} {\bibfield  {journal}
  {\bibinfo  {journal} {Rep. Prog. Phys.}\ }\textbf {\bibinfo {volume} {82}},\
  \bibinfo {pages} {124501} (\bibinfo {year} {2019})}\BibitemShut {NoStop}%
\bibitem [{\citenamefont {Bylander}\ \emph {et~al.}(2011)\citenamefont
  {Bylander}, \citenamefont {Gustavsson}, \citenamefont {Yan}, \citenamefont
  {Yoshihara}, \citenamefont {Harrabi}, \citenamefont {Fitch}, \citenamefont
  {Cory}, \citenamefont {Nakamura}, \citenamefont {Tsai},\ and\ \citenamefont
  {Oliver}}]{BylanderNP2011}%
  \BibitemOpen
  \bibfield  {author} {\bibinfo {author} {\bibfnamefont {J.}~\bibnamefont
  {Bylander}}, \bibinfo {author} {\bibfnamefont {S.}~\bibnamefont
  {Gustavsson}}, \bibinfo {author} {\bibfnamefont {F.}~\bibnamefont {Yan}},
  \bibinfo {author} {\bibfnamefont {F.}~\bibnamefont {Yoshihara}}, \bibinfo
  {author} {\bibfnamefont {K.}~\bibnamefont {Harrabi}}, \bibinfo {author}
  {\bibfnamefont {G.}~\bibnamefont {Fitch}}, \bibinfo {author} {\bibfnamefont
  {D.~G.}\ \bibnamefont {Cory}}, \bibinfo {author} {\bibfnamefont
  {Y.}~\bibnamefont {Nakamura}}, \bibinfo {author} {\bibfnamefont {J.-S.}\
  \bibnamefont {Tsai}},\ and\ \bibinfo {author} {\bibfnamefont {W.~D.}\
  \bibnamefont {Oliver}},\ }\bibfield  {title} {\bibinfo {title} {{Noise
  spectroscopy through dynamical decoupling with a superconducting flux
  qubit}},\ }\href {https://doi.org/10.1038/nphys1994} {\bibfield  {journal}
  {\bibinfo  {journal} {Nat. Phys.}\ }\textbf {\bibinfo {volume} {7}},\
  \bibinfo {pages} {565} (\bibinfo {year} {2011})}\BibitemShut {NoStop}%
\bibitem [{\citenamefont {Ikram}\ \emph {et~al.}(2007)\citenamefont {Ikram},
  \citenamefont {Li},\ and\ \citenamefont {Zubairy}}]{IkramPRA2007}%
  \BibitemOpen
  \bibfield  {author} {\bibinfo {author} {\bibfnamefont {M.}~\bibnamefont
  {Ikram}}, \bibinfo {author} {\bibfnamefont {F.-l.}\ \bibnamefont {Li}},\ and\
  \bibinfo {author} {\bibfnamefont {M.~S.}\ \bibnamefont {Zubairy}},\
  }\bibfield  {title} {\bibinfo {title} {{Disentanglement in a two-qubit system
  subjected to dissipation environments}},\ }\href
  {https://doi.org/10.1103/PhysRevA.75.062336} {\bibfield  {journal} {\bibinfo
  {journal} {Phys. Rev. A}\ }\textbf {\bibinfo {volume} {75}},\ \bibinfo
  {pages} {062336} (\bibinfo {year} {2007})}\BibitemShut {NoStop}%
\bibitem [{\citenamefont {Leggett}\ \emph {et~al.}(1987)\citenamefont
  {Leggett}, \citenamefont {Chakravarty}, \citenamefont {Dorsey}, \citenamefont
  {Fisher}, \citenamefont {Garg},\ and\ \citenamefont
  {Zwerger}}]{LeggettRMP87A}%
  \BibitemOpen
  \bibfield  {author} {\bibinfo {author} {\bibfnamefont {A.~J.}\ \bibnamefont
  {Leggett}}, \bibinfo {author} {\bibfnamefont {S.}~\bibnamefont
  {Chakravarty}}, \bibinfo {author} {\bibfnamefont {A.~T.}\ \bibnamefont
  {Dorsey}}, \bibinfo {author} {\bibfnamefont {M.~P.~A.}\ \bibnamefont
  {Fisher}}, \bibinfo {author} {\bibfnamefont {A.}~\bibnamefont {Garg}},\ and\
  \bibinfo {author} {\bibfnamefont {W.}~\bibnamefont {Zwerger}},\ }\bibfield
  {title} {\bibinfo {title} {{Dynamics of the dissipative two-state system}},\
  }\href {https://doi.org/10.1103/RevModPhys.59.1} {\bibfield  {journal}
  {\bibinfo  {journal} {Rev. Mod. Phys.}\ }\textbf {\bibinfo {volume} {59}},\
  \bibinfo {pages} {1} (\bibinfo {year} {1987})}\BibitemShut {NoStop}%
\bibitem [{\citenamefont {Wang}\ \emph {et~al.}(2019)\citenamefont {Wang},
  \citenamefont {Wu},\ and\ \citenamefont {Wang}}]{WangPRA2019}%
  \BibitemOpen
  \bibfield  {author} {\bibinfo {author} {\bibfnamefont {Z.}~\bibnamefont
  {Wang}}, \bibinfo {author} {\bibfnamefont {W.}~\bibnamefont {Wu}},\ and\
  \bibinfo {author} {\bibfnamefont {J.}~\bibnamefont {Wang}},\ }\bibfield
  {title} {\bibinfo {title} {{Steady-state entanglement and coherence of two
  coupled qubits in equilibrium and nonequilibrium environments}},\ }\href
  {https://doi.org/10.1103/PhysRevA.99.042320} {\bibfield  {journal} {\bibinfo
  {journal} {Phys. Rev. A}\ }\textbf {\bibinfo {volume} {99}},\ \bibinfo
  {pages} {042320} (\bibinfo {year} {2019})}\BibitemShut {NoStop}%
\bibitem [{\citenamefont {Chuang}\ and\ \citenamefont
  {Nielsen}(1997)}]{ChuangJMOpt1997}%
  \BibitemOpen
  \bibfield  {author} {\bibinfo {author} {\bibfnamefont {I.~L.}\ \bibnamefont
  {Chuang}}\ and\ \bibinfo {author} {\bibfnamefont {M.~A.}\ \bibnamefont
  {Nielsen}},\ }\bibfield  {title} {\bibinfo {title} {{Prescription for
  experimental determination of the dynamics of a quantum black box}},\ }\href
  {https://doi.org/10.1080/09500349708231894} {\bibfield  {journal} {\bibinfo
  {journal} {J. Mod. Opt.}\ }\textbf {\bibinfo {volume} {44}},\ \bibinfo
  {pages} {2455} (\bibinfo {year} {1997})}\BibitemShut {NoStop}%
\bibitem [{\citenamefont {Schuch}\ and\ \citenamefont
  {Siewert}(2003)}]{SchuchPRA2003}%
  \BibitemOpen
  \bibfield  {author} {\bibinfo {author} {\bibfnamefont {N.}~\bibnamefont
  {Schuch}}\ and\ \bibinfo {author} {\bibfnamefont {J.}~\bibnamefont
  {Siewert}},\ }\bibfield  {title} {\bibinfo {title} {{Natural two-qubit gate
  for quantum computation using the $\mathrm{XY}$ interaction}},\ }\href
  {https://doi.org/10.1103/PhysRevA.67.032301} {\bibfield  {journal} {\bibinfo
  {journal} {Phys. Rev. A}\ }\textbf {\bibinfo {volume} {67}},\ \bibinfo
  {pages} {032301} (\bibinfo {year} {2003})}\BibitemShut {NoStop}%
\bibitem [{\citenamefont {McKay}\ \emph {et~al.}(2017)\citenamefont {McKay},
  \citenamefont {Wood}, \citenamefont {Sheldon}, \citenamefont {Chow},\ and\
  \citenamefont {Gambetta}}]{McKayPRA2017}%
  \BibitemOpen
  \bibfield  {author} {\bibinfo {author} {\bibfnamefont {D.~C.}\ \bibnamefont
  {McKay}}, \bibinfo {author} {\bibfnamefont {C.~J.}\ \bibnamefont {Wood}},
  \bibinfo {author} {\bibfnamefont {S.}~\bibnamefont {Sheldon}}, \bibinfo
  {author} {\bibfnamefont {J.~M.}\ \bibnamefont {Chow}},\ and\ \bibinfo
  {author} {\bibfnamefont {J.~M.}\ \bibnamefont {Gambetta}},\ }\bibfield
  {title} {\bibinfo {title} {{Efficient $Z$ gates for quantum computing}},\
  }\href {https://doi.org/10.1103/PhysRevA.96.022330} {\bibfield  {journal}
  {\bibinfo  {journal} {Phys. Rev. A}\ }\textbf {\bibinfo {volume} {96}},\
  \bibinfo {pages} {022330} (\bibinfo {year} {2017})}\BibitemShut {NoStop}%
\bibitem [{\citenamefont {Breuer}\ \emph {et~al.}(2016)\citenamefont {Breuer},
  \citenamefont {Laine}, \citenamefont {Piilo},\ and\ \citenamefont
  {Vacchini}}]{BreuerRMP2016}%
  \BibitemOpen
  \bibfield  {author} {\bibinfo {author} {\bibfnamefont {H.-P.}\ \bibnamefont
  {Breuer}}, \bibinfo {author} {\bibfnamefont {E.-M.}\ \bibnamefont {Laine}},
  \bibinfo {author} {\bibfnamefont {J.}~\bibnamefont {Piilo}},\ and\ \bibinfo
  {author} {\bibfnamefont {B.}~\bibnamefont {Vacchini}},\ }\bibfield  {title}
  {\bibinfo {title} {{Colloquium: Non-Markovian dynamics in open quantum
  systems}},\ }\href {https://doi.org/10.1103/RevModPhys.88.021002} {\bibfield
  {journal} {\bibinfo  {journal} {Rev. Mod. Phys.}\ }\textbf {\bibinfo {volume}
  {88}},\ \bibinfo {pages} {021002} (\bibinfo {year} {2016})}\BibitemShut
  {NoStop}%
\bibitem [{\citenamefont {Rivas}\ \emph {et~al.}(2014)\citenamefont {Rivas},
  \citenamefont {Huelga},\ and\ \citenamefont {Plenio}}]{RivasRPP2014}%
  \BibitemOpen
  \bibfield  {author} {\bibinfo {author} {\bibfnamefont {{\'A}.}~\bibnamefont
  {Rivas}}, \bibinfo {author} {\bibfnamefont {S.~F.}\ \bibnamefont {Huelga}},\
  and\ \bibinfo {author} {\bibfnamefont {M.~B.}\ \bibnamefont {Plenio}},\
  }\bibfield  {title} {\bibinfo {title} {{Quantum non-Markovianity:
  characterization, quantification and detection}},\ }\href
  {https://doi.org/10.1088/0034-4885/77/9/094001} {\bibfield  {journal}
  {\bibinfo  {journal} {Rep. Prog. Phys.}\ }\textbf {\bibinfo {volume} {77}},\
  \bibinfo {pages} {094001} (\bibinfo {year} {2014})}\BibitemShut {NoStop}%
\bibitem [{\citenamefont {Wilhelm}\ \emph {et~al.}()\citenamefont {Wilhelm},
  \citenamefont {Kirchhoff}, \citenamefont {Machnes}, \citenamefont {Wittler},\
  and\ \citenamefont {Sugny}}]{WilhelmARXIV2020}%
  \BibitemOpen
  \bibfield  {author} {\bibinfo {author} {\bibfnamefont {F.~K.}\ \bibnamefont
  {Wilhelm}}, \bibinfo {author} {\bibfnamefont {S.}~\bibnamefont {Kirchhoff}},
  \bibinfo {author} {\bibfnamefont {S.}~\bibnamefont {Machnes}}, \bibinfo
  {author} {\bibfnamefont {N.}~\bibnamefont {Wittler}},\ and\ \bibinfo {author}
  {\bibfnamefont {D.}~\bibnamefont {Sugny}},\ }\href@noop {} {\bibinfo {title}
  {{An introduction into optimal control for quantum technologies}}},\ \Eprint
  {https://arxiv.org/abs/2003.10132} {arXiv:2003.10132} \BibitemShut {NoStop}%
\bibitem [{\citenamefont {Tanimura}(2014)}]{Tanimura2014}%
  \BibitemOpen
  \bibfield  {author} {\bibinfo {author} {\bibfnamefont {Y.}~\bibnamefont
  {Tanimura}},\ }\bibfield  {title} {\bibinfo {title} {{Reduced hierarchical
  equations of motion in real and imaginary time: Correlated initial states and
  thermodynamic quantities}},\ }\href {https://doi.org/10.1063/1.4890441}
  {\bibfield  {journal} {\bibinfo  {journal} {J. Chem. Phys.}\ }\textbf
  {\bibinfo {volume} {141}},\ \bibinfo {pages} {044114} (\bibinfo {year}
  {2014})}\BibitemShut {NoStop}%
\bibitem [{\citenamefont {Nakamura}\ and\ \citenamefont
  {Tanimura}(2018)}]{Nakamura18PRA}%
  \BibitemOpen
  \bibfield  {author} {\bibinfo {author} {\bibfnamefont {K.}~\bibnamefont
  {Nakamura}}\ and\ \bibinfo {author} {\bibfnamefont {Y.}~\bibnamefont
  {Tanimura}},\ }\bibfield  {title} {\bibinfo {title} {{Hierarchical
  Schr\"odinger equations of motion for open quantum dynamics}},\ }\href
  {https://doi.org/10.1103/PhysRevA.98.012109} {\bibfield  {journal} {\bibinfo
  {journal} {Phys. Rev. A}\ }\textbf {\bibinfo {volume} {98}},\ \bibinfo
  {pages} {012109} (\bibinfo {year} {2018})}\BibitemShut {NoStop}%
\bibitem [{\citenamefont {Nakamura}\ and\ \citenamefont
  {Tanimura}(2021)}]{NakamuraJCP2021}%
  \BibitemOpen
  \bibfield  {author} {\bibinfo {author} {\bibfnamefont {K.}~\bibnamefont
  {Nakamura}}\ and\ \bibinfo {author} {\bibfnamefont {Y.}~\bibnamefont
  {Tanimura}},\ }\bibfield  {title} {\bibinfo {title} {{Optical response of
  laser-driven charge-transfer complex described by Holstein–Hubbard model
  coupled to heat baths: Hierarchical equations of motion approach}},\ }\href
  {https://doi.org/10.1063/5.0060208} {\bibfield  {journal} {\bibinfo
  {journal} {J. Chem. Phys.}\ }\textbf {\bibinfo {volume} {155}},\ \bibinfo
  {pages} {064106} (\bibinfo {year} {2021})}\BibitemShut {NoStop}%
\bibitem [{\citenamefont {Li}\ \emph {et~al.}(2010)\citenamefont {Li},
  \citenamefont {Zou},\ and\ \citenamefont {Shao}}]{LiPRA2010}%
  \BibitemOpen
  \bibfield  {author} {\bibinfo {author} {\bibfnamefont {J.-G.}\ \bibnamefont
  {Li}}, \bibinfo {author} {\bibfnamefont {J.}~\bibnamefont {Zou}},\ and\
  \bibinfo {author} {\bibfnamefont {B.}~\bibnamefont {Shao}},\ }\bibfield
  {title} {\bibinfo {title} {{Entanglement evolution of two qubits under noisy
  environments}},\ }\href {https://doi.org/10.1103/PhysRevA.82.042318}
  {\bibfield  {journal} {\bibinfo  {journal} {Phys. Rev. A}\ }\textbf {\bibinfo
  {volume} {82}},\ \bibinfo {pages} {042318} (\bibinfo {year}
  {2010})}\BibitemShut {NoStop}%
\bibitem [{\citenamefont {Nakatsukasa}\ \emph {et~al.}(2018)\citenamefont
  {Nakatsukasa}, \citenamefont {S\`{e}te},\ and\ \citenamefont
  {Trefethen}}]{NakatsukasaSIAM2018}%
  \BibitemOpen
  \bibfield  {author} {\bibinfo {author} {\bibfnamefont {Y.}~\bibnamefont
  {Nakatsukasa}}, \bibinfo {author} {\bibfnamefont {O.}~\bibnamefont
  {S\`{e}te}},\ and\ \bibinfo {author} {\bibfnamefont {L.~N.}\ \bibnamefont
  {Trefethen}},\ }\bibfield  {title} {\bibinfo {title} {{The AAA Algorithm for
  Rational Approximation}},\ }\href {https://doi.org/10.1137/16M1106122}
  {\bibfield  {journal} {\bibinfo  {journal} {SIAM J. Sci. Comput.}\ }\textbf
  {\bibinfo {volume} {40}},\ \bibinfo {pages} {A1494} (\bibinfo {year}
  {2018})}\BibitemShut {NoStop}%
\bibitem [{\citenamefont {Ikeda}\ and\ \citenamefont
  {Scholes}(2020)}]{IkedaJCP2020}%
  \BibitemOpen
  \bibfield  {author} {\bibinfo {author} {\bibfnamefont {T.}~\bibnamefont
  {Ikeda}}\ and\ \bibinfo {author} {\bibfnamefont {G.~D.}\ \bibnamefont
  {Scholes}},\ }\bibfield  {title} {\bibinfo {title} {{Generalization of the
  hierarchical equations of motion theory for efficient calculations with
  arbitrary correlation functions}},\ }\href
  {https://doi.org/10.1063/5.0007327} {\bibfield  {journal} {\bibinfo
  {journal} {J. Chem. Phys.}\ }\textbf {\bibinfo {volume} {152}},\ \bibinfo
  {pages} {204101} (\bibinfo {year} {2020})}\BibitemShut {NoStop}%
\bibitem [{\citenamefont {Lubich}\ \emph {et~al.}(2015)\citenamefont {Lubich},
  \citenamefont {Oseledets},\ and\ \citenamefont
  {Vandereycken}}]{LubichSIAMJNA2015}%
  \BibitemOpen
  \bibfield  {author} {\bibinfo {author} {\bibfnamefont {C.}~\bibnamefont
  {Lubich}}, \bibinfo {author} {\bibfnamefont {I.~V.}\ \bibnamefont
  {Oseledets}},\ and\ \bibinfo {author} {\bibfnamefont {B.}~\bibnamefont
  {Vandereycken}},\ }\bibfield  {title} {\bibinfo {title} {{Time Integration of
  Tensor Trains}},\ }\href {https://doi.org/10.1137/140976546} {\bibfield
  {journal} {\bibinfo  {journal} {SIAM J. Numer. Anal.}\ }\textbf {\bibinfo
  {volume} {53}},\ \bibinfo {pages} {917} (\bibinfo {year} {2015})}\BibitemShut
  {NoStop}%
\end{thebibliography}%
\end{document}